\documentclass[useAMS,usenatbib]{mn2e}
\usepackage{times}
\usepackage{graphicx}
\usepackage{psfig}
\usepackage{fancyhdr,rotating,aas_macros,amssymb,amsmath}

\newcommand{\valqrwind}{\mbox{$0.015$}}	
\newcommand{\valqrshell}{\mbox{$0.200$}}

\newcommand{\valkonewind}{\mbox{$0.014$}}
\newcommand{\valktwowind}{\mbox{$2.152$}}
\newcommand{\valkoneshell}{\mbox{$0.223$}}
\newcommand{\valktwoshell}{\mbox{$0.925$}}

\newcommand{\lya}{\mbox{$\rmn{Ly}\alpha$}}
\newcommand{\lyaint}{\mbox{$L_{Ly\alpha,0}$}}
\newcommand{\ha}{\mbox{$\rmn{H}\alpha$}}
\newcommand{\galform}{\texttt{GALFORM}}

\newcommand{\mocalata}{\texttt{MoCaLaTA}}

\newcommand{\fesc}{\mbox{$\rm{f}_{esc}$}}

\newcommand{\llya}{\mbox{$L_{\rmn{Ly}\alpha}$}}						%Lum. abbreviation
\newcommand{\lunits}{\mbox{$[\rmn{erg} \ \rmn{s}^{-1} \ \rmn{h}^{-2}]$}}		%Luminosity units
\newcommand{\logllya}{\mbox{${\rm log}(L_{Ly\alpha}\lunits)$}}
\newcommand{\vect}[1]{\boldsymbol{#1}}

\title[Can galactic outflows explain \lya\ emitters?]
{Can galactic outflows explain the properties of \lya\ emitters?} 
\author [A. Orsi et al.]
{Alvaro Orsi$^{1,2,3}$\thanks{Email: aaorsi@astro.puc.cl},  Cedric G. Lacey$^3$ and Carlton M. Baugh$^3$\\
1. Departamento de Astronom\'ia y Astrof\'isica, Pontificia Universidad Cat\'olica, Av. Vicu\~na Mackenna 4860, Santiago, Chile. \\
2. Centro de Astro-Ingenier\'ia, Pontificia Universidad Cat\'olica, Av. Vicu\~na Mackenna 4860, Santiago, Chile.\\
3. Institute for Computational Cosmology, Department of Physics, University of Durham, South Road, Durham DH1 3LE, UK. \\
}
\begin{document}
\maketitle
\begin{abstract}
We study the properties of \lya\ emitters in a cosmological framework by computing the escape of \lya\ photons 
through galactic outflows. We combine 
the \galform\ semi-analytical model of galaxy formation with a Monte Carlo \lya\ radiative transfer code. 
The properties of \lya\ emitters at $0<z<7$ are predicted using two outflow geometries: 
a Shell of neutral gas and a Wind ejecting material, both expanding at constant velocity.
We characterise the differences in the \lya\ line profiles predicted by the two outflow geometries in terms 
of their width, asymmetry and shift from the line centre for a set of outflows with different
hydrogen column densities, expansion velocities and metallicities. 
In general, the \lya\ line profile of the Shell geometry is broader and more asymmetric, 
and the \lya\ escape fraction is lower than with the Wind geometry for the same set of parameters. 
In order to implement the outflow geometries in the semi-analytical model \galform,
a number of free parameters in the outflow model are set by matching the luminosity function of \lya\ 
emitters over the whole observed redshift range. 
The resulting neutral hydrogen column densities of the outflows for 
observed \lya\ emitters are predicted to be in the range $\sim 10^{18}-10^{23} [{\rm cm^{-2}}]$.
The models are consistent with the observationally inferred \lya\ escape fractions, 
equivalent width distributions and with the shape of the \lya\ line from composite spectra. 
Interestingly, our predicted UV luminosity function of \lya\ emitters and the fraction 
of \lya\ emitters in Lyman-break galaxy samples at high redshift are in partial agreement
with observations. Attenuation of the \lya\ line by the presence of a neutral intergalactic 
medium at high redshift could be responsible for this disagreement.
We predict that \lya\ emitters constitute a subset of the galaxy population with 
lower metallicities, lower instantaneous star formation rates and larger sizes than the overall
population at the same UV luminosity.

\end{abstract}


\section{Introduction}

Over the past 10 years the \lya\ line has proved to be a successful tracer of  galaxies in the redshift 
range $2<z<7$ \citep[e.g. ][]{cowie98,kudritzki00,rhoads00,hu02,gronwall07,ouchi08,nilsson09b,shimasaku06,
kashikawa06,hu10,guaita10}.
More recently, samples of \lya\ emitters at $z\sim 0.2$ obtained with the GALEX satellite
\citep{deharveng08,cowie10}, have allowed us to study this galaxy population over an even broader range of redshifts.

Star-forming galaxies emit \lya\ radiation when ionizing photons produced by massive 
young stars are absorbed by atomic hydrogen (HI) regions in the interstellar medium (ISM).
These hydrogen atoms then recombine leading to the emission of \lya\ photons.  
Therefore, \lya\ emission is, in principle, closely related to the star formation rate (SFR) of galaxies. 
However, in general only a small fraction of \lya\ photons manage to escape from galaxies \citep[e.g.][]{hayes11}. 
This makes it difficult to relate \lya\ emitters to other star forming galaxy populations at high redshift, such as Lyman-break galaxies (LBGs) 
or sub-millimetre galaxies (SMGs).

The physical properties of galaxies selected by their \lya\ emission are inferred
from spectral and photometric data 
\citep{gawiser07,gronwall07,nilsson09b,guaita11}. Furthermore, \lya\ emitters are currently used to
study the kinematics of the ISM in high redshift galaxies \citep{shapley03,steidel10,steidel11,kulas11},
to trace the large scale structure of the Universe 
\citep{shimasaku06,gawiser07,kovac07,orsi08,francke09,ouchi10}, to constrain the epoch of
reionization  \citep{kashikawa06,dayal11,ouchi10,stark10,schenker11,pentericci11} 
and to test galaxy formation models \citep{ledelliou05,ledelliou06,kobayashi07,nagamine10,dayal10a}.

Despite this progress, understanding the physical mechanisms which drive the escape of \lya\ 
radiation from a galaxy remains a challenge.
\lya\ photons undergo resonant scattering when interacting with hydrogen atoms, 
resulting in an increase of the path length that photons need to travel before escaping the medium. Therefore, the probability
of photons being absorbed by dust grains is greatly enhanced, making the escape of \lya\ photons very sensitive to even small amounts of dust. 
Furthermore, as a result of the complex radiative transfer, the frequency of the escaping \lya\ photons generally departs from 
the line centre as a consequence of the large number of scatterings with many hydrogen atoms. 

Recent observational studies have used various methods to infer the escape fraction of \lya\ photons, \fesc\ 
\citep{atek08,atek09,ostlin09,kornei10,hayes10b,hayes11}. 
This is generally done either by comparing the observed line ratio between \lya\ and other hydrogen 
recombination lines, such as \ha\ and H$\beta$, 
or by comparing the star formation rate derived 
from the \lya\ luminosity to that obtained from the ultraviolet continuum.
The first method is the more direct, since the intrinsic fluxes of the comparison lines can be inferred 
after correcting the observed fluxes for extinction.
Then, the departure from case B recombination of the ratio of the \lya\ intensity to another hydrogen recombination line 
is attributed to the escape fraction of \lya\ differing from unity.
The second method, on the other hand, relies heavily on the assumed stellar population model 
used, the choice of the stellar initial mass function (IMF) and the attenuation of the ultraviolet continuum by dust,
and is therefore more uncertain.

These measurements have revealed that the escape fraction of \lya\ emitters can 
be anything between $10^{-3}$ and $1$. The observational data
listed above also suggest a correlation between the value of the escape fraction and the dust extinction.
The large scatter found in this relation suggests there is a range of physical parameters which determine the value of \fesc. 

Early theoretical models of \lya\ emission from galaxies were based on a static ISM \citep[see, e.g.][]{neufeld90,charlot93}. 
These models explained the difficulty of observing \lya\ in emission due to its very high sensitivity to dust in such a medium.
Moreover, the first observations of \lya\ emission in local starburst dwarf galaxies suggested a strong correlation between metallicity and \lya\ luminosity 
\citep{meier81,hartmann84,hartmann88}, leading to the conclusion that 
metallicity, which supposedly traces the amount of dust in galaxies, 
is the most important factor driving the visibility of the \lya\ line.
 
However, subsequent observational studies showed only
a weak correlation between \lya\ luminosity and metallicity, suggesting instead the importance of the 
neutral gas distribution and its kinematics \citep[e.g., ][]{giavalisco96}. 
Further analysis of metal lines in local starbursts revealed the presence of 
outflows which allow the escape of \lya\ photons. The observed asymmetric P-Cygni \lya\ line profiles are 
consistent with \lya\ photons escaping from an expanding
shell of neutral gas  \citep{thuan97,kunth98,mas-hesse03}. This established outflows as the main mechanism responsible for
the escape of \lya\ photons from galaxies. Furthermore, observations at higher redshifts reveal \lya\ line profiles which also resemble 
those expected when photons escape through a galactic outflow \citep{shapley03,kashikawa06,kornei10,hu10}.
 
In the last few years there has been significant progress in the modelling 
of \lya\ emitters in a cosmological setting.
The first consistent hierarchical galaxy formation model which included \lya\ emission is the one described by
\citet{ledelliou05,ledelliou06} and \citet{orsi08}, which makes use of the \galform\ semi-analytical
model. In this model, the simple assumption of a fixed escape fraction, $\fesc = 0.02$, regardless
of any galaxy property or redshift, allowed us to predict remarkably well the abundances and clustering of \lya\ emitters
over a wide range of redshifts and luminosities.

\citet{nagamine06,nagamine10} modelled \lya\ emitters in cosmological SPH simulations.
In order to match the abundances of \lya\ emitters at different redshifts, they were forced to 
introduce a tunable escape fraction and a duty cycle parameter.
\citet{kobayashi07,kobayashi10} developed a simple phenomenological model to compute \fesc\ in a
semianalytical model. Their analytical prescription for \fesc\ distinguishes between outflows produced 
in starbursts and static media in quiescent galaxies. 
Dayal et. al use an SPH simulation to study \lya\ emitters at high redshift and their attenuation by
the neutral intergalactic medium (IGM). However, they assume the  \lya\ escape fraction is related to the escape 
of UV continuum photons \citep{dayal08,dayal10a,dayal11}.
\citet{tilvi09,tilvi10} make predictions for \lya\ emitters using an N-body simulation and 
the uncertain assumption that the \lya\ luminosity is proportional to the halo mass accretion rate.
More recently, \citet{forero-romero11} presented a model for high redshift \lya\ emitters based on 
a hydrodynamic simulation, approximating \lya\ photons to escape from 
homogeneous and clumpy gaseous, static slabs.

Motivated by observational evidence showing that \lya\ photons escape through outflows, we 
present a model that incorporates a more physical treatment of the \lya\ propagation than previous work, 
whilst at the same time being computationally efficient, so as to allow its application to a large sample of galaxies
at different redshifts.

Such a physical approach to modelling the escape of \lya\ photons requires a treatment of the radiative transfer 
of photons through an HI region. The scattering and destruction of \lya\ photons have been extensively studied 
due to their many applications in astrophysical media. \citet{harrington73} studied analytically the emergent 
spectrum from an optically thick, homogeneous static 
slab with photons generated at the line centre.
This result was generalised by \citet{neufeld90}, to include photons generated at any 
frequency, and to provide an analytical expression for \fesc\ in this configuration.

Numerical methods, on the other hand, allow us to study the line profiles and escape fractions of \lya\ photons in a wider 
variety of configurations.
The standard approach is to use a Monte Carlo algorithm, in which the paths of a set of photons are followed 
one at a time through many scattering
events, until the photon either escapes or is absorbed by a dust grain. 
Such calculations have been applied successfully to study the properties of 
\lya\ emitters in different 
scenarios \citep[see, e.g.][]{ahn00,zheng02,ahn03,ahn04,verhamme06,dijkstra06,laursen07,laursen09a}.
 
Recently, \citet{zheng10,zheng11} combined a 
Monte Carlo \lya\ radiative transfer model with a cosmological reionization simulation at $z\sim 6$, obtaining 
extended \lya\ emission due to spatial diffusion. Their simulation box is, however, too small to be evolved to 
the present day without density fluctuations becoming nonlinear on the box scale, 
and does not have a volume large enough to sample a wide range of environments.

Previous work has not studied \lya\ emitters in a framework 
that at the same time spans the galaxy formation and evolution process over a broad range of redshifts 
and includes \lya\ radiative transfer. 
The need for such hybrid approach is precisely the motivation for this paper.

Given the observational evidence that outflows facilitate the escape of \lya\ photons from galaxies, 
here we study the nature of \lya\ emitters by computing the escape of photons from galaxies 
in an outflow of material by using a Monte Carlo \lya\ radiative transfer model. 
Galactic outflows in our model are defined
according to predicted galaxy properties in a simple way. This makes our modelling feasible on a 
cosmological scale, whilst retaining all the complexity of \lya\ radiative transfer. 
Following our previous work, we use the semianalytical model \galform. This paper represents a  
significant improvement over the treatment of \lya\ emitters in hierarchical galaxy formation models 
initially described in \citet{ledelliou05,ledelliou06} and \citet{orsi08}, which all assumed
a constant escape fraction.

The outline of this paper is as follows: Section \ref{sec.model} describes the galaxy formation 
and radiative transfer models used. Also, we describe the two 
versions of galactic outflow geometries we use and explain how these are constructed in terms
of galaxy parameters that can be extracted from our semi-analytical model. 
In Section \ref{sec.comparison} we present the properties of the \lya\ line profiles 
and escape fractions predicted by our outflow geometries. In Section \ref{sec.combining}
we describe the galaxy properties that are relevant to our \lya\ modelling and describe how we 
combine the \lya\ radiative transfer model with the semi-analytical model. 
In Section \ref{sec.results} we present our main results, where we compare with 
observational measurements whenever possible, spanning the redshift range $0<z<7$.
In Section \ref{sec.nature} we present the implications of our modelling for 
the properties of galaxies selected by their \lya\ emission, compared
to the bulk of the galaxy population. Finally, in Section \ref{sec.conclusions} 
we discuss our main findings. 
In the Appendix we give details of the implementation of the Monte Carlo radiative 
transfer model, the effect of the UV background on the outflow geometries and the 
numerical strategy followed to compute
the escape fraction of \lya\ photons from galaxies.

\section{Model description}
\label{sec.model}

Our approach to modelling \lya\ emitters in a 
cosmological framework involves combining two independent
codes. The backbone of our calculation is the \galform\ semi-analytical model of 
galaxy formation, outlined in Section \ref{sec.galform}, from which all relevant galaxy properties can be obtained, 
including the intrinsic \lya\ luminosity. The second is
the Monte Carlo \lya\ radiative transfer code to compute both 
the frequency distribution of \lya\ photons and the \lya\ escape fraction.
This code is described briefly in Section \ref{sec.MCintro}, with further
information and tests presented in Appendix \ref{app.lyart}.
In Section \ref{sec.outflows} we describe the outflow geometries and the way
these are defined in terms of galaxy properties.

\subsection{Galaxy formation model}
\label{sec.galform}
We use the semi-analytical model of galaxy formation, \galform, to
predict the properties of galaxies and their
abundance as a function of redshift. The \galform\ model is fully
described in \citet{cole00} and \citet{benson03} \citep[see also the review by][]{baugh06}.
The variant used here was introduced by \citet{baugh05}, and is also 
described in detail in \citet{lacey08}.  
The model computes
star formation and galaxy merger histories for the whole galaxy population, following
the hierarchical evolution of the host dark matter haloes.

The \citet{baugh05} model used here is the same \galform\ variant used in our 
previous work on \lya\ emitters \citep{ledelliou05,ledelliou06,orsi08}. 
A critical assumption of the 
\citet{baugh05} model is that stars which form
in starbursts have a top-heavy initial mass function (IMF). The
IMF is given by ${\rm d}N/{\rm d} \ln (m) \propto m^{-x}$ with
$x=0$ in this case. Stars formed quiescently in discs have a solar neighbourhood
IMF, with the form proposed by \citet{kennicutt83} with $x=0.4$ for $m < 1
M_\odot$ and $x = 1.5$ for $m> 1 M_\odot$. Both IMFs cover the mass
range $0.15 M_\odot < m < 125 M_\odot$. Within the framework of
$\Lambda$CDM, Baugh et~al.  argued that the top-heavy IMF is essential
to match the counts and redshift distribution of galaxies detected
through their sub-millimetre emission, whilst retaining the match to
galaxy properties in the local Universe, such as the optical and
far-IR luminosity functions, galaxy gas fractions and
metallicities. \citet{lacey08} showed that the 
model predicts galaxy evolution in the IR in good agreement with
 observations from {\em Spitzer}. Moreover, \citet{lacey11} 
also showed that the Baugh et~al. model predicts the 
abundance of Lyman-break galaxies (LBGs) remarkably well 
over the redshift range $3<z<10$ \citep[see also][]{gonzalez11}.

In \galform, the suppression of gas cooling from ionizing radiation produced by stars
and active galactic nuclei (AGNs) during the epoch of reionization is modelled
in a simple way: after the redshift of reionization, taken to be $z_{\rm reion} = 10$, 
photoionization of the IGM completely suppresses the cooling and collapse
of gas in haloes with circular velocities below $V_{\rm crit} = 30$[km/s].

The Baugh et~al. model assumes two distinct modes of feedback
from supernovae, a {\it reheating mode},  in which cold gas 
is expelled back to the hot halo, and a {\it superwind mode}, in which cold gas is ejected out of the galaxy halo. 
We describe both modes of supernova feedback in more detail in section \ref{sec.wind} \citep[see also][]{lacey08}.

Unlike the \citet{bower06} variant of \galform, the model used here does not 
incorporate feedback from an AGN. The superwind mode of feedback 
produces similar consequences to the 
quenching of gas cooling by the action of an AGN, 
as both mechanisms suppress the bright end of the LF. However, the \citet{bower06} model does not
reproduce the observed abundances of LBGs or submillimetre galaxies (SMGs) at high redshift, 
which span a redshift range which overlaps with that of the LFs 
of \lya\ emitters considered in this paper. 
Therefore, we do not use this \galform\ variant to make predictions for \lya\ emitters.

In \galform, the intrinsic \lya\ emission is computed as follows: 

(i) The composite stellar spectrum of the galaxy is calculated, based on
its predicted star formation history, including the effect of the metallicity with which new stars 
are formed, and taking into account the IMFs adopted for
different modes of star formation. 

(ii) The rate of production of
Lyman continuum (Lyc) photons is computed by integrating over the
composite stellar spectrum. We assume that all of these ionizing photons are
absorbed by neutral hydrogen within the galaxy. 
The mean number of \lya\ photons produced per Lyc photon, 
assuming a gas temperature of $10^4$ K for the ionized gas, is approximately 
0.677, assuming Case B recombination \citep{osterbrock89}.

(iii) The observed \lya\ flux depends on the fraction of \lya\ photons which escape from
the galaxy ($f_{\rm{esc}}$). Previously we
assumed this to be constant and
independent of galaxy properties. 
The escape fraction was fixed at $\fesc = 0.02$, resulting in a remarkably good match 
to the observed \lya\ LFs over the redshift range $3<z<6$.
In our new model, we make use of a Monte Carlo radiative transfer model of \lya\ photons to compute \fesc\ in a more physical way. 
This model is briefly described in the next subsection.

\subsection{Monte Carlo Radiative Transfer Model}
\label{sec.MCintro}
In order to compute the \lya\ escape fraction and line profile we construct 
a Monte Carlo radiative transfer model for \lya\ photons. This simulates the 
escape of a set of photons from a source as they travel through an expanding 
HI region, which may contain dust, by following the scattering histories of individual photons. 
By following a large number of photons we can 
compute properties such as the \lya\ spectrum and the escape fraction.

Our Monte Carlo radiative transfer code works on a 3D grid in which each cell 
contains information about the neutral hydrogen density, $n_H$, the temperature 
of the gas, $T$, and the bulk velocity, $v_{\rm bulk}$.
Once a \lya\ photon is created, a random direction and frequency are assigned to it, 
and the code follows its trajectory and computes each scattering event of the photon 
as it crosses the HI region until it either escapes or is 
absorbed by a dust grain.  If the photon escapes, then its final frequency is recorded, 
a new photon is generated and the procedure is repeated.

Clearly, to get an accurate description of the escape of \lya\ photons from a given 
geometrical setup, many photons must be followed. The number of photons needed 
to achieve convergence will depend on the properties of the medium, but also on the 
quantity in which we are interested. For example, for most of the outflows studied 
here, only a few thousand  photons are needed to compute an accurate escape
fraction. However, tens of thousand of photons are needed to obtain a smooth line profile.

Our radiative transfer model of \lya\ photons is similar to previous models in the literature
\citep[e.g.][]{zheng02,ahn03,ahn04,verhamme06,dijkstra06,laursen09a,barnes10}. 
We describe the numerical implementation of our Monte Carlo model 
and the validation tests applied to it in Appendix \ref{app.lyart}.

\subsection{Outflow geometries}
\label{sec.outflows}
In our model, the physical conditions in the medium used to compute the escape of \lya\ photons 
depend on several properties of galaxies
predicted by \galform. Below we describe two 
outflow geometries for the HI region surrounding the source of \lya\ photons.
They differ in their geometry and the way
the properties of galaxies from \galform\ are used. We assume the temperature
of the medium to be constant at $T = 10^4$ K, which sets the thermal velocity dispersion of atoms, 
following a Maxwell-Boltzmann distribution, to be $v_{\rm th} = 12.85 \ {\rm km s^{-1}}$
(see Eq. \ref{eq.vth} in Appendix \ref{app.lyart} for more details). 
For simplicity, the source of \lya\ photons is taken to be at rest in the frame of the galaxy, 
in the centre of the outflow, and emits photons at the line centre only, $\lambda = 1215.68$ \AA{}. 

\subsubsection{Expanding thin shell}

Previous radiative transfer studies of \lya\ line profiles 
adopted an expanding shell in the same way as we used here 
\citep[see, e.g. ][and references therein]{ahn03,ahn04,verhamme06,schaerer07,verhamme08}. 
This model, hereafter named ``Shell'', 
consists of a homogeneous, expanding, isothermal spherical shell, in which dust and gas are uniformly mixed. 
The shell, although thin, is described by an inner and outer radius $R_{\rm in}$ and $R_{\rm out}$, 
which satisfy $R_{\rm in} = f_{\rm th} R_{\rm out}$. We set $f_{\rm th} = 0.9$. In addition, the medium
is assumed to be expanding radially with constant velocity $V_{\rm exp}$. The column density through the Shell is given by

\begin{equation}
\label{eq.nh_shell}
 N_H(r) = \frac{ X_H M_{\rm shell} }{4\pi m_H R_{\rm out}^2},
\end{equation}
where $X_H = 0.74$ is the fraction of hydrogen in the cold gas and $m_H$ is the mass of a hydrogen atom.
In \galform, the \lya\ luminosity originates in the disk (in quiescent galaxies) 
or the bulge (in starbursts). Some galaxies may also have contributions from both 
components. Therefore, $M_{\rm shell}$, $R_{\rm out}$ and $V_{\rm exp}$ 
are taken to be proportional to the mass of cold gas $M_{\rm cold}$, half-mass radius $R_{1/2}$ and circular 
velocity $V_{\rm circ}$, respectively, i.e., 
\begin{eqnarray}
\label{eq.f_M}
M_{\rm shell} & = & f_M \langle M_{\rm gas} \rangle, \\
\label{eq.f_R}
R_{\rm out}   &  = & f_R \langle R_{1/2} \rangle, \\
\label{eq.f_V}
V_{\rm exp}  &  = & f_V \langle V_{\rm circ} \rangle,
\end{eqnarray}
where $f_M$, $f_R$ and $f_V$ are free parameters. To take into account the contribution from both components
of a galaxy, we define
%\begin{eqnarray}
%\label{eq.w_mgas}
 %\langle M_{\rm gas} \rangle  &=&  \frac{1}{L_{\rm Ly\alpha}^{\rm total}} \left( M_{\rm gas}^{\rm disk} L_{\rm Ly\alpha}^{\rm disk} + 
%M_{\rm gas}^{\rm bulge} L_{\rm Ly\alpha}^{\rm bulge} \right), \\
%\label{eq.w_r}
%\langle R_{1/2} \rangle & = & \frac{1}{L_{\rm Ly\alpha}^{\rm total}} \left( R_{1/2}^{\rm disk} L_{\rm Ly\alpha}^{\rm disk} + 
%R_{1/2}^{\rm bulge} L_{\rm Ly\alpha}^{\rm bulge} \right), \\
%\label{eq.w_v}
%\langle V_{\rm circ} \rangle & = & \frac{1}{L_{\rm Ly\alpha}^{\rm total}} \left( V_{\rm circ}^{\rm disk} L_{\rm Ly\alpha}^{\rm disk} + 
%V_{\rm circ}^{\rm bulge} L_{\rm Ly\alpha}^{\rm bulge} \right),
%\end{eqnarray}
\begin{eqnarray}
\label{eq.w_mgas}
 \langle M_{\rm gas} \rangle  &=&  F_{\rm Ly\alpha}^{\rm disk} M_{\rm gas}^{\rm disk} +  
 (1-F_{\rm Ly\alpha}^{\rm disk}) M_{\rm gas}^{\rm bulge}, \\
\label{eq.w_r}
\langle R_{1/2} \rangle & = &  F_{\rm Ly\alpha}^{\rm disk} R_{1/2}^{\rm disk} +  
 (1-F_{\rm Ly\alpha}^{\rm disk}) R_{1/2}^{\rm bulge}, \\
\label{eq.w_v}
\langle V_{\rm circ} \rangle & = &  F_{\rm Ly\alpha}^{\rm disk} V_{\rm circ}^{\rm disk} +  
 (1-F_{\rm Ly\alpha}^{\rm disk}) V_{\rm circ}^{\rm bulge}, \\
 F_{\rm Ly\alpha}^{\rm disk} & \equiv &\frac{L_{\rm Ly\alpha}^{\rm disk}}{L_{\rm Ly\alpha}^{\rm total}}.
\end{eqnarray}
In each term, the superscript indicates the contribution from the disk, the bulge or the total (the sum of the two).
In most galaxies, however, either the disk or bulge term completely dominates.

Likewise, the metallicity of the shell $Z_{\rm out}$ is taken to be the metallicity of the cold gas $Z_{\rm cold}$ 
weighted by a combination of the mass of cold gas and the \lya\ luminosity of each component, i.e., 
\begin{equation}
\label{eq.zout}
 Z_{\rm out} = \frac{ M_{\rm gas}^{\rm disk} L_{\rm Ly\alpha}^{\rm disk} Z_{\rm cold}^{\rm disk}+ 
M_{\rm gas}^{\rm bulge} L_{\rm Ly\alpha}^{\rm bulge}Z_{\rm cold}^{\rm bulge}}{M_{\rm gas}^{\rm disk}
 L_{\rm Ly\alpha}^{\rm disk} + M_{\rm gas}^{\rm bulge} L_{\rm Ly\alpha}^{\rm bulge}}.
\end{equation}
In order to compute the dust content in the outflow we assume that the mass of dust in the outflow, $M_{\rm dust}$, 
is proportional to the gas mass and metallicity, i.e.
\begin{equation}
 M_{\rm dust} = \frac{\delta_*}{Z_{\odot}} M_{\rm shell} Z_{\rm out},
\end{equation}
where the dust-to-gas ratio is taken to be $\delta_* = 1/110$ at the solar metallicity $Z_{\odot} = 0.02$ \citep{granato00}. The extinction optical depth at the wavelength of \lya\ can be written as 
\begin{equation}
 \tau_d = \frac{E_{\odot}}{Z_{\odot}} N_H Z_{\rm out},
\label{eq.taud}
\end{equation}
where $E_{\odot} = 1.77 \times 10^{-21} [{\rm cm^2}]$ is the ratio $\tau_d/N_H$ for solar metallicity at the wavelength of \lya\ (1216 \AA{}). Throughout this work we 
use the extinction curve and albedo from \citet{silva98}, which are fit to the observed extinction and albedo in the Galactic ISM. 
For a dust albedo $A$, the optical depth for absorption is simply
\begin{equation}
 \tau_a = (1-A) \tau_d.
\label{eq.taua}
\end{equation}
At the wavelength of \lya,  $A_{Ly\alpha} = 0.39$.

\subsubsection{Galactic Wind}
\label{sec.wind}
Supernovae heat and accelerate the ISM through shocks and hence generate outflows from galaxies 
\citep[see, e.g. ][]{frank99,strickland02}. Here we develop an outflow model, hereafter called 
``Wind'', which relates the density of the outflow to the mass ejection rate from galaxies due to supernovae. 
In \galform, this mass ejection rate is given by
\begin{equation}
\label{eq.m_ej}
 \dot{M}_{\rm ej} = \left[ \beta_{\rm reh}(V_{\rm circ}) + \beta_{\rm sw}(V_{\rm circ}) \right] \psi,
\end{equation}
where
\begin{eqnarray}
\label{eq.betareh}
\beta_{\rm reh} &=& \left(\frac{V_{\rm circ}}{V_{\rm hot}}\right)^{-\alpha_{\rm hot}}, \\
\label{eq.betasw}
\beta_{\rm sw} & = & f_{\rm sw} {\rm min}[1,(V_{\rm circ}/V_{\rm sw})^{-2}].
\end{eqnarray}
The terms $\beta_{\rm reh}$ and $\beta_{\rm sw}$ define the two different modes of supernova feedback (the {\it reheating} and {\it superwind}), 
and the constants $\alpha_{\rm hot}, V_{\rm hot}, f_{sw}$ and $V_{sw}$ are free parameters of \galform, chosen by fitting the model predictions
to observed galaxy LFs. 
The instantaneous star formation rate, $\psi$ is obtained as
\begin{equation}
\label{eq.sfr}
 \psi = \frac{M_{\rm gas}}{\tau_*},
\end{equation}
where $\tau_*$ is the star formation time-scale, which is different for quiescent galaxies and starbursts.
For a detailed description of the star formation and supernova feedback
processes in this variant of \galform, see \citet{baugh05} and \citet{lacey08}.  Since star formation can occur in the disk 
and in the bulge, there is a mass ejection rate term $\dot{M}_{\rm ej}$ associated with each component.

We construct the wind as an isothermal, spherical flow expanding at constant velocity $V_{\rm exp}$, 
and inner radius $R_{\rm wind}$ (the wind is empty inside).
In a steady-state spherical wind, the mass ejection rate is related to the velocity and density at any point of the 
Wind via the equation of mass
continuity, i.e.
\begin{equation}
\label{eq.mdot}
 \dot{M}_{\rm ej} = 4\pi r^2 V_{\rm exp} \rho(r),
\end{equation}
where $\rho(r)$ is the mass density of the medium, and $V_{\rm exp}$ is calculated following equation \eqref{eq.f_V}.
Since star formation in \galform\ can occur in both the disk and bulge of galaxies, $\dot{M_{\rm ej}}$ in equation \eqref{eq.mdot}
corresponds to the sum of the ejection rate from the disk and the bulge.

Following equation \eqref{eq.mdot}, the number density profile $n_{HI}(r)$ in the Wind geometry varies according to
\begin{equation}
\label{eq.rho}
 n_{HI}(r) = \left\{ \begin{array}{ll}
               0	 & {\rm r < R_{\rm wind}} \\
	       \dfrac{X_H\dot{M}_{\rm ej}}{4\pi m_{H} r^2 V_{\rm exp}} & {\rm r \ge R_{\rm wind}.}
              \end{array}
\right.
\end{equation}
The column density $N_H$ of the outflow is
\begin{equation}
 N_H = \dfrac{X_H\dot{M}_{\rm ej}}{4\pi m_{H} R_{\rm wind} V_{\rm exp}},
\label{eq.nh_wind}
\end{equation}
where the inner radius of the wind, $R_{\rm wind}$, is computed in an analogous 
way to $R_{\rm out}$ in the Shell geometry (Eq. \ref{eq.f_R}).
Note that both $f_R$ and $f_V$ in this case are different free parameters 
and independent of those used in the Shell geometry.
The metallicity of the Wind, $Z_{\rm wind}$, corresponds to the metallicity of the cold gas 
in the disk and bulge weighted by their respective mass ejection rates. 

For computational reasons, the radiative transfer code requires us to 
define an outer radius, $R_{\rm outer}$, for the Wind. 
However, since the number density of atoms decreases as $\sim r^{-2}$, 
we expect that at some point away from the centre the optical depth becomes 
so small that the photons will be able to escape regardless of the exact extent 
of the outflow. We find that an outer radius $R_{\rm outer} = 20 R_{\rm wind}$ 
yields converged results, i.e. the escape fraction of \lya\ photons does not vary 
if we increase the value of $R_{\rm outer}$ further. 

\section{Outflow properties}
\label{sec.comparison}
In this section we explore the properties of the \lya\ 
radiative transfer in our outflow geometries prior to coupling these to 
the galaxy formation model \galform. We do this by running our 
Monte Carlo radiative
transfer model over a grid of configurations spanning a wide range
of hydrogen column densities, expansion velocities and metallicities.
In order to obtain well-defined \lya\ line profiles, we run each 
configuration using  $5\times 10^4$ photons.
Therefore, the minimum \lya\ escape fraction our models can compute
in this case is $\fesc = 2\times 10^{-5}$.

\begin{figure*}
\centering
\includegraphics[width=15cm,angle=90]{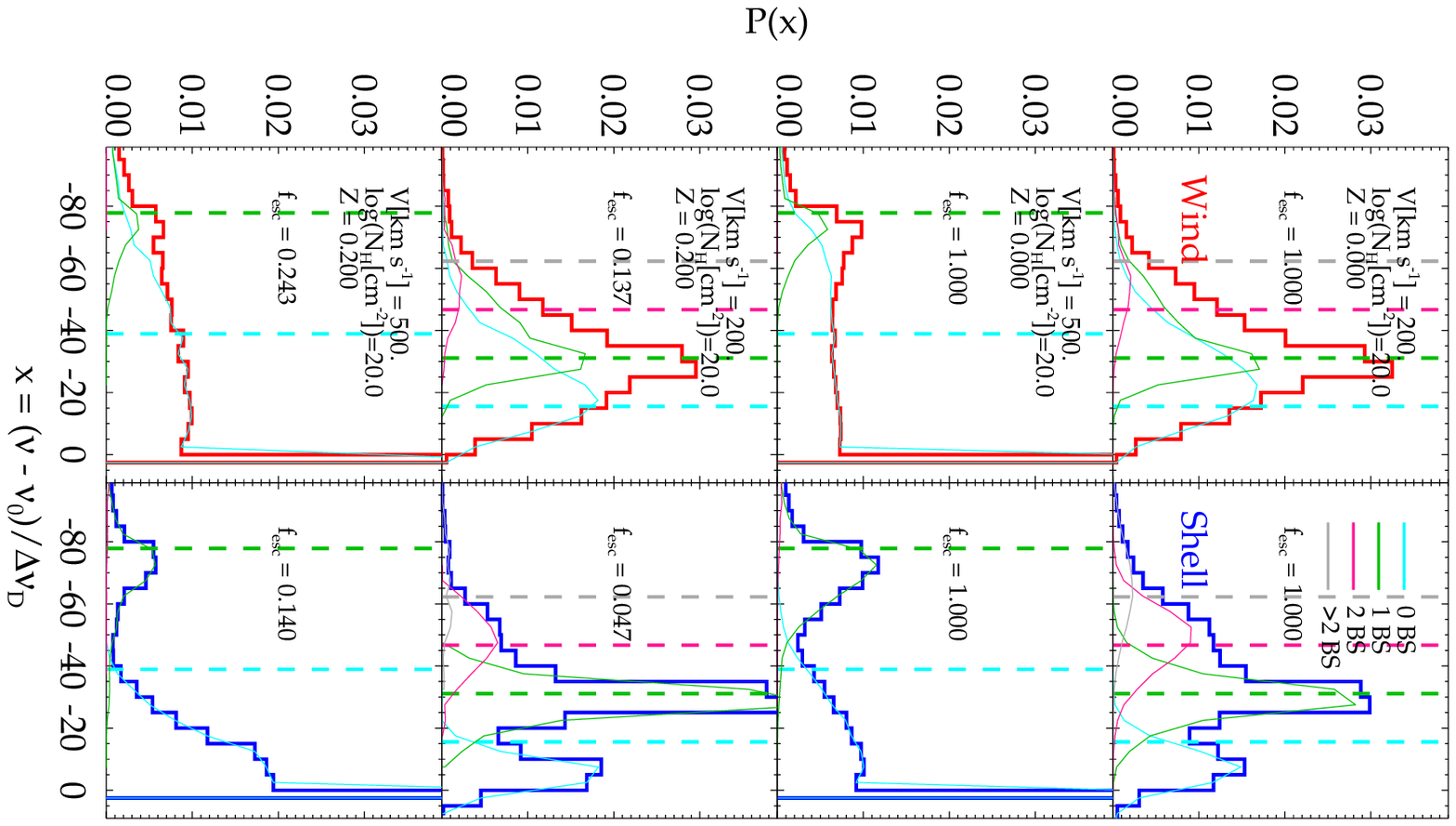}
\includegraphics[width=15cm,angle=90]{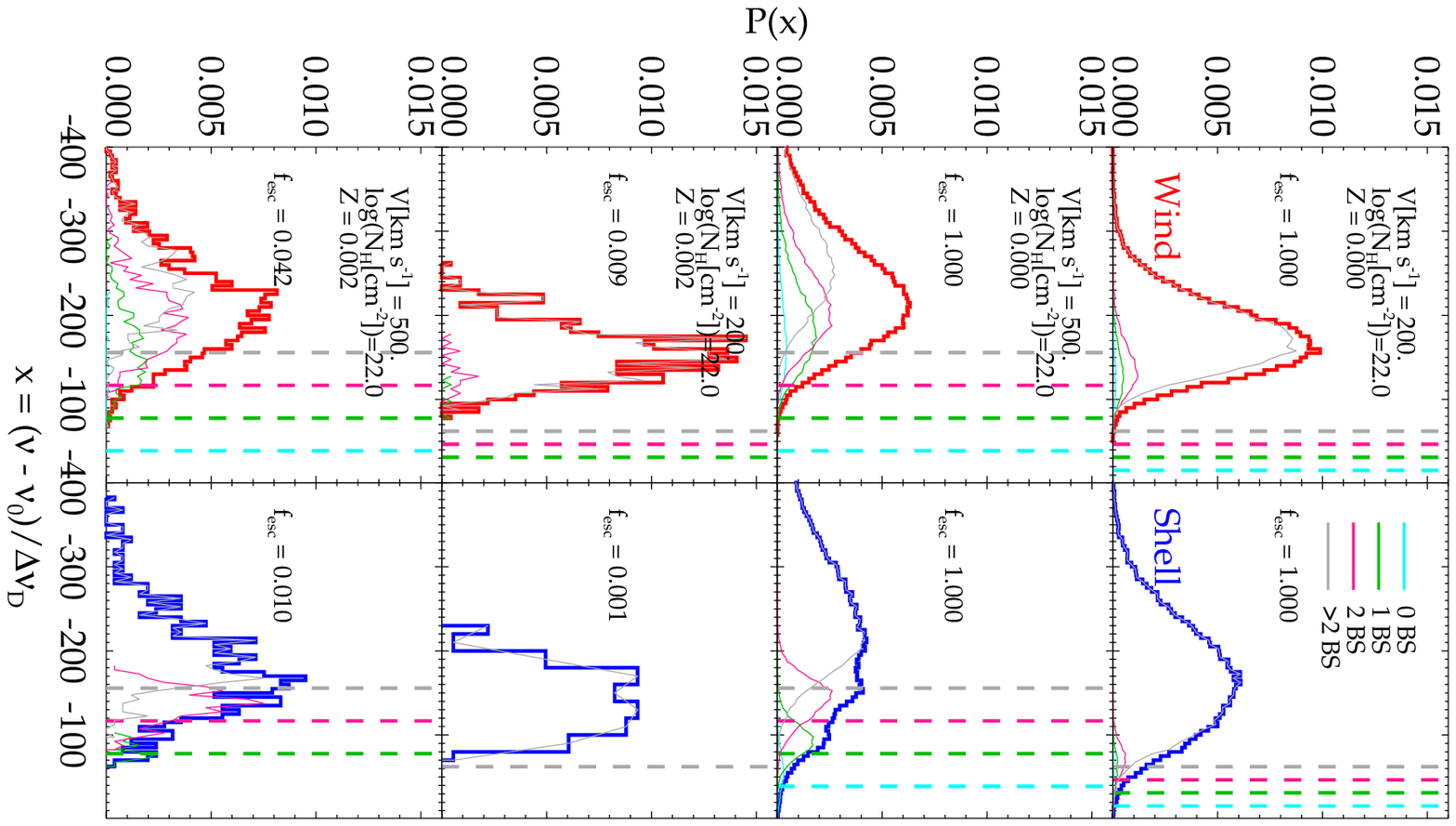}
\caption{Comparison of the \lya\ line profile obtained with the Wind 
and Shell geometries for $N_H = 10^{20} {\rm [cm^{-2}]}$ (leftmost 
two columns) and $N_H = 10^{22} {\rm [cm^{-2}]}$ (rightmost 
two columns). The red (blue) histogram shows the full spectrum 
obtained for the Wind (Shell) outflow geometry. 
The cyan, green, pink and gray lines show the spectra of photons 
which experienced 0, 1, 2 and 3 or more back-scatterings before 
escaping, respectively. Likewise, the same colours are used 
to plot vertical dashed lines showing the frequencies corresponding to
$-x_{bs}, -2x_{bs}, -3x_{bs}$ and $-4x_{bs}$, respectively (see text for details). 
Each row displays a different configuration, characterised 
by a given $N_H, V_{\rm exp}$ and $Z$, as indicated in the legend. 
The top two rows correspond to dust-free configurations ($Z = 0$), 
whereas the bottom two have different metallicities, chosen to have equal 
dust optical depth, although having different column densities 
of hydrogen. 
The escape fraction of \lya\ photons, \fesc, is indicated in each box. 
The \lya\ profiles shown are normalized to the total number of escaping 
photons (instead of the total number of photons run), to ease the 
comparison between the dusty and dust-free cases. Note the different
range in the x-axis between the left and right panels.}
\label{fig.comp_spectra_nh}
\end{figure*}

\subsection{\lya\ line properties}

One important difference between our two outflow geometries is the way the column 
densities are computed. In the Shell geometry the column density is a function 
of the total cold gas mass of a galaxy and its size, whereas in the Wind 
geometry it depends on the mass ejection rate given by the supernova 
feedback model along with the size and the circular velocity of the galaxy. 
This difference translates into different predicted properties for \lya\ 
emitters, as will be shown in the next section. However, even when the 
column density, velocity of expansion and outflow metallicity are the same, 
the two models will give different escape fractions and line profile shapes due to their
different geometries.

Fig. \ref{fig.comp_spectra_nh} shows the \lya\ line profiles 
obtained with the two models when matching the key properties for outflows at 
the same column densities. In order to make a fair comparison between the Shell 
and Wind geometries, we compare configurations with the same column density, 
expansion velocity and metallicity. In addition, the inner radius in the Wind geometry 
is chosen to be the same as its counterpart, the outer radius in the Shell geometry. 
The two panels of Fig. \ref{fig.comp_spectra_nh} display a set of configurations 
with column densities of $N_H = 10^{20} [{\rm cm^{-2}}]$ (left panel) and 
$N_H = 10^{22}[{\rm cm^{-2}}]$ (right panel).
In the following, we express the photon's frequency in terms of $x$, 
the shift in frequency around the line centre $\nu_0$, in units of the 
thermal width, i.e.
\begin{equation}
\label{eq.x}
 x \equiv \frac{(\nu - \nu_0)}{\Delta \nu_D},
\end{equation}
where $\Delta \nu_D = v_{\rm th}\nu_0/c$, and $c$ is the speed of light,
and $v_{\rm th}$ is the thermal velocity discussed in Section \ref{sec.outflows}.

As a general result, outflows with a column density of $N_H = 10^{20} 
[{\rm cm^{-2}}]$, regardless of the other properties, produce 
multiple peaks at frequencies redward of the line centre in both geometries, 
with different levels of asymmetry. 
The Shell geometry generates more prominent peaks than the 
Wind geometry. The frequency of the main peak is, however, the same in both 
outflow geometries.

\begin{figure*}
\centering
\includegraphics[width=5.2cm,angle=90]{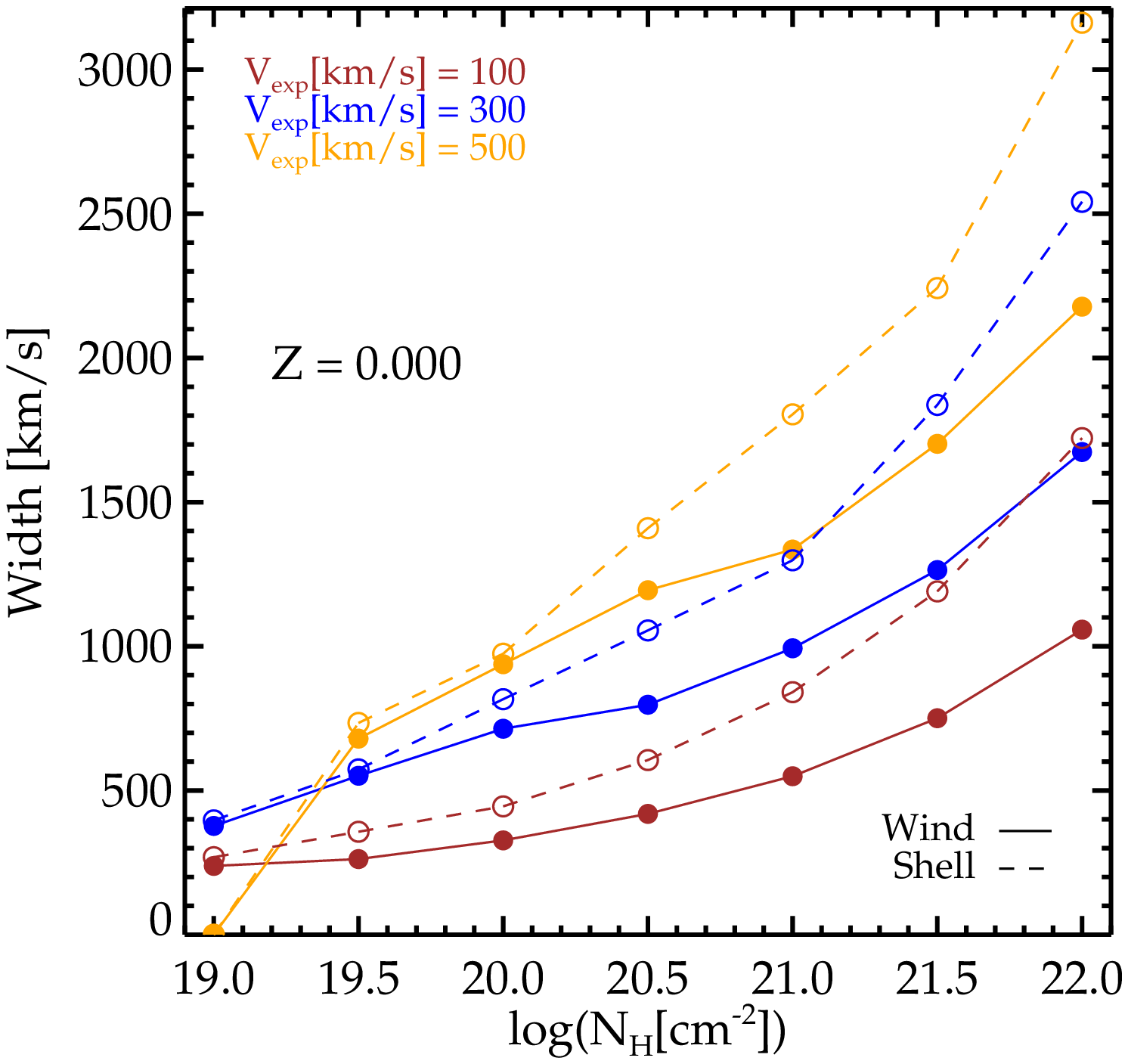}
\includegraphics[width=5.2cm,angle=90]{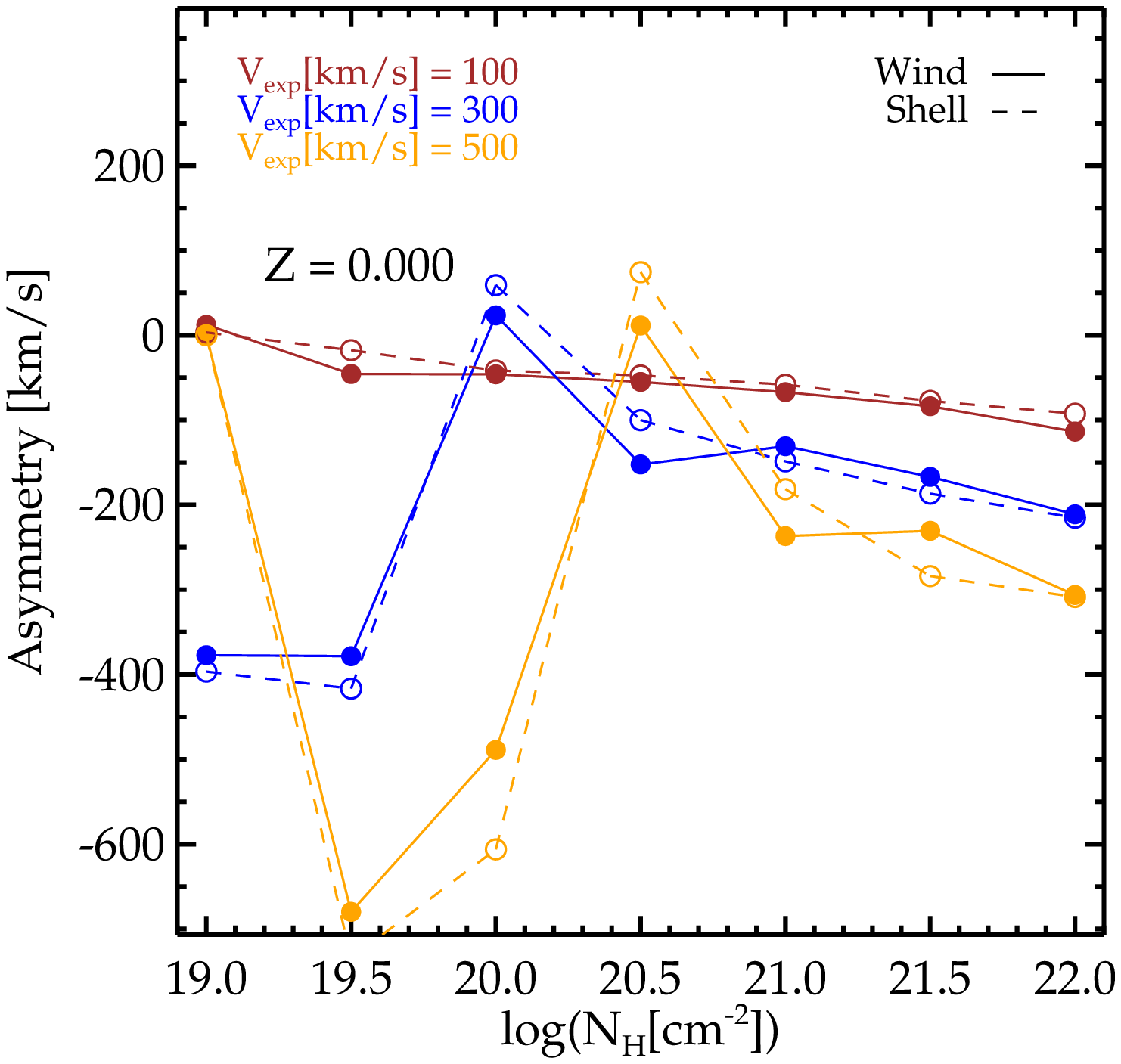}
\includegraphics[width=5.2cm,angle=90]{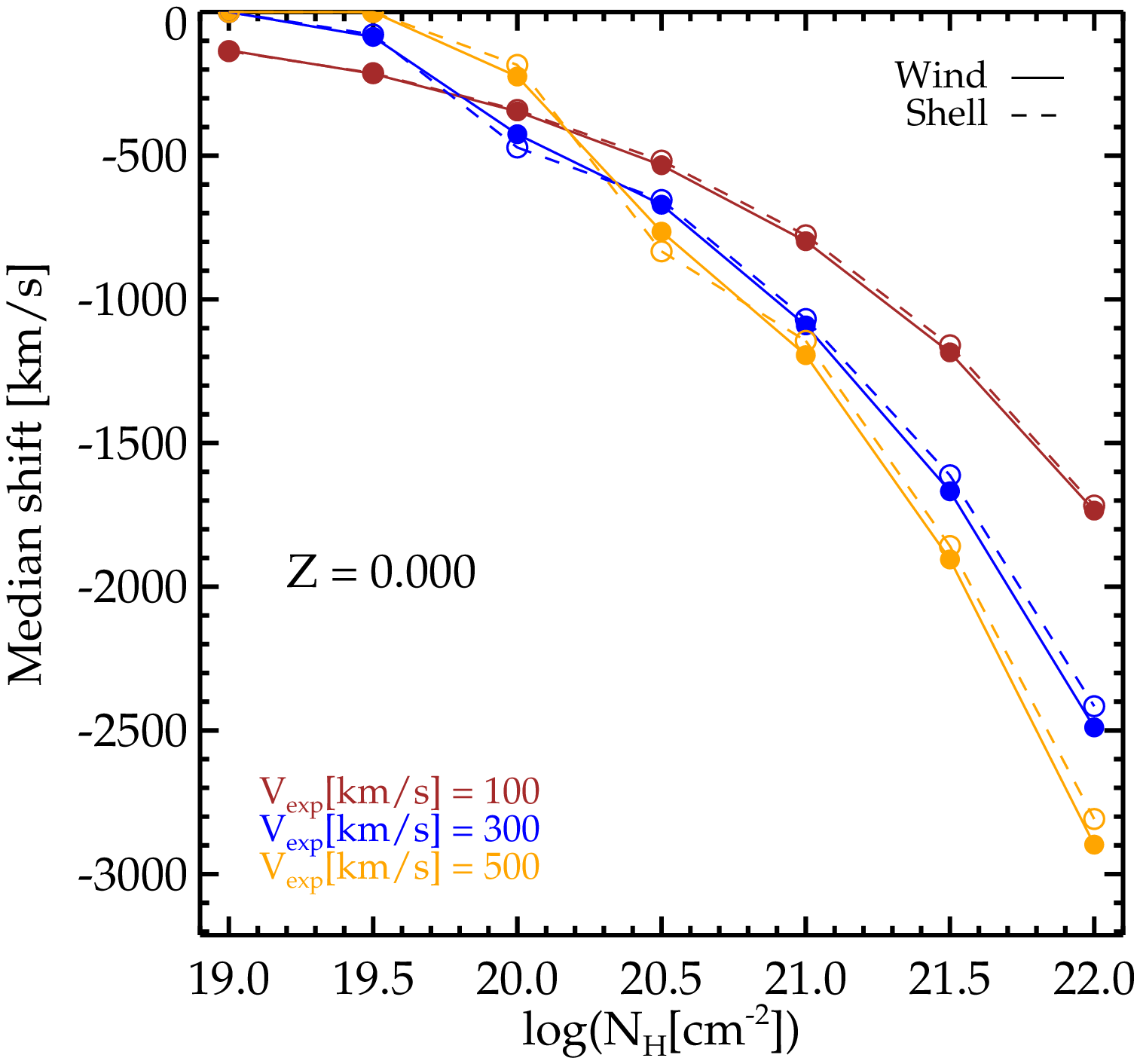}
\includegraphics[width=5.2cm,angle=90]{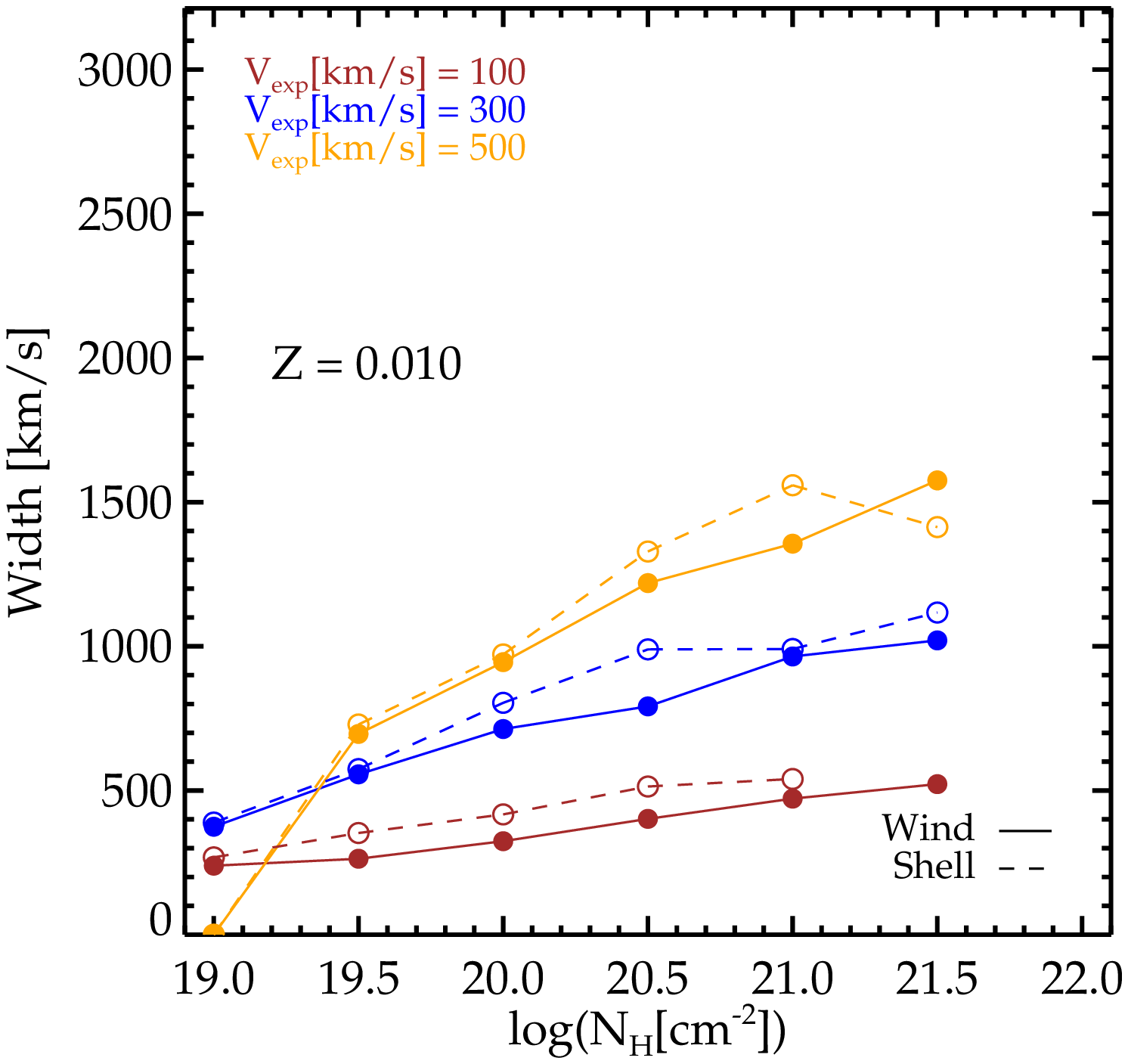}
\includegraphics[width=5.2cm,angle=90]{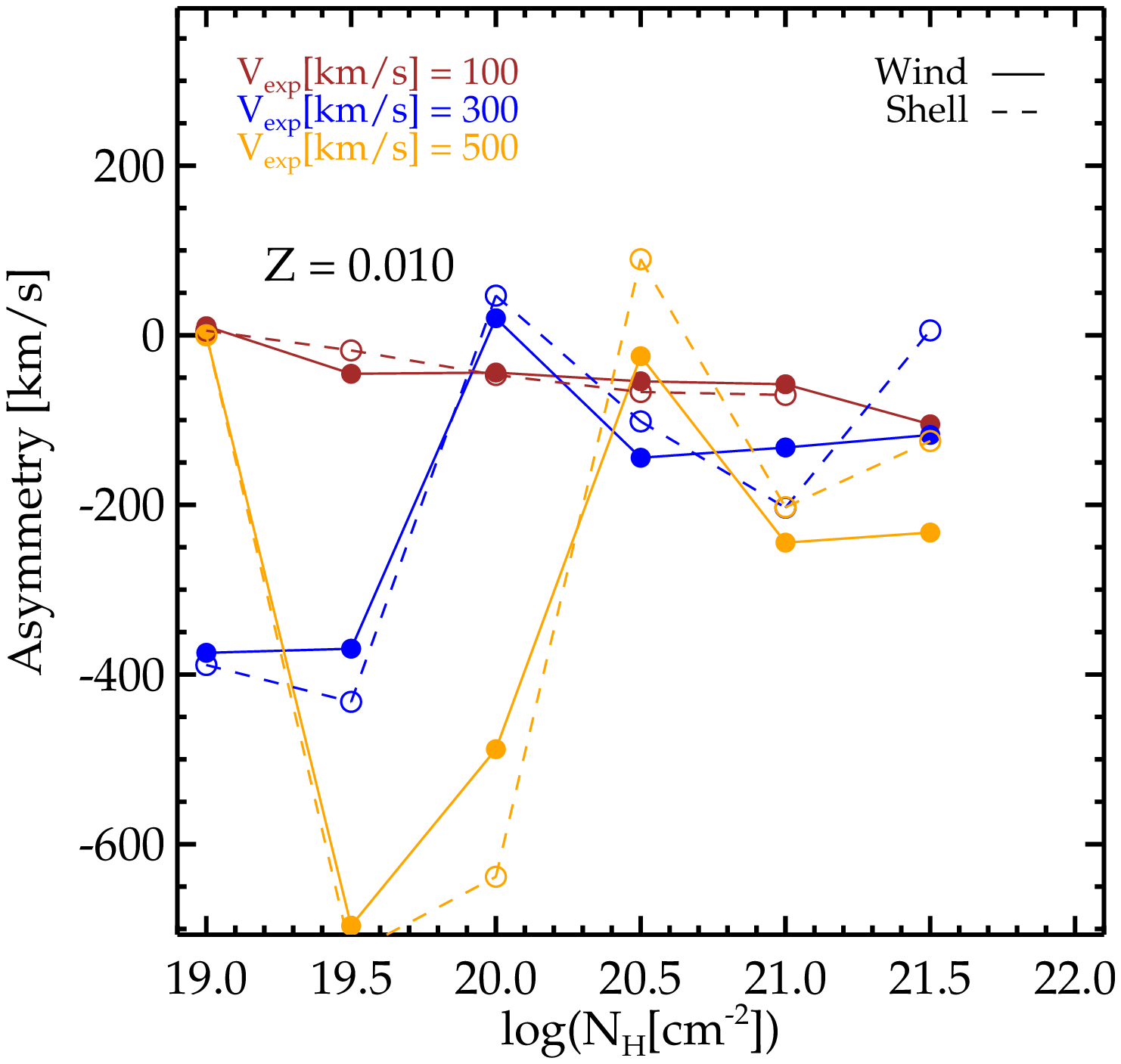}
\includegraphics[width=5.2cm,angle=90]{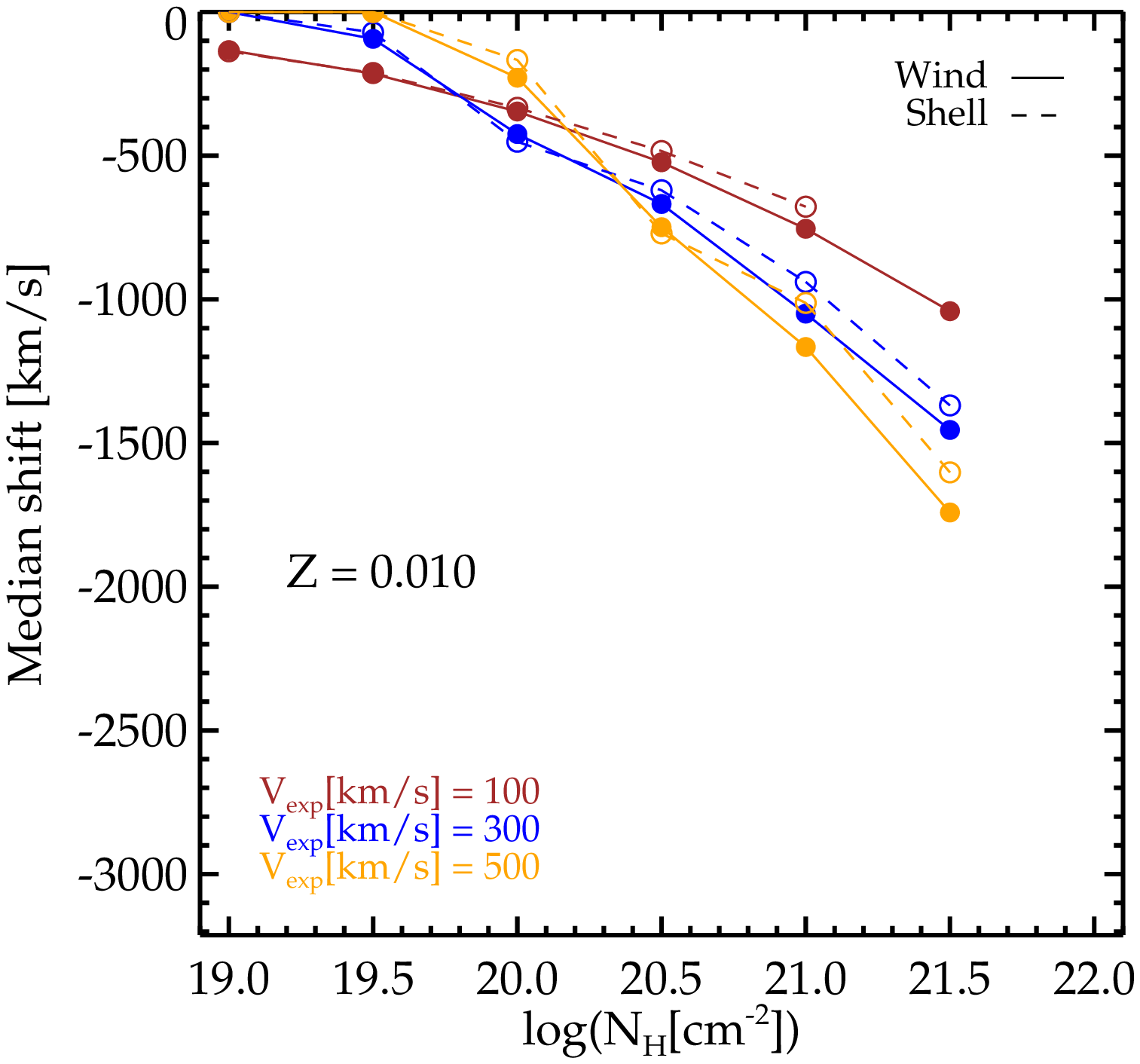}
\caption{The properties of the \lya\ profiles predicted by the two
outflow geometries for a given column density, 
expansion velocity and metallicity. 
Left panels: the  width of the profiles, defined as the difference between the 
90 and 10 percentiles of the frequency distribution of escaping \lya\ photons, 
measured in km/s. 
Middle panels: the asymmetry, defined in the text, and measured in km/s.
Right panels: the shift of the median of the \lya\ profiles with respect to the 
line centre, measured in km/s. Note that negative velocities indicate redshifting.
The top row shows the \lya\ profile properties in dust-free outflows. The bottom
row shows outflows with a metallicity $Z = 0.01$. Different expansion velocities
spanning the range $100-500$ km/s are shown with different colours, 
as shown by the labels in each panel. The Wind geometry is shown with solid
lines and filled circles. The Shell geometry is shown with dashed lines
and open circles.}
\label{fig.prop_profiles}
\end{figure*}
On the other hand, outflows with $N_H = 10^{22} [{\rm cm^{-2}}]$ display
broader \lya\ profiles. As opposed to the lower column density case, these
profiles display a single peak, also redward of the line centre. The position 
of this peak is also the same for both geometries. 

The effect of a large expansion velocity is shown in the 2nd and 4th rows in both 
panels of Fig. \ref{fig.comp_spectra_nh}.
Here the configurations have $V_{\rm exp} = 500 {\rm [km/s]}$.  
The differences with the configurations where $V_{\rm exp} = 200 {\rm [km/s]}$
are evident: the profiles are broader, and the position 
of the peaks are displaced to redder frequencies. Furthermore,
in the left panel of Fig. \ref{fig.comp_spectra_nh}, where $N_H = 10^{20} [{\rm cm^{-2}}]$,
it is shown that a significant fraction of photons escape at the line centre. 
The high expansion velocity in this case
makes the medium optically thin, allowing many photons to escape without
undergoing any scattering.

For the configurations with $N_H = 10^{20} [{\rm cm^{-2}}]$, the optical 
depth that a photon at the line centre ($x = 0$) sees when traveling 
along the radial direction is $\tau_0 = 3.63$ and $\tau_0 = 0.57$ 
for expansion velocities $V_{\rm exp} = 200 [{\rm km/s}]$ and
$V_{\rm exp} = 500 [{\rm km/s}]$, respectively.  Accordingly, the configurations
with $N_H = 10^{22} [{\rm cm^{-2}}]$ have optical depths a factor $100$
higher.
On the other hand, a static medium with $N_H = 10^{20} [{\rm cm^{-2}}]$
has a much higher optical depth, $\tau_0 = 3.31 \times 10^6$.
This illustrates the strong effect of the expansion velocity in reducing the 
optical depth of the medium, thus allowing photons to escape.

The \lya\ line profiles obtained can be characterised by the frequency distribution 
of photons split according to the number of backscatterings they experience before escaping
(i.e. the number of times photons bounce back to the inner, empty region). 
When photons interact for the first time
with the outflow, a fraction of them will experience a backscattering. These events
are significant since the distribution of scattering angles is dipolar 
(see Eq. \ref{eq.dipolar} in Appendix \ref{app.lyart}). 
The frequency of a photon after a scattering event, in the observer's frame, is given by Eq. \eqref{eq.xf}.
Depending on the direction of the photon after the scattering event, its frequency will fall within the range 
$x = [-2 x_{bs},0]$, where $x_{bs} \equiv V_{\rm exp}/v_{\rm th}$ \citep[see also][]{ahn03,verhamme06}.
Photons that do not experience a backscattering, or escape directly, form the cyan curves in Fig. \ref{fig.comp_spectra_nh}.
If the photon is backscattered exactly backwards, then its frequency will be $x = -2x_{bs}$. The cross-section for
scattering is significantly reduced for these photons, and so a fraction escape without undergoing any 
further interaction with the outflow. Photons backscattered once form the green curves in Fig. \ref{fig.comp_spectra_nh}.
If a photon experiences a second backscatter in the exact opposite direction, then its frequency will become $x = -4x_{bs}$, and the
cross-section for a further scattering will again reduce significantly. The magenta curves show the distribution of photons
that experienced 2 backscatterings. Finally, photons that experience 3 or more backscatterings are shown in gray.

In detail, the geometrical 
differences between the Shell and Wind models 
are translated into each backscattering peak contributing
in a different proportion and with a different shape to the overall spectrum for each geometry. 
Previous studies have also found this relation between the peaks of 
backscattered photons and $x_{\rm bs}$ in media with column densities of the order 
of $N_H \sim 10^{20} [{\rm cm^{-2}}]$ \citep{ahn03,ahn04,verhamme06}, 
although they did not study the line profiles for higher optical depths as we do here. 
For $N_H = 10^{22} [{\rm cm^{-2}}]$, we find that 
the peaks are displaced considerably from their expected position based on the simple argument above.
This is not surprising, since in outflows with very large optical depths
the number of scatterings broadens the profiles and reddens the 
peak positions.

In the Wind geometry, the contribution of photons featuring no backscatterings
dominates most of the total profile for $N_H = 10^{20} [{\rm cm^{-2}}]$, 
whereas in the overall line profile for the Shell 
geometry there is a clear distinction between a region dominated by photons suffering 
no backscatterings and those backscattered once (green curve). This
illustrates again how the \lya\ line profile in the Shell geometry features clear 
multiple peaks from one or more backscatterings, whereas in the Wind 
geometry the secondary peaks are less obvious. 

Fig. \ref{fig.comp_spectra_nh} also shows the effect of including 
dust in the outflows. Overall, dust absorption has more effect on the 
redder side of the line profiles than at frequencies 
closer to the line centre, where the probability of scattering with hydrogen 
atoms is significantly higher than the probability of interaction with dust. 
In detail, the sensitivity of \lya\ photons to dust 
is the result of an interplay between the optical depths of hydrogen 
and dust at the photon frequency (the cross-section of scattering 
is significantly reduced away from the line centre, and hence so is the 
average number of scatterings) and the bulk velocity of the gas, 
reducing the number of scatterings further, and thus also 
the probability for a photon to be absorbed. 

A less obvious outflow property for determining the \lya\ attenuation 
by dust is its geometry. Despite the 
similarities between the outflow geometries, the Shell geometry 
consistently gives a lower \lya\ escape fraction than the Wind
geometry in the configurations studied in Fig. \ref{fig.comp_spectra_nh}. 

As a final comparison, in Fig. \ref{fig.comp_spectra_nh} we chose a metallicity of $Z = 0.2$ 
for the configurations with $N_H = 10^{20}[{\rm cm^{-2}}]$ and 
$Z = 0.002$ for the configurations with $N_H = 10^{22}[{\rm cm^{-2}}]$. 
Although the metallicities are different, 
the optical depth of dust, given by Eq. \eqref{eq.taud}, is the same in both cases, 
$\tau_d = 1.77$.
By matching the optical depth of dust, we
can study the effect of $N_H$ on the \lya\ escape fraction.
Fig. \ref{fig.comp_spectra_nh} shows that 
even when the optical depth of dust is the same, outflows with 
$N_H = 10^{22}[{\rm cm^{-2}}]$ have \lya\ escape fractions up to an 
order of magnitude lower than those with $N_H = 10^{20}[{\rm cm^{-2}}]$. 
This occurs since, in the former case, the average number of scatterings is about 
two orders of magnitude larger than in the latter, and hence
the probability of photons interacting with a dust grain increases accordingly. 
This illustrates the complexity of the
\lya\ radiative transfer process.

\begin{figure}
\centering
\includegraphics[width=7cm,angle=90]{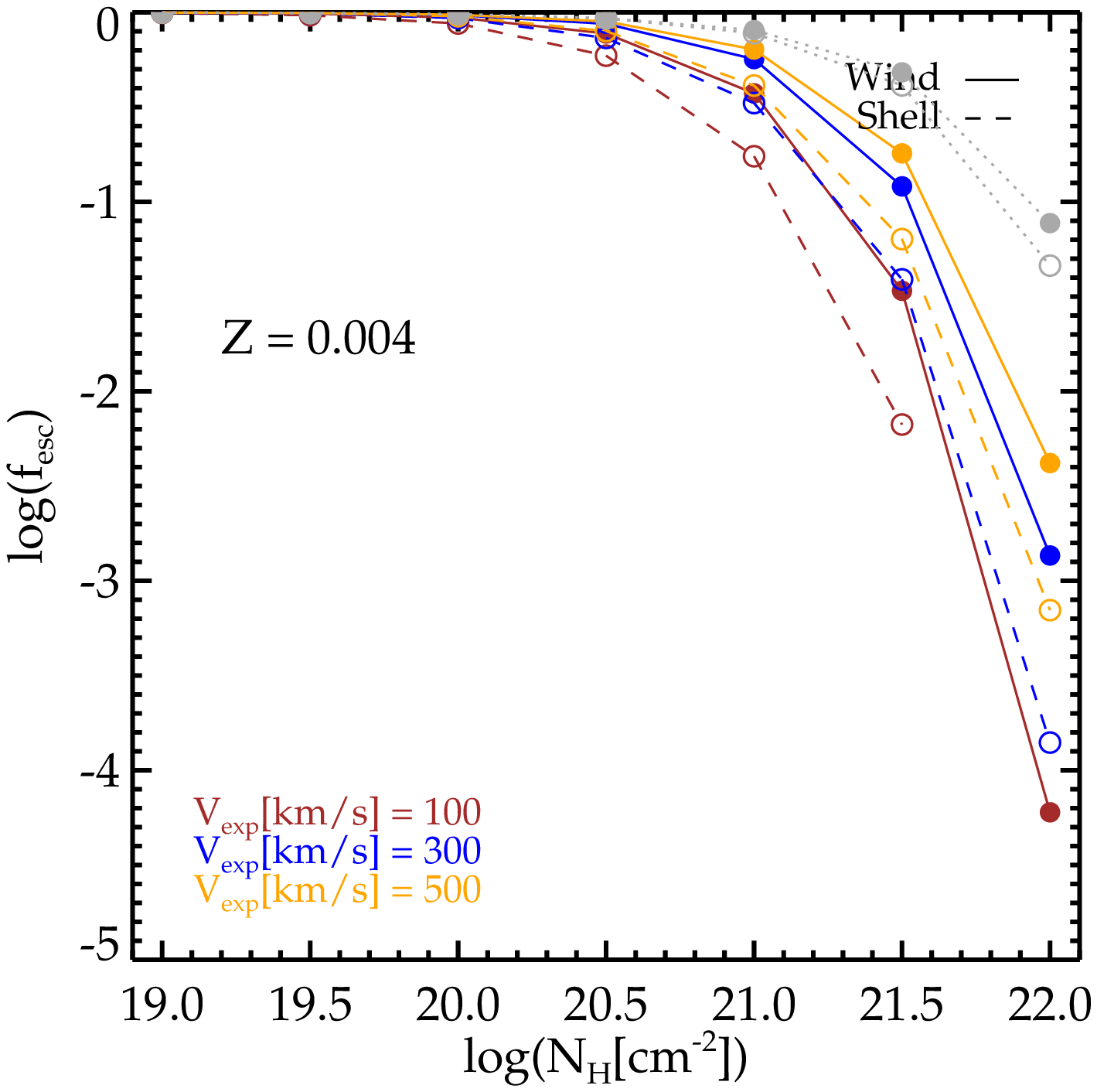}
\includegraphics[width=7cm,angle=90]{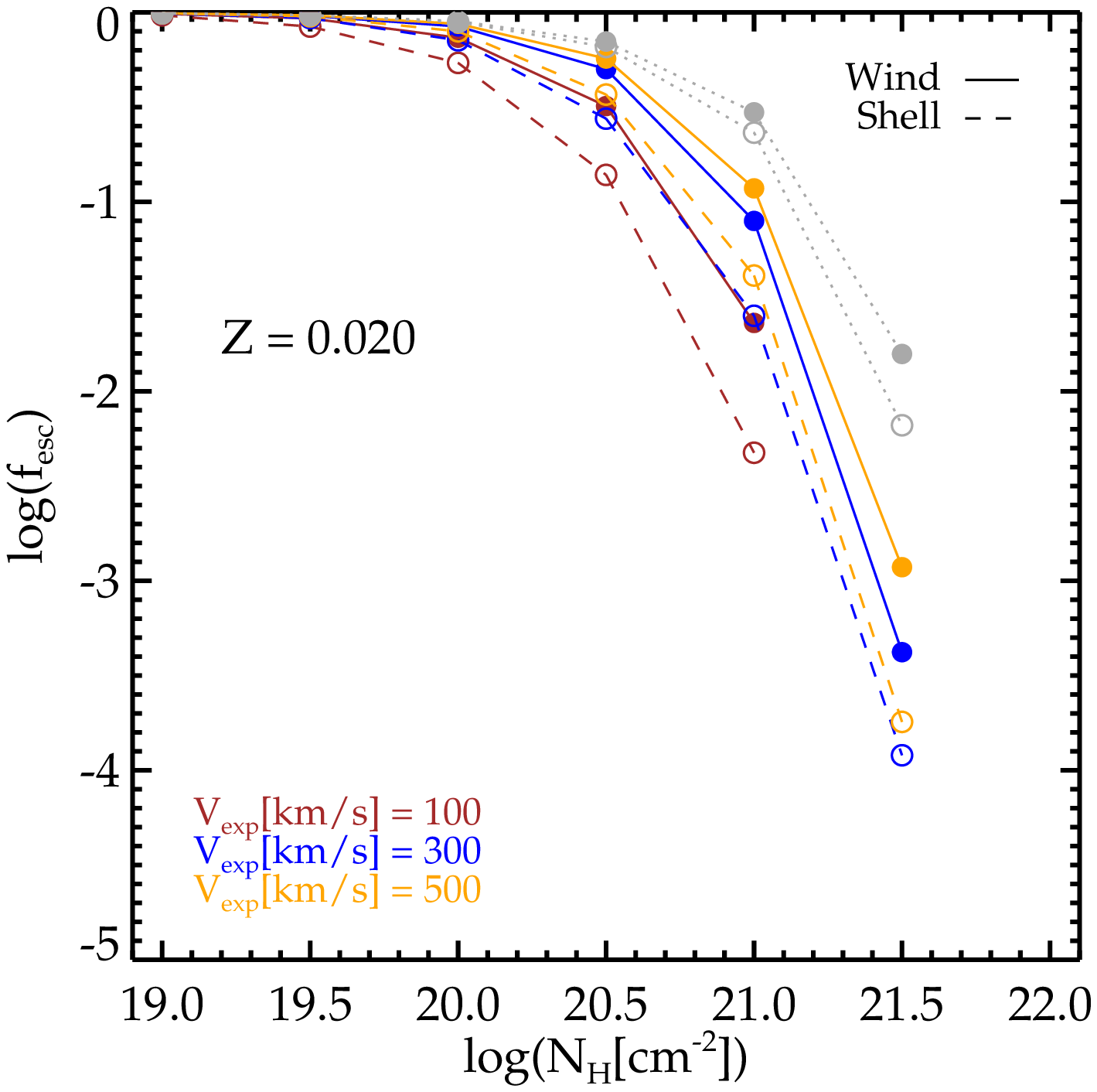}        
\caption{The \lya\ escape fraction predicted by the two
outflow geometries as a function of column density, 
expansion velocity and metallicity. The top panel shows configurations
with $Z = 0.004$. In the bottom panel $Z = 0.02$. Expansion velocities
spanning the range $100-500$ km/s are shown with different colours, 
as shown in the legend of each box. The Wind geometry is shown with solid
lines and filled circles. The Shell geometry is shown with dashed lines
and open circles. The gray circles show the result of switching off the 
scattering of \lya\ photons due to hydrogen atoms in the model, making
photons interact only with dust grains.}
\label{fig.prop_fesc}
\end{figure}

To gain more insight into the difference in the \lya\ profiles generated by 
different configurations, we show in Fig. \ref{fig.prop_profiles} a measure
of the width, asymmetry and average frequency shift of the \lya\ profiles for a set of 
configurations spanning a range of 
expansion velocities of $V_{\rm exp} [{\rm km/s}] = 100-500$, column 
densities $N_H [{\rm cm ^{-2}}] = 10^{19}-10^{22}$, and also 
two metallicities, $Z = 0$ and $Z=0.01$. 
To measure the width of the profiles we compute the difference in frequency
between the 90 and 10 percentiles, expressed in velocity units. The asymmetry
is computed as the difference between the blue side of the profile, $P_{90}-P_{50}$
and the red side, $P_{50}-P_{10}$, where $P_{10}, P_{50}$ and $P_{90}$ are
the 10, 50 (median) and 90 percentile of the frequency distribution. 
The shift of the median is simply the position of the median of the \lya\ profiles
in velocity units.

Overall, by examining the top and bottom rows of Fig.\ref{fig.prop_profiles} 
it becomes clear that the width, asymmetry and median shift of the \lya\ profiles 
are fairly insensitive to the presence of dust in both outflow geometries.
The only important effect of dust is to limit the range of column densities 
where the \lya\ profiles are appreciable. 
Configurations with $\log(N_H[{\rm cm^{-2}}])> 21.5 $ 
do not feature data points on the plot, since all photons used to compute the 
\lya\ profiles were absorbed by dust in this case. The optical 
depth of absorption, $\tau_a$, increases in proportion with the 
column density, as shown by Eqs. \eqref{eq.taud} and \eqref{eq.taua}.

The small contribution of dust in shaping the \lya\ profiles limits
the information that can be extracted observationally from the \lya\ line profile. 
If the outflow geometries studied here are good approximations
to the outflows in \lya\ emitters, then the \lya\ escape fraction cannot be usefully
constrained by using the shape of the spectrum.

Despite the above, the shape of the \lya\ profiles in our outflow geometries 
is sensitive to other properties.
Fig. \ref{fig.prop_profiles} shows a clear increase in the width of the \lya\ 
profiles, with both increasing column density and expansion velocity. Also,
the Shell geometry generates consistently broader profiles than the Wind geometry, 
for all the configurations studied. Note that this is illustrated in Fig. 
\ref{fig.comp_spectra_nh} as well.

The broadening of the \lya\ profiles is, however, less evident 
in the configurations with $N_H = 10^{19} [{\rm cm ^{-2}}]$.
At this low column density, and even at the lower expansion velocities, 
many photons manage to escape at the line centre without being
scattered due to the reduced optical depth, 
as explained above. Hence, the \lya\ profiles are narrower.
In particular, the width of the \lya\ profile, when $N_H = 10^{19} 
[{\rm cm ^{-2}}]$ and $V_{\rm exp} = 500 [{\rm km/s}]$ is zero, 
showing that nearly all photons escaped directly.

At higher column densities, the expansion velocity plays an increasingly
important role at broadening the profiles. In the outflows with
$N_H = 10^{20} [{\rm cm ^{-2}}]$, without dust, the width of the profiles 
increases from $\sim 300 [{\rm km/s}]$ to $\sim 1000 [{\rm km/s}]$
for expansion velocities of $V_{\rm exp} = 100 [{\rm km/s}]$ and 
$V_{\rm exp} = 500 [{\rm km/s}]$, respectively.
When $N_H = 10^{22} [{\rm cm ^{-2}}]$, the width of the profile depends
strongly on the geometry of the outflow. The width for the Wind geometry spans a range of
$\sim 1000 [{\rm km/s}]$ to $\sim 2200 [{\rm km/s}]$
for expansion velocities of $V_{\rm exp} = 100 [{\rm km/s}]$ and 
$V_{\rm exp} = 500 [{\rm km/s}]$, respectively, whereas for
the Shell geometry it spans a much larger range of $\sim 1700 [{\rm km/s}]$ to 
$\sim 3200 [{\rm km/s}]$ for the same range of expansion velocities.

The effect of dust on the width of the profiles is negligible, 
except for the Shell geometry with $N_H = 10^{21} [{\rm cm^{-2}}]$, 
where the width is reduced and is, therefore, closer to the width 
of the profiles of the Wind geometry. 

The asymmetry of the profiles is somewhat more complicated to characterise
in terms of the column density and expansion velocity of the outflows.
As a general result, the \lya\ profiles are asymmetric towards the red-side of the 
spectrum. This is not surprising, since atoms in an outflow "see" the 
\lya\ photons redshifted, and the change in the direction of the photons 
due to the scattering event makes photons appear redder
in the observer's frame. Overall, as Fig. \ref{fig.prop_profiles}
shows, the asymmetry is larger in the configurations with the highest
expansion velocities when $N_H \geq 10^{20} [{\rm cm^{-2}}]$.

Finally, Fig. \ref{fig.prop_profiles} also shows the shift of the median 
of the profiles for the two outflow geometries. As expected, the 
median is redder with increasing column density and expansion velocity.
The column density has a greater impact than the expansion velocity, 
since higher values of $N_H$ imply a larger number of scatterings, thus
increasing the reddening of the profiles.

\subsection{Escape fractions}

Fig. \ref{fig.prop_fesc} compares the predicted \lya\ escape fractions in both
outflow geometries for a set of column densities, expansion velocities and metallicities, similar to
those considered in Fig. \ref{fig.prop_profiles}.

As expected, both outflow geometries predict that the \lya\ escape fraction 
decreases rapidly with increasing $N_H$, as
the medium becomes optically thicker. 
High values of the expansion velocity reduce the 
optical depth of the medium, hence enhancing the 
\lya\ escape fraction. 

For the range of properties studied here, we find that the Shell geometry 
predicts consistently lower \lya\ escape fractions than the Wind geometry, for the 
same set of parameters. 
This demonstrates that in the detailed interplay of physical conditions  
shaping \fesc, the outflow geometry plays an important role.

The difference is less obvious in outflows with $N_H < 10^{21} [{\rm cm^{-2}}]$,
as is the influence of different expansion velocities. At larger column densities, 
even slightly different expansion velocities can lead to significant differences in the 
resulting \lya\ escape fraction.

Also, in Fig.\ref{fig.prop_fesc} we show the effect on \fesc\ of removing the scattering of 
photons by hydrogen atoms. We achieve this setting the \lya\ scattering cross 
section to zero as well (i.e. $H(x) = 0$ in Eq. \ref{eq.tau_x} of Appendix \ref{app.lyart}). 
The predicted \fesc\ is much higher than when considering 
the scattering by H atoms, and is also independent of expansion velocity, 
since in our modelling the optical depth of dust depends only on the metallicity and the 
column density of hydrogen, as shown in Eq. \eqref{eq.taud}. Fig.\ref{fig.prop_fesc} shows
the effect of the resonant scattering resulting from the high cross-section at the line centre increasing 
the path length, and hence makes the resulting \fesc\ lower.

It is interesting to note that even in the case of removing the scatterings by H atoms, the
Shell geometry is more sensitive to dust than the Wind geometry, hence showing again
the key role of the geometry of the outflows in determining the \lya\ \fesc. 

\section{The model for \lya\ emitters}
\label{sec.combining}
In order to understand the nature of the predictions of our model for \lya\ emitters, we 
study in this section the galaxy properties predicted by \galform\ that are relevant in 
determining the properties of \lya\ emitters (Section \ref{sec.galprop}), prior to describing how we combine
the radiative transfer model for \lya\ photons with \galform\ (Section \ref{sec.tuning}).

\begin{figure*}
\centering
\includegraphics[width=5.6cm,angle=90]{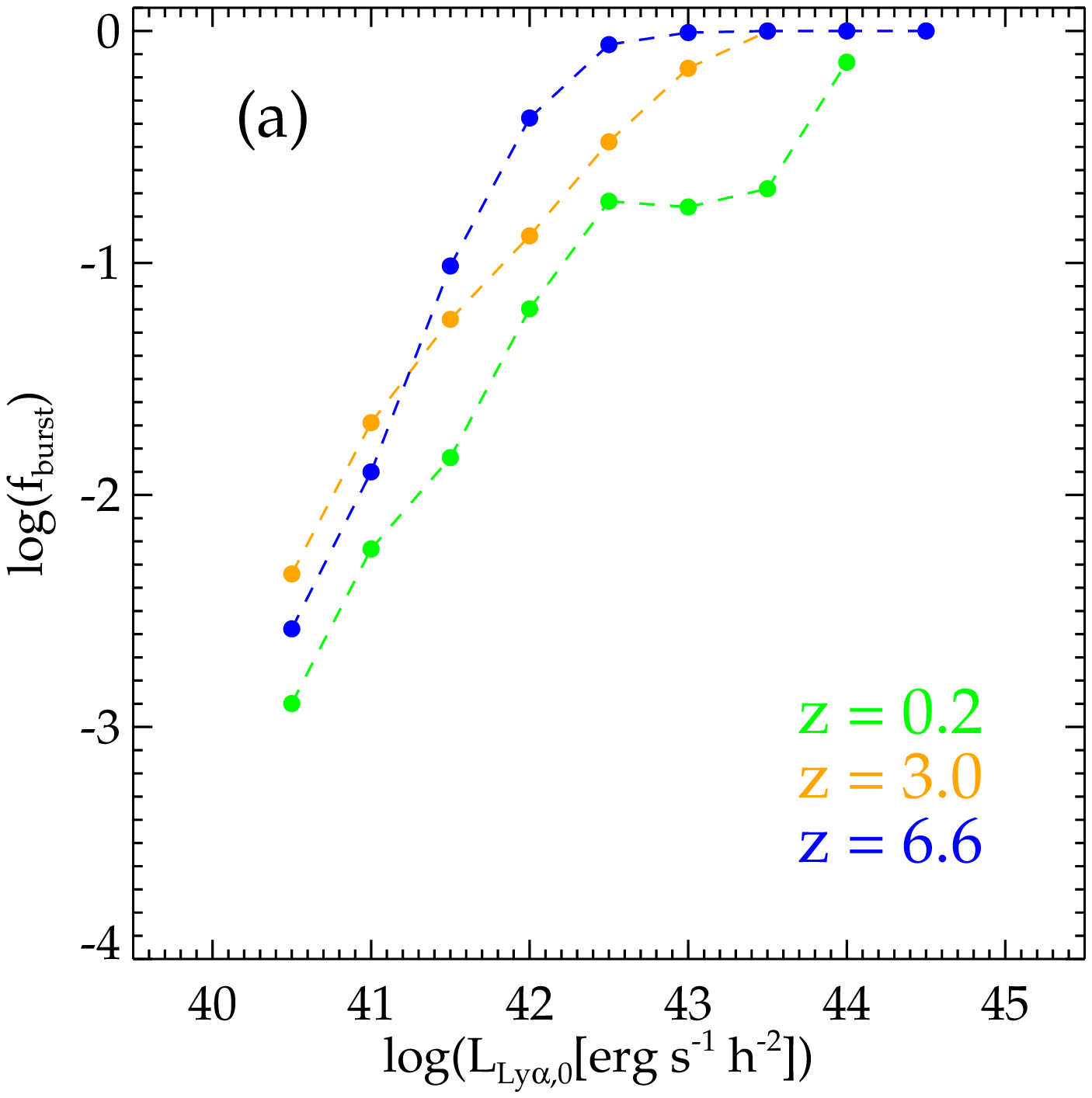}
\includegraphics[width=5.6cm,angle=90]{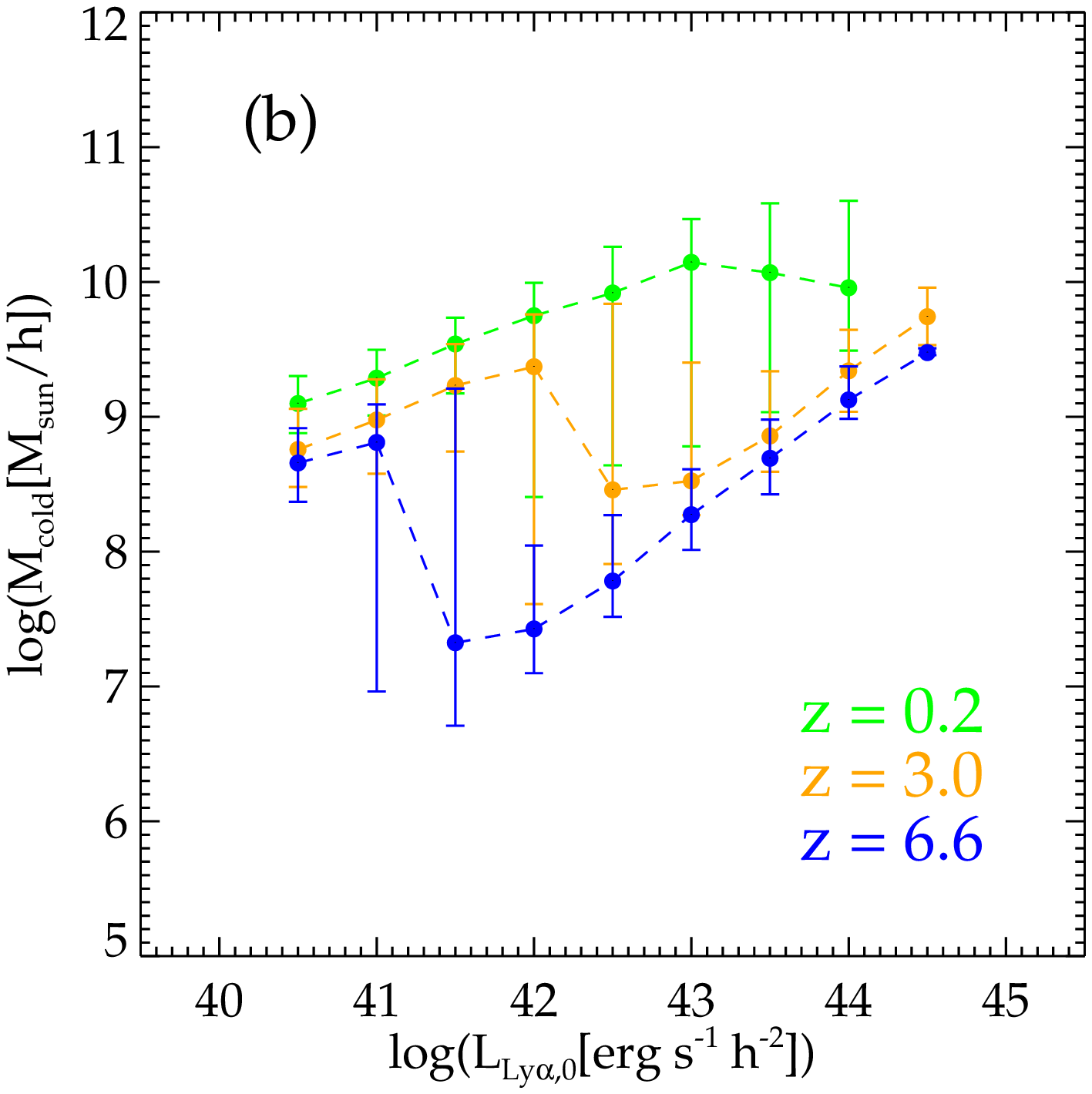}
\includegraphics[width=5.6cm,angle=90]{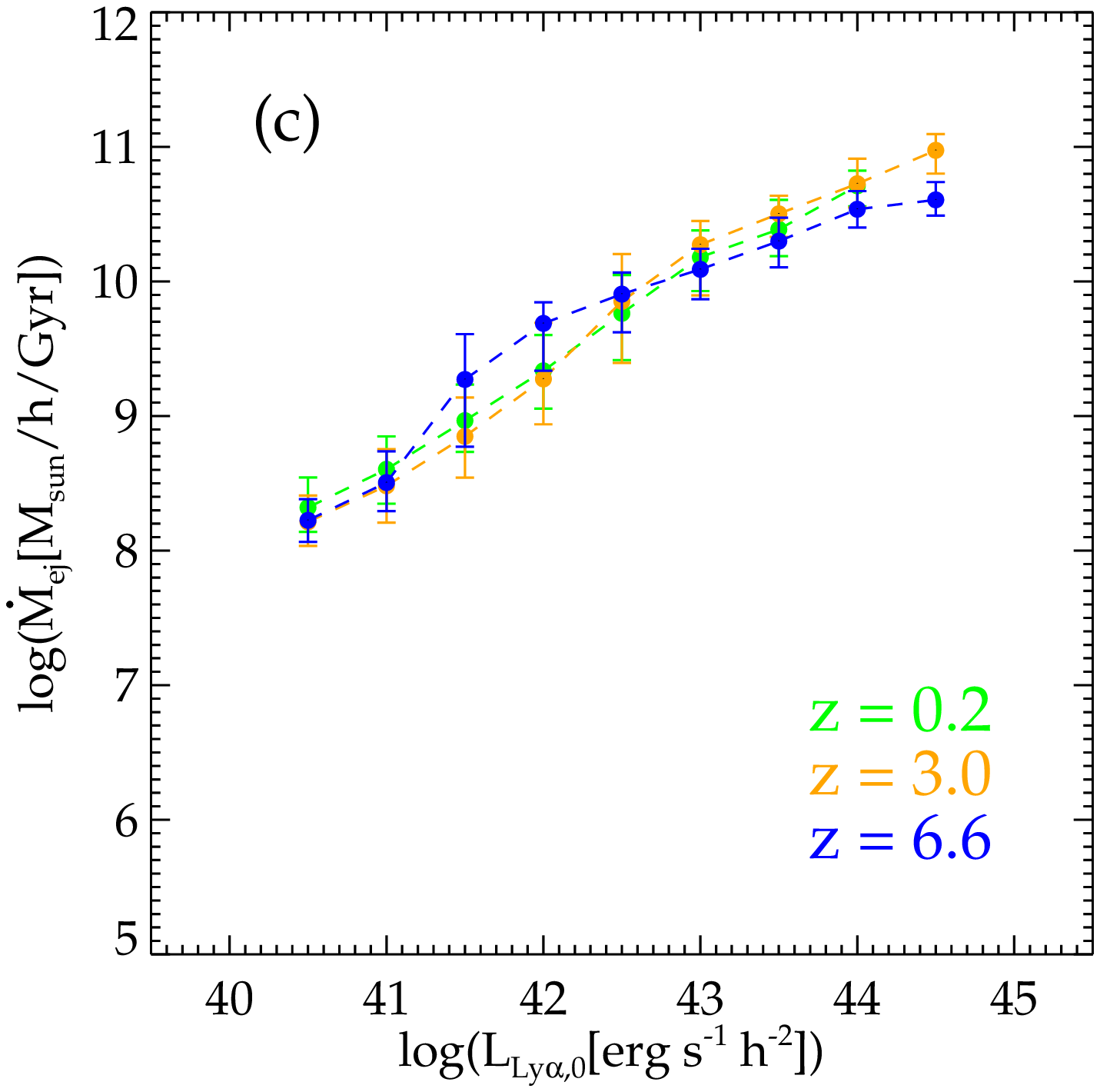}
\includegraphics[width=5.6cm,angle=90]{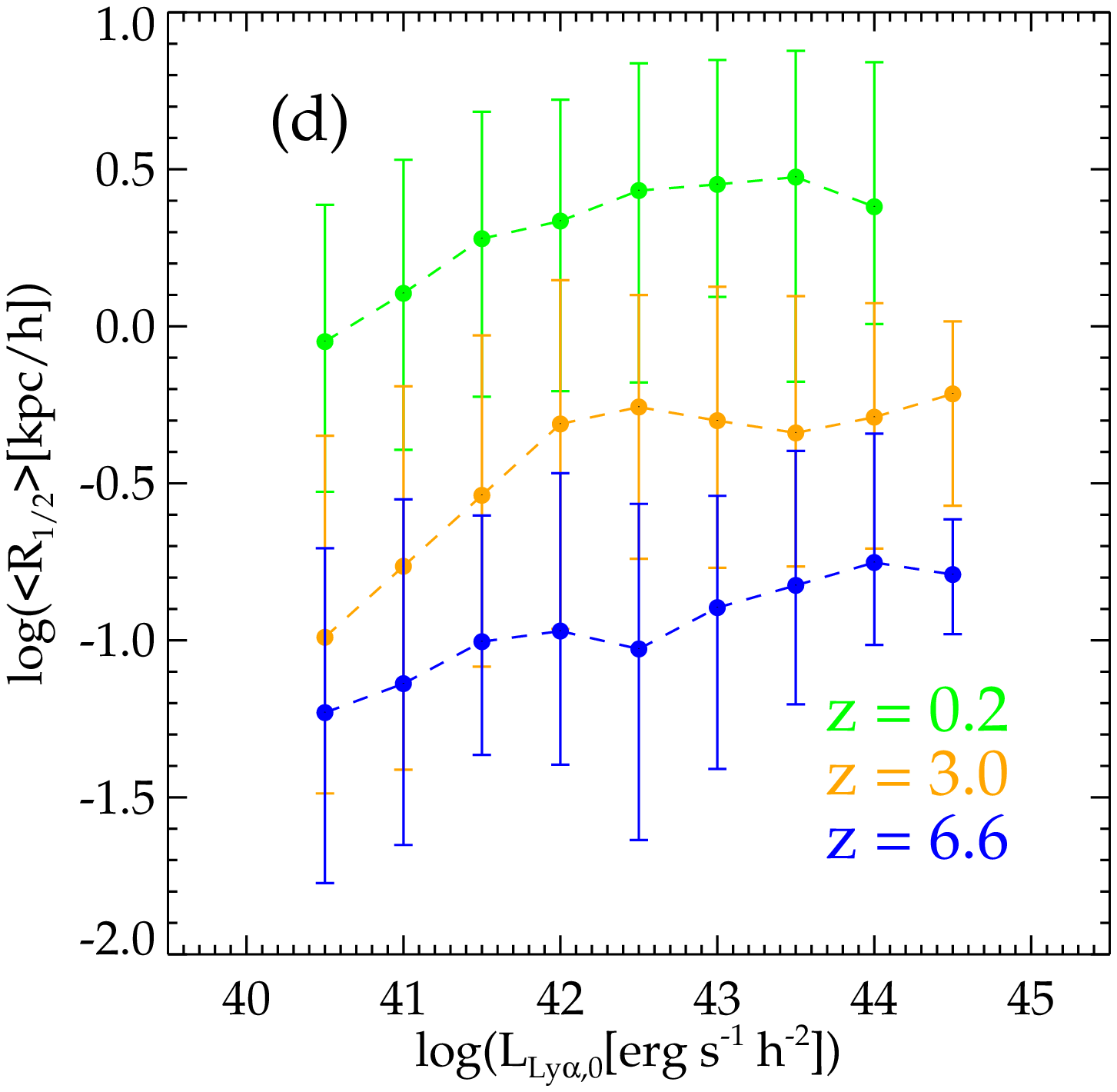}
\includegraphics[width=5.6cm,angle=90]{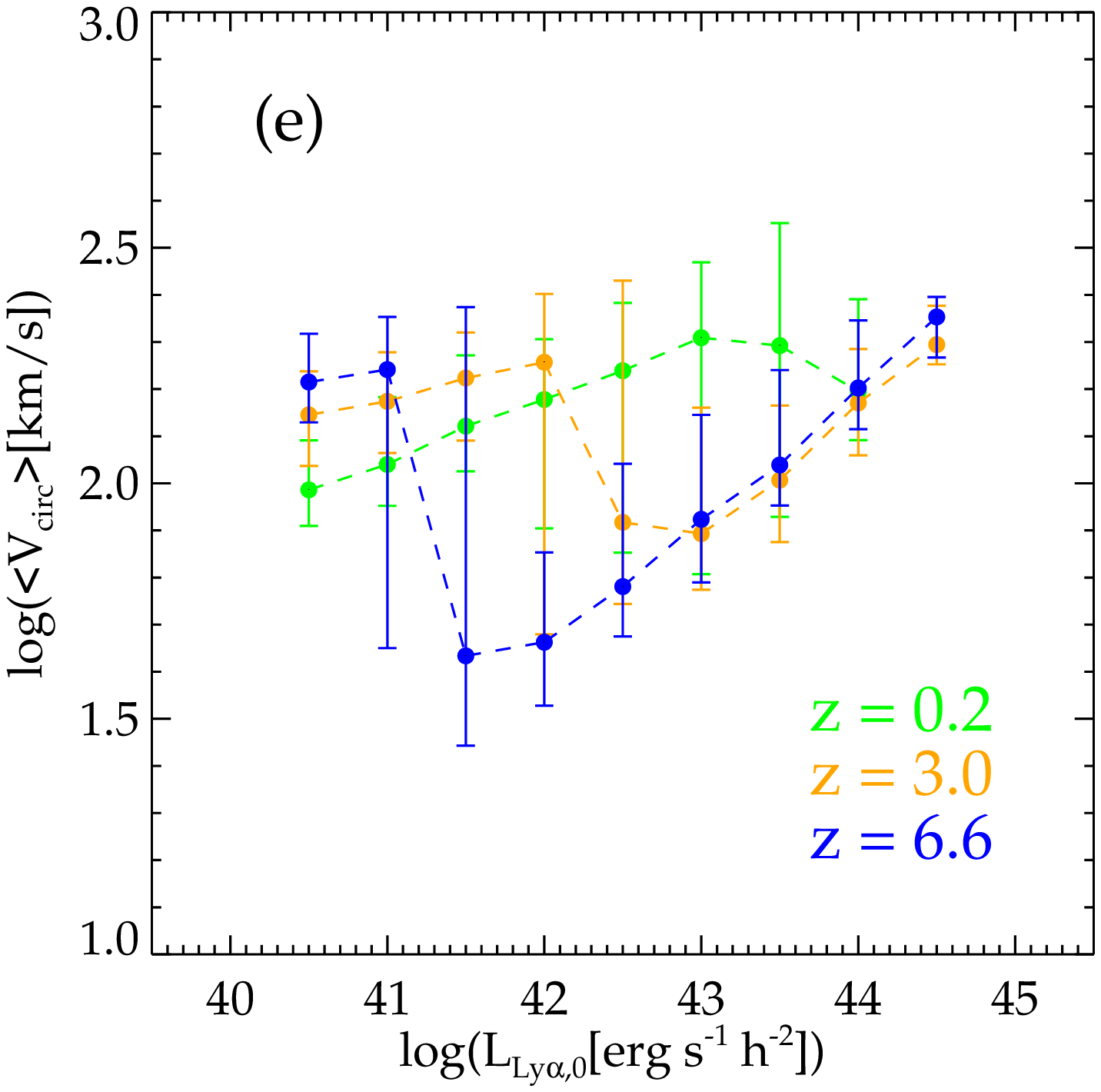}        
\includegraphics[width=5.6cm,angle=90]{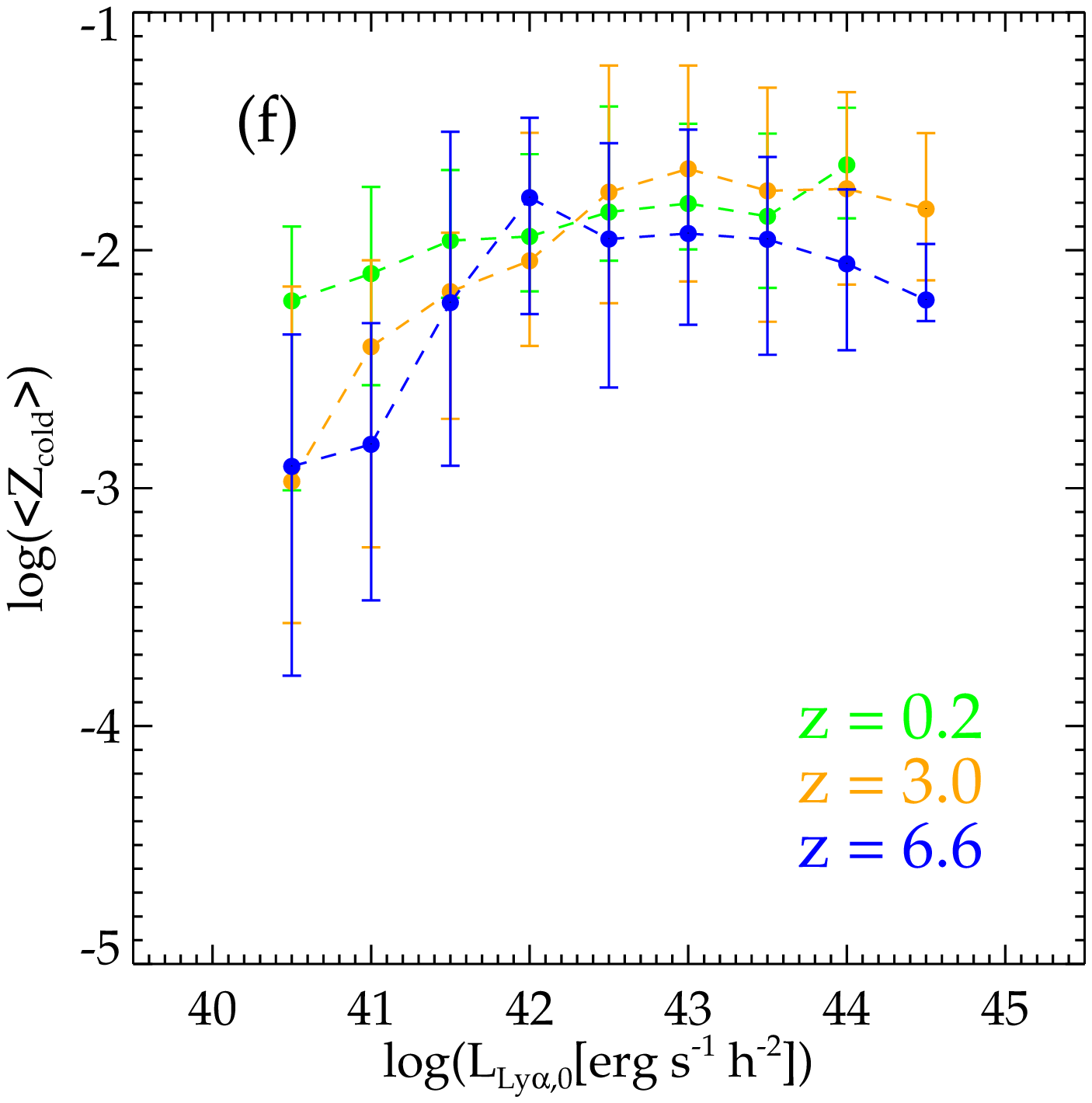}        
\caption{The evolution of the galaxy properties predicted by GALFORM which 
are used as inputs to the outflow geometries for $z=0.2$ (green), 
$z=3.0$ (orange) and $z=6.6$ (blue). The panels show, as a function of intrinsic 
\lya\ luminosity, median values for (a) the fraction of starbursts; (b) the cold gas mass of the galaxy, $M_{\rm cold}$; 
(c) the mass ejection rate, $\dot{M}_{\rm ej}$; (d) the luminosity-weighted 
half-mass radius, $\langle R_{1/2}\rangle$; (e) the luminosity-weighted circular 
velocity, $\langle V_{\rm circ} \rangle$ and (f) the (mass and \lya\ luminosity)-weighted 
metallicity of the gas $\langle Z_{\rm gas}\rangle$ (Eq. \ref{eq.zout}). Error bars show the 10-90 
percentile range of the predicted distributions.}
\label{fig.props}
\end{figure*}

\subsection{Galaxy properties}
\label{sec.galprop}
Each of our outflow geometries requires a series of galaxy properties, provided 
by \galform, to compute the escape of \lya\ photons. We consider the 
contribution of the disk and the bulge to compute averaged quantities,
as described above.
These are (i) the half-mass radius, $R_{1/2}$, (ii) the circular velocity, $V_{\rm circ}$, 
(iii) the metallicity of the cold gas, $Z_{\rm cold}$, and (iv) the mass 
of cold gas of the galaxy, $M_{\rm gas}$. 
In addition, the Wind geometry requires the mass ejection rate due to supernovae, 
$\dot{M}_{\rm ej}$ (see Eqns. \ref{eq.m_ej} and \ref{eq.mdot}).

Fig. \ref{fig.props} shows the evolution of the galaxy 
properties listed above in the redshift range $0 < z < 7$, as a function of 
the intrinsic \lya\ luminosity, $L_{Ly\alpha,0}$. It is worth noting that these properties are 
extracted directly from \galform, so they do not depend 
on the details of the outflow model.

In Fig. \ref{fig.props}(a) we show the fraction of starbursts as a function of 
intrinsic \lya\ emission. Naturally, given the form of the IMF adopted, starbursts 
dominate in the brightest luminosity bins, regardless of redshift. However, 
the transition between quiescent and starburst \lya\ emitters
shifts towards fainter luminosities as we go to lower redshifts. An important 
consequence of this trend is that the nature of \lya\ emitters,  
even at a fixed intrinsic \lya\ luminosity, is redshift dependent. Moreover, 
one might expect that the environment in which 
\lya\ photons escape in quiescent galaxies and starbursts is different. 
Although our fiducial outflow geometries do not make such a distinction, 
in section \ref{sec.tuning} we make our outflow geometries scale differently 
with redshift depending on whether galaxies are quiescent or 
starbursts.

In Fig. \ref{fig.props}(b) we show the dependence of the cold gas 
mass on the intrinsic \lya\ luminosity. As expected, in general
the cold gas mass increases with \lyaint\  at a given redshift, 
since the latter is directly proportional to the star formation rate of galaxies,  
and, in this variant of \galform, the star formation rate is directly proportional 
to the cold gas mass, as shown in Eq. \eqref{eq.sfr}. Note the star formation
timescale is different for quiescent and starburst galaxies (see \citealt{baugh05}
for details on how the star formation is calculated in this variant of \galform, 
and \citealt{lagos11} for an alternative model). 

At low \lyaint\, quiescent galaxies dominate, whereas bright 
galaxies are predominately starbursts. 
However, there is a luminosity range in which quiescent 
galaxies and starbursts both contribute. This is shown in Fig. 
\ref{fig.props}(b) by a break in the cold gas mass-luminosity relation, 
which occurs in the luminosity range where low mass starbursts are 
as common as massive quiescent galaxies. This luminosity 
corresponds to $\llya \lunits \sim 10^{43}$ at $z\sim 0$, and
it shifts towards fainter \lya\ luminosities at higher redshifts. 
At $z\sim 6$, both quiescent and starbursts contribute at $\llya \lunits \sim 10^{41.5}$.

The total (disk $+$ bulge) mass ejection rate is found to correlate strongly with the intrinsic 
\lya\ luminosity, as shown in Fig. \ref{fig.props}(c).
However, no significant evolution is found with redshift. 
Since the mass ejection rate due to supernovae is directly 
proportional to the star formation rate, as shown in Eqns. 
\eqref{eq.betareh} and \eqref{eq.betasw},  
and the conversion between the star formation rate and $L_{Ly\alpha,0}$ 
depends on the production rate of Lyc 
photons, but not on redshift, 
it is not surprising that the mass ejection rate does not evolve 
with redshift.

The half-mass radius $\langle R_{1/2} \rangle$ is the parameter 
that is found to have the strongest evolution with redshift, as shown 
in Fig. \ref{fig.props}(d). Galaxies at $z=0.2$ typically have half-mass 
radii of a few ${\rm kpc/h}$. The median size of galaxies decreases rapidly with 
increasing redshift, falling by an order of magnitude or more by 
$z=6.6$. In contrast, $\langle R_{1/2} \rangle$ varies only weakly with \lyaint.

The circular velocity of galaxies depends both on their mass and 
half-mass radius. The sizes of galaxies are computed based on angular 
momentum conservation, centrigugal equilibrium for disks and virial equilibrium and energy
conservation in mergers for spheroids, as described in detail by \citet{cole00}. 
We find that the sizes correlate only weakly
with the intrinsic \lya\ luminosity, as shown in Fig. \ref{fig.props}(d), 
and although the total mass of the galaxy is not necessarily related to the cold gas mass,
the circular velocity $\langle V_{\rm circ} \rangle$ has similar form to the dependence of  
$\langle M_{\rm gas} \rangle$ on $L_{Ly\alpha,0}$, as shown in Fig. \ref{fig.props}(e). 
This explains the break in this relation at $z=6.6$, which corresponds to the 
switch from quiescent to burst galaxies, as was the case for the cold gas mass.

Finally, the metallicity $\langle Z_{\rm cold} \rangle$ correlates strongly with 
$L_{Ly\alpha,0}$ but fairly weakly with redshift, as shown in 
Fig. \ref{fig.props}(f). The metallicity of galaxies with  
$\llya \lunits > 10^{42}$ is found to be around $\sim 10^{-2}$.

The quantities shown in Fig. \ref{fig.props} are fed into the Monte Carlo radiative transfer code to calculate the 
\lya\ escape fraction \fesc\ and the line profile. The net \lya\ luminosity of the galaxy is then simply 
$L_{\rm Ly\alpha} = \fesc  L_{\rm Ly\alpha,0}$. 
The value of \fesc\ depends, of course, on the outflow geometry.

\subsection{Choosing the outflow parameters}
\label{sec.tuning}

\begin{figure*}
\centering
\includegraphics[width=8.5cm,angle=90]{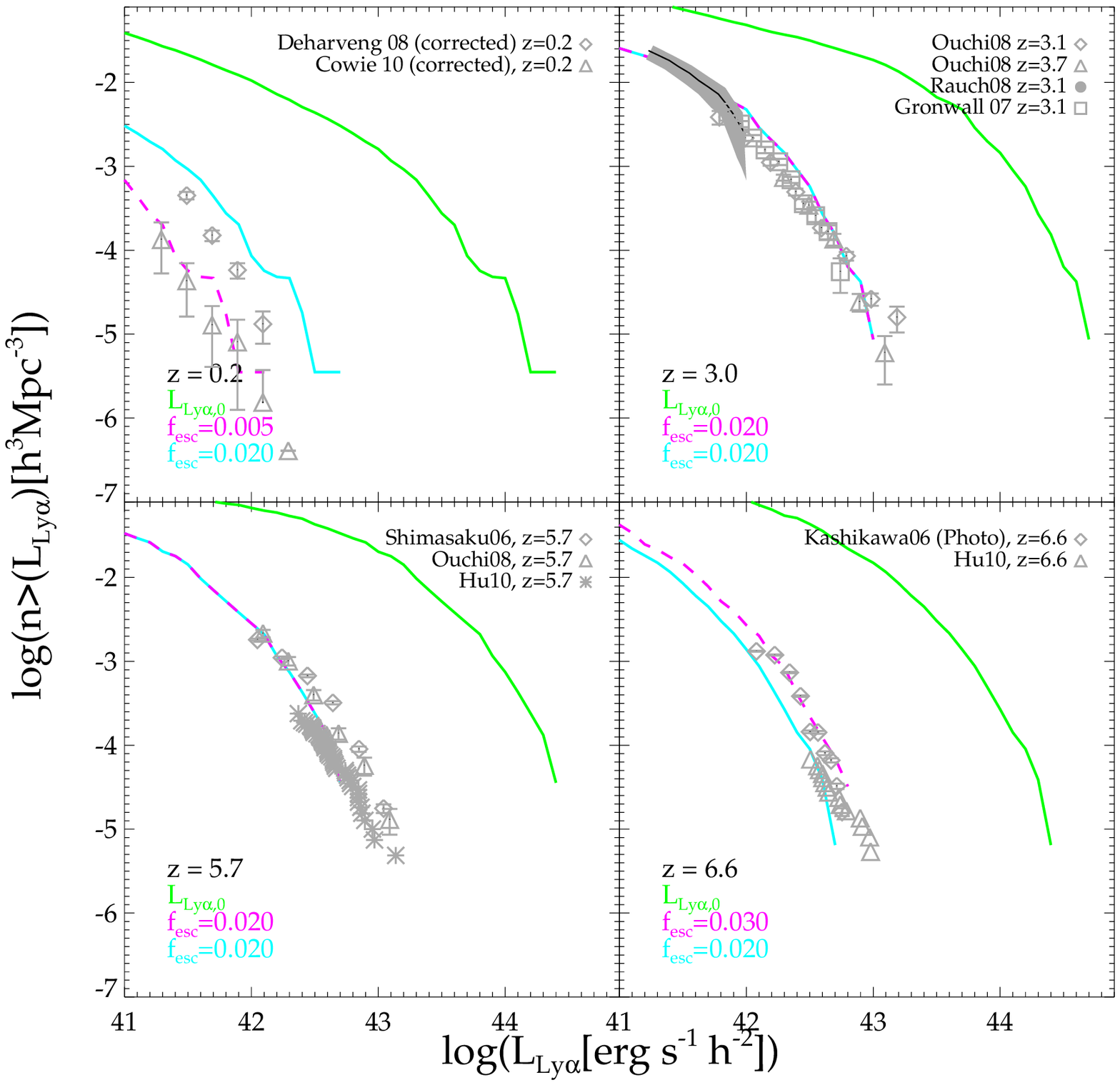}
\includegraphics[width=8.5cm,angle=90]{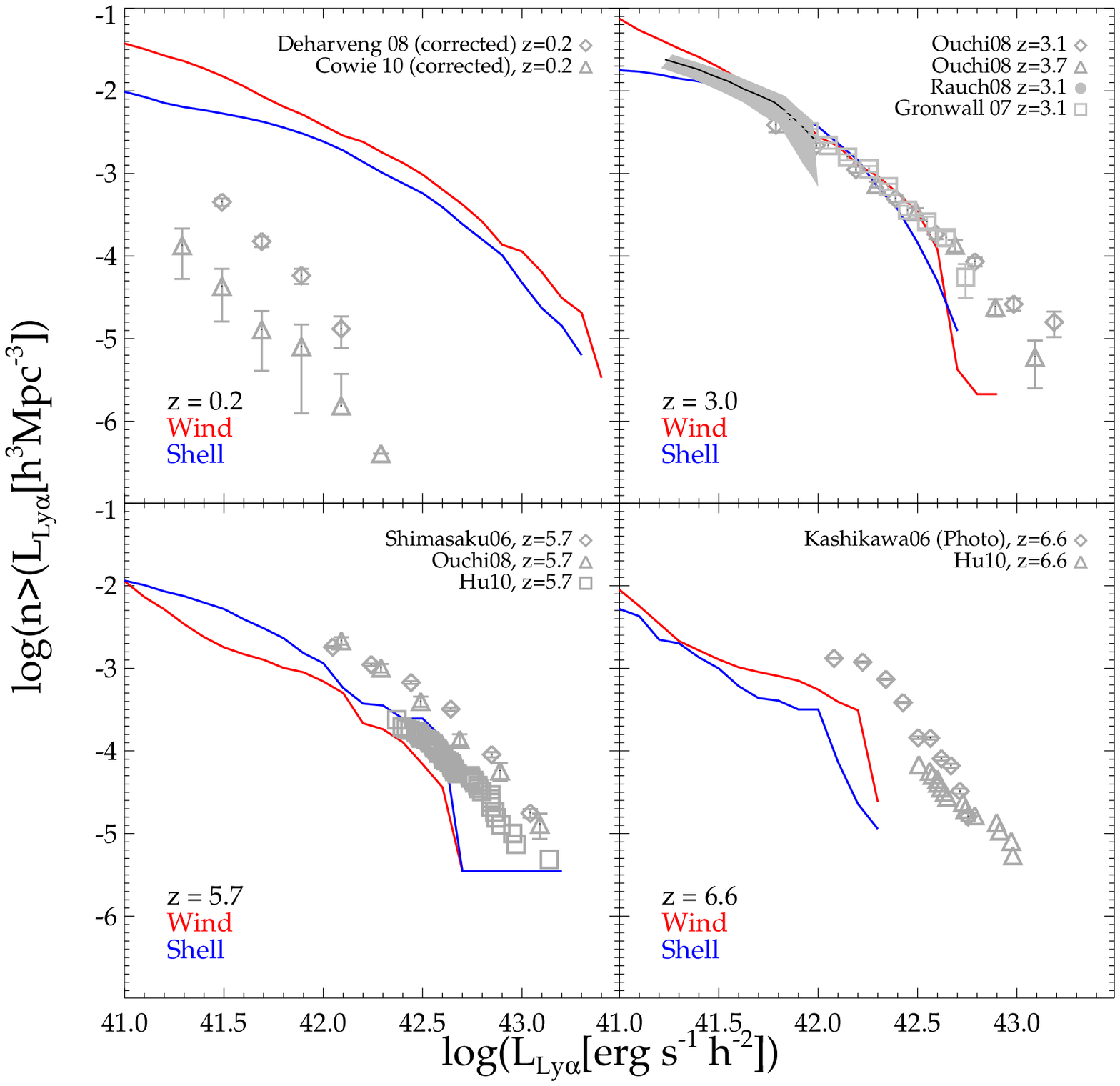}
\caption{The cumulative luminosity function of \lya\ emitters at redshifts $z=0.2,3.0,5.7$ and $6.6$. 
The green lines in the left panel show the CLF obtained using the intrinsic (without attenuation) \lya\
luminosity at different redshifts. The cyan lines show the CLF obtained when applying a fixed global 
escape fraction of $\fesc = 0.02$ to all galaxies. The magenta dashed lines show 
the effect of applying an escape fraction that varies with redshift, to 
match the observational data. The values chosen are written in the labels.
The right panel shows with solid red (blue) lines the CLFs obtained with the Wind (Shell) geometry
when choosing the free parameters to match the observational CLF at $z=3$, and then the same
parameters were then used to predict the CLF at the other redshifts (see the text for details).
In both panels, the observational CLFs at each redshift are shown with gray symbols: At $z=0.2$ 
data is taken from \citet{deharveng08} and \citet{cowie10}; 
at $z=3.0$ from \citet{gronwall07,ouchi08} and \citet{rauch08}; at $z=5.7$ from 
\citet{shimasaku06,ouchi08} and \citet{hu10}; and 
at $z=6.6$ from \citet{kashikawa06} and \citet{hu10}.}
\label{fig.first_cumlf}
\end{figure*}

%\begin{figure}
%\centering
%\includegraphics[width=8cm,angle=90]{figs/cum_lf2.eps}
%\caption{The cumulative luminosity function of \lya\ emitters at redshifts $z=0.2,3.0,5.7$ and $6.6$. 
%The green line shows the CLF obtained using the intrinsic (without attenuation) \lya\
%luminosity. The cyan line shows the CLF obtained when applying a global fixed
%escape fraction of $\fesc = 0.02$ to all galaxies. The magenta dashed line shows 
%the effect of applying a fixed escape fraction but varying with redshift, to 
%match the observational data.
%Observational CLFs at each redshift are shown with gray symbols: At $z=0.2$ 
%data is taken from \citet{deharveng08} and \citet{cowie10}; 
%at $z=3.0$ from \citet{gronwall07,ouchi08} and \citet{rauch08}; at $z=5.7$ from 
%\citet{shimasaku06,ouchi08} and \citet{hu10}; and 
%at $z=6.6$ from \citet{kashikawa06} and \citet{hu10}.}
%\label{fig.fixed_fesc_cumlf}
%\end{figure}

In the following, we outline the procedure to choose 
the value of the free parameters of our outflow models.
Our strategy to set the value of these parameters ($[f_M,f_V,f_R]$ 
for the Shell geometry, and $[f_V,f_R]$ for the Wind geometry) consists 
of matching the observed cumulative \lya\ luminosity functions (CLFs)
in the redshift range $0<z<7$.
Then, we use the values obtained to make the predictions 
in the remainder of the paper.
  
%\begin{figure}
%\centering
%\includegraphics[width=8cm,angle=90]{figs/run_final_cum_lf.eps}
%\caption{The cumulative luminosity function of \lya\ emitters at redshifts $z=0.2,3.0,5.7$ and $6.6$,
%similar to Fig. \ref{fig.fixed_fesc_cumlf}.
%The CLF predicted using the Wind model is shown with 
%a solid red curve, and the Shell model is shown in blue.}
%\label{fig.simple_cumlf}
%\end{figure}  
  
A similar strategy was followed in previous modelling of 
\lya\ emitters \citep{ledelliou05,ledelliou06,nagamine06,kobayashi07}.
However, it is worth pointing out that, to date, the model presented here
is the only one that attempts to match the recently observed abundances of 
\lya\ emitters at $z = 0.2$ and at higher redshifts at the same time. 

As a first step towards our definitive model, we show in 
the left panel of Fig. \ref{fig.first_cumlf} the cumulative luminosity
function (CLF) predicted by \galform\ alone and applying a fixed \fesc, 
i.e. without using the Monte Carlo radiative transfer code. 

The CLF constructed using the intrinsic \lya\ luminosity of galaxies
greatly overpredicts the observed estimates 
in the redshift range $0.2<z<6.6$. This is not surprising, since, 
as discussed earlier, \lya\ photons are expected to suffer an important
attenuation due to the presence of dust, and therefore, to have small
escape fractions. Fig. \ref{fig.first_cumlf} shows also the results of the
approach followed by \citet{ledelliou05,ledelliou06} and \citet{orsi08}, 
in which $\fesc = 0.02$ is adopted for all 
galaxies, regardless of their physical properties or redshift. This method
is equivalent to a global shift in the intrinsic \lya\ CLF faintwards.
This assumption is able to match remarkably well 
the observed CLF of \lya\ emitters at $z=3$.  Also, it is found to provide
a good fit to observational estimates at $z=5.7$ and to slightly underpredict the
observed CLF at $z=6.6$. However, the largest difference occurs at $z=0.2$, 
where the fixed $\fesc=0.02$ scenario overpredicts the observational CLF by a factor
of $\sim 5$ in \lya\ luminosity. Note that the $z=0.2$ CLF estimate was not available
at the time Le Delliou et al. studied \lya\ emitters with a constant value for \fesc, 
and therefore these authors were not aware of this disagreement.

Fig. \ref{fig.first_cumlf} also shows the effect of choosing a value for the 
\lya\ escape fraction that varies with redshift. At $z=0.2$, we find that a 
value of $\fesc = 0.005$ is needed to match the observational data. 
At $z=6.6$, a value of $\fesc = 0.03$ provides a better fit to 
the observational data than $\fesc = 0.02$. 

This second method, i.e. varying \fesc\ with redshift to find the best fitting value, 
has been also used in previous works
\citep[e.g.][]{nagamine10}. 
Although it reproduces the observed abundances of \lya\ emitters
at different redshifts, it lacks physical motivation. Therefore, 
we now turn to implementing our Monte Carlo radiative transfer 
model to compute \fesc.

In our implementation of the \lya\ radiative transfer code in \galform,
the code can take anything 
from a few seconds up to minutes to run for a single configuration.
Hence, the task of finding the best combination of parameters that 
match the observational CLFs shown in Fig. \ref{fig.first_cumlf}  
could be computationally infeasible 
if we were to run the code on each galaxy from a \galform\ run
(which could be in total a few hundred thousand or even 
millions of galaxies). To tackle this practical issue we 
construct a grid of configurations for a particular choice of
parameters values. This grid spans the
values of the physical properties predicted by \galform\ that are relevant to constructing
the outflows. The grid is constructed in such way that the number of grid points
is significantly smaller (a factor 10 or more) than the number of galaxies
used to construct it.
An efficient way to construct the grid allows us to run the 
Monte Carlo radiative transfer code over each of the grid points, 
hence obtaining a value for \fesc\ in a reasonable time. 
Finally, we interpolate 
the values of \fesc\ in this multidimensional grid for each 
galaxy in \galform, according to their physical 
parameter values, thus obtaining a value of \fesc\ for each
galaxy. 
The methodology to construct
the grid and to interpolate the value of \fesc\ for each galaxy
is described in detail in Appendix \ref{appendix.grid}.

\begin{figure*}
\centering
\includegraphics[width=12cm,angle=90]{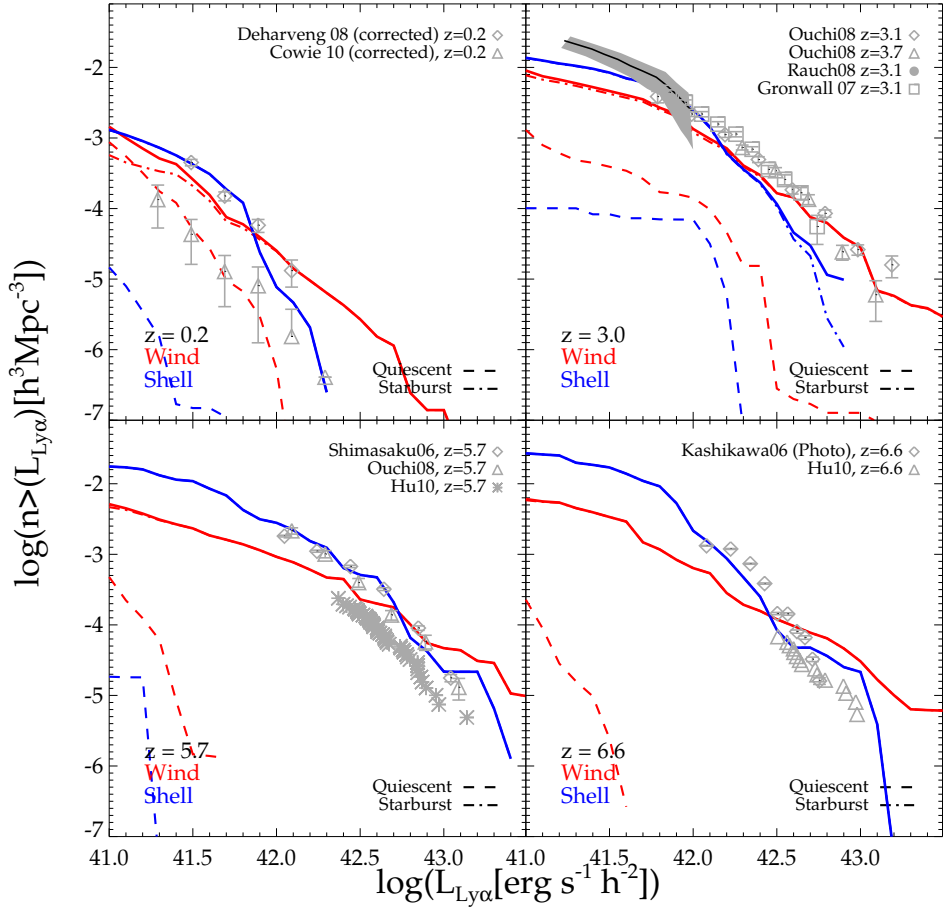}
\caption{The cumulative luminosity function of \lya\ emitters at redshifts $z=0.2,3.0,5.7$ and $6.6$. 
The CLF predicted using the Wind geometry is shown with 
a solid red curve, and the Shell geometry is shown in blue. 
The contribution of quiescent and starbursts to the model CLFs are shown 
with dashed and dot-dashed curves, respectively. Note that at z>3 starbursts
completely dominate the total CLF.
Observational CLFs at each redshift are shown with gray symbols, like in Fig. \ref{fig.first_cumlf} }
\label{fig.cumlf}
\end{figure*}

Ideally, we would like to find a single set of parameter values to reproduce 
the observed CLFs at all redshifts. Fig. \ref{fig.first_cumlf} shows the 
result of choosing the best combination of parameters to match the 
observational CLF at $z=3$. We chose this particular redshift 
because the observational measurements of the CLF at this redshift
span a broad range of \lya\ luminosities, and the scatter between the
different observational estimates of the CLF is smaller than at 
other redshifts.

For simplicity, we have chosen $f_V = 1.0$ 
in both geometries. Also, $f_M = 0.1$ was used for the Shell geometry.
Hence, the only truly adjustable parameter in each model is $f_R$. We have
found that a value of $f_R = 0.85$ for the Shell geometry, and $f_R = 0.15$
for the Wind geometry are needed to match the observed CLF of \lya\ emitters
at $z=3$. 

Despite finding a reasonable fit to the observational data at this redshift, 
neither of our outflow geometries is able to match the observed CLFs over the whole redshift range 
$0<z< 7$ if one fixed set of parameters is used. 

As shown in Fig. \ref{fig.first_cumlf}, the observed CLF at $z=0.2$ is particularly difficult to match when 
the model parameters have been set to match the CLFs at higher redshift. 
This occurs because the number density of \lya\ emitters at $z=0.2$ 
reported observationally \citep{deharveng08,cowie10} 
is much lower than what our model predicts, 
implying \fesc\ is also
considerably lower than what our model suggests. 
On the other hand, the observed CLF at $z=6.6$ implies a higher abundance of 
\lya\ emitters than what our model for computing the \lya\ escape fraction
predicts.

%Therefore, this suggests an 
%important role of the physical quantities 
%varying strongly with redshift in determining the properties of \lya\ emitters. 
%From these, the most relevant is the half-mass radius, as discussed before and shown in Fig. \ref{fig.props}.

An improved strategy to reproduce the observed CLFs of \lya\ emitters at $0<z<7$ is to choose the free
parameters of the models in the following way:

(i)  The expansion velocity of the outflows is set to be equal to the \lya-weighted circular velocity of the
galaxies (i.e. $f_V =1$ in both geometries).

(ii) In the Shell geometry, the fraction of cold gas mass in the outflow is fixed at $f_M = 0.1$.

(iii) Given the strong evolution of the half-mass radius of galaxies with redshift, 
the value of $f_R$ is allowed to evolve with redshift.

Galaxies at low redshift are predicted to be larger in size than galaxies at higher redshifts, so 
if $f_R$ is fixed like the other parameters, then outflows at lower redshift would 
have, on average, larger sizes than at high redshifts, making the associated column densities
smaller (see Eqs. \ref{eq.nh_shell} and \ref{eq.nh_wind}). If the other outflow properties
do not evolve strongly with redshift, then galaxies 
at low redshift would have larger 
\lya\ escape fractions than at higher redshifts. 
As discussed above, this is the opposite trend needed to reproduce the observed CLFs.
 
However, there is growing evidence that star 
forming regions in local ultraluminous infrared galaxies (ULIRGS) 
are significantly smaller than similarly luminous ULIRGS and submillimetre galaxies (SMGs) 
at higher redshifts \citep[see ][for a comparison of sizes]{iono09,rujopakarn11}. 
If the outflow radius in starbursts is assumed to scale with the 
size of the star forming regions instead of the full galaxy size, 
then $f_R$ would have a natural redshift dependence in the direction we need. 

\begin{table}
\centering
\begin{tabular}{@{}l|cccccc}
% \hline
% \hline
%     & (1)  & (2) \\
\hline
%                &       {\bf Wind} & {\bf Shell} \\
 & $f_M$	&$f_V$	& $f_R^q$	&$f_{R,0}^b$	&$\gamma$ \\
\hline
Shell & 0.10 & 1.00 & \valqrshell & \valkoneshell & \valktwoshell \\
Wind & --    & 1.00 & \valqrwind  & \valkonewind & \valktwowind \\
\hline
\end{tabular}
\caption{Summary of the parameter values of the Shell and Wind geometries used to 
fit the \lya\ cumulative luminosity function at different redshifts (see the text for details).}
\label{table.pars}
\end{table}

Hence, we employ a simple phenomenological evolution of $f_R$ with redshift. We call $f_R^b$ 
the radius parameter of starbursts and allow it to scale with redshift like a power-law:
\begin{equation}
 f_R^b = f_{R,0}^b(1 + z)^{\gamma},
\label{eq.f_Rb}
\end{equation}
where $f_{R,0}^b$ and $\gamma$ are free parameters. Since there is no equivalent observational evidence for the size of 
star forming regions in quiescent galaxies scaling with redshift,
we set $f_R^q$, the radius parameter for quiescent galaxies, to be an adjustable parameter but fixed (i.e. independent of redshift).

Table \ref{table.pars} summarizes a suitable choice of the parameters values used in our model. 
Also, Fig. \ref{fig.cumlf} shows the predicted cumulative luminosity functions obtained with that choice of parameters.

Our previous modelling of \lya\ emitters used the simple assumption of a constant \lya\ escape fraction, with $\fesc = 0.02$ 
being a suitable value to reproduce the \lya\ CLFs at $3<z<7$ \citep{ledelliou05,ledelliou06,orsi08}. 
As shown in Fig. \ref{fig.first_cumlf}, this simple model overestimates the CLF at $z=0.2$, but it reproduces remarkably
well the CLFs at higher redshifts. Our outflow geometries, on the other hand, are consistent with the observational CLFs 
at all redshifts, although they fail to reproduce their full shape. 
This may be surprising at first, since the intrinsic \lya\ CLF (shown in 
green in Fig. \ref{fig.first_cumlf}) roughly reproduces the shape of the observed CLFs, 
although displaced to brighter luminosities (which is 
why the constant escape fraction scenario works well at reproducing the CLFs). 
However, in our model, the escape fraction
in each galaxy is the result of a complex interplay between 
several physical properties, and this in turn modifies the 
resulting shape of the CLF. 

The sizes of the outflows predicted by our model, compared to the 
extent of the galaxies themselves (quantified by the half-mass radius 
$\langle R_{1/2} \rangle$) are very different between the two outflow geometries.
In quiescent galaxies these are $1.5$ and $20$ percent of the half-mass 
radius of the galaxies, in the Wind and Shell geometries respectively. 
Similarly, at $z=0.2$, outflows in starbursts are $2$ and $26$ percent of the 
half-mass radius in the Wind and Shell geometries. 
The rather small size of outflows in the Wind geometry appears to 
be in contradiction with observations of \lya\ in local starbursts 
which display galactic-scale outflows 
\citep[see, e.g. ][]{giavalisco96,thuan97,kunth98,mas-hesse03,ostlin09,mas-hesse09}.
However, local starburst samples are sparse and still probably not large enough to 
characterise the nature (in a statistical sense) of \lya\ emitters at low redshifts.

At higher redshifts, our model keeps the sizes of outflows in 
quiescent galaxies unchanged (with respect to their half-mass radius). 
However, outflows in starbursts grow in radius, relative to their host galaxy, 
according to a power law, 
as given by Eq. \eqref{eq.f_Rb}, with the best fitting values listed in 
Table \ref{table.pars}, so at $z=3$ their sizes are $27$ and $80$ 
percent of the half-mass radius 
for the Wind and Shell geometries, respectively. By $z=6.6$, the sizes are 
$110$ and $145$ percent of the
half-mass radius. Therefore, at $z\gtrsim 3$, all outflows in \lya-emitting 
starbursts are galactic-scale according to our models. 

A consequence of setting the free parameters in the model to reproduce the low-$z$ data is that
the CLF of \lya\ emitters at $z=0.2$ has a contribution from both quiescent and starburst galaxies. 
Despite the details over which component is dominant at this redshift 
(which is something somewhat arbitrary given the freedom to adjust the other free parameters of the models), 
the CLF of \lya\ emitters at high redshifts is invariably dominated by starbursts, and only
a negligible fraction of \lya\ emitters are quiescent galaxies.

\section{Properties of \lya\ emission}
\label{sec.results}
Having chosen the parameters in our outflow geometries we proceed 
to study the predictions of our hybrid model for the properties of 
\lya\ emitters. Whenever possible, we compare our predictions with 
available observational data.

\begin{figure}
\centering
\includegraphics[width=11cm,angle=90]{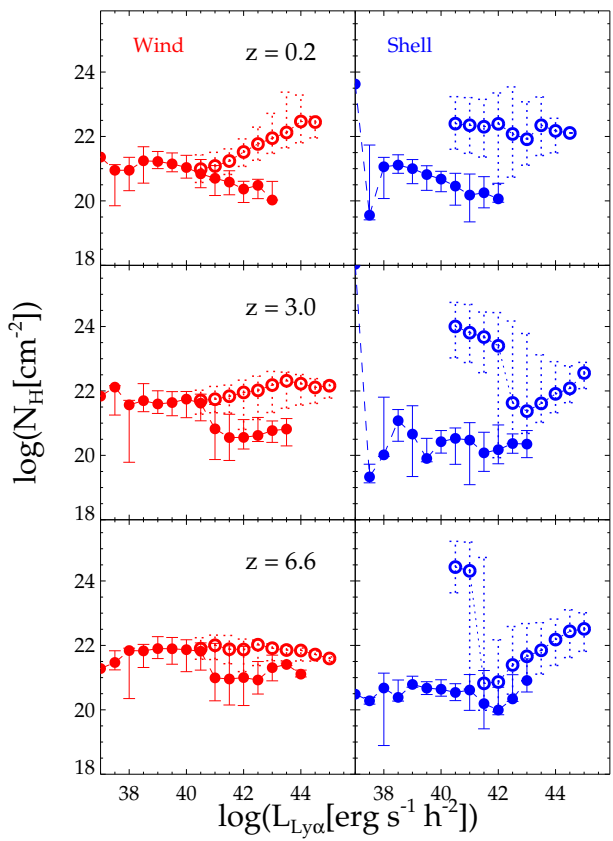}
 \caption{The neutral hydrogen column density of outflows for galaxies as a function of \lya\ luminosity at redshifts 
$z=0.2$ (top), $z=3.0$ (middle) and $z=6.6$ (bottom).
The Wind geometry is shown in red (left), and the Shell geometry in blue (right). 
Filled circles show the median of the column density 
distribution as a function of the attenuated \lya\ luminosity. Open circles show the same but as a function of the intrinsic \lya\
luminosity. Error bars show the 10-90 percentiles of the column density distribution at each 
\lya\ luminosity bin.}
\label{fig.coldens}
 \end{figure}
\begin{figure}
  \centering
 \includegraphics[width=11cm,angle=90]{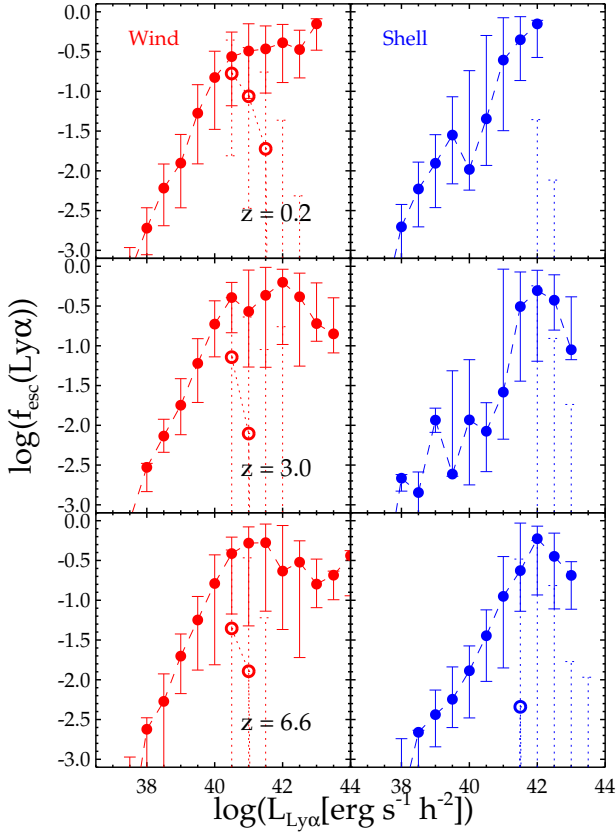}
 \caption{The escape fraction as a function of \lya\ luminosity at $z=0.2$ (top), $z=3.0$ 
 (middle) and $z=6.6$ (bottom). The Wind geometry predictions are shown in red (left), 
 and the Shell geometry in blue (right). Solid circles show the median of the escape fraction 
 at different attenuated \lya\ luminosity bins. Open circles show the same relation as a 
 function of intrinsic \lya\ luminosity instead. Error bars represent the 10-90 percentiles 
 around the median of the escape fraction distribution
at a given \lya\ luminosity bin.}
\label{fig.fesc_llya}
\end{figure}

\subsection{Column densities}
An immediate consequence of the choice of parameters shown in Table \ref{table.pars} 
is the distribution of predicted column densities of the outflows of galaxies. 
The derived hydrogen column density is shown in Fig. \ref{fig.coldens}, as a function 
of \lya\ luminosity for different redshifts. Here we study this distribution both as a 
function of intrinsic and attenuated \lya\ luminosity. 

When studied as a function of intrinsic \lya\ luminosity, our model displays a large 
range of column densities, varying from $N_H \sim 10^{21} [{\rm cm^{-2}}]$ 
to $N_H \sim 10^{24}[{\rm cm^{-2}}]$, depending on the outflow geometry and 
the redshift. Typically, in both outflow geometries, the column density increases 
with intrinsic \lya\ luminosity, which is due to the overall increase in the mass of cold 
gas and the mass ejection rate with intrinsic \lya\ luminosity.
However, in some cases, there is 
a noticeable decrease of the column density with increasing intrinsic \lya\ luminosity, reflecting
the rather complicated relation between the quantities which affect the column density and the
intrinsic \lya\ luminosity, as shown in Fig. \ref{fig.props}. 

When including attenuation due to dust, this relation is modified, as shown by the
solid circles in Fig. 
\ref{fig.coldens}. The number of scatterings scales
with the column density of the medium, hence the escape fraction is low 
(or effectively zero in some cases, according to our
model) for outflows with large column densities. As a result, 
the column density distribution as a function of attenuated \lya\ luminosity
spans $N_H \sim 10^{19-22} [{\rm cm^{-2}}]$.  Also, 
galaxies with brighter (attenuated) \lya\ luminosities tend to have smaller column densities.

The column densities predicted by our models are similar to those
inferred by \citet{verhamme08} on fitting their models
to a small sample of high redshift galaxies
with high resolution spectra. These authors fit \lya\ 
line profiles with a Monte Carlo radiative transfer model using 
a geometry identical to our Shell geometry. In addition, Verhamme et~al. 
present a compilation of results for observationally-measured column 
densities of local starbursts showing \lya\ emission,
with values between $10^{19}-10^{22} {[\rm cm^{-2}]}$, 
which are also consistent with our model predictions.
\subsection{\lya\ escape fractions}
A fundamental prediction of our models is the distribution of \lya\ escape fractions, 
shown in Fig. \ref{fig.fesc_llya}. In terms of the attenuated
\lya\ emission, we find that \fesc\ grows monotonically from $10^{-3}$ 
in the faintest \lya\ emitters (with $\llya \sim 10^{38} [{\rm erg \ s^{-1} h^{-2}}]$) 
to $\fesc \sim 0.3$ at $\llya \sim 10^{40}[{\rm erg \ s^{-1} h^{-2}}]$. 
Brighter galaxies have in general  escape fractions ranging from $0.1$ to $\sim 1$, 
depending on the redshift and outflow geometry. Perhaps not
surprisingly, both models predict similar distributions of escape fractions as a 
function of \lya\ luminosity, since the models are forced to
match the observed CLFs (Section \ref{sec.tuning}). 

On the other hand, in terms of the intrinsic \lya\ emission, Fig. 
\ref{fig.fesc_llya} shows that only in some cases it is possible to obtain 
a median \fesc\ above zero. This reflects the complicated interplay of 
physical conditions which shape the escape fraction of \lya\ photons. In other
words, the value of the intrinsic \lya\ luminosity in a galaxy does not determine
\fesc\ and, therefore, its attenuated \lya\ luminosity.

Fig. \ref{fig.fesc_llya} illustrates further the contrast betweetn using a physical 
model to compute the \lya\ escape fraction, and the 
constant \fesc\ scenario. In the former, galaxies seen with high \lya\ 
luminosities have high escape fractions, so 
their intrinsic \lya\ luminosities are similar to the attenuated ones. 
Fainter galaxies, on the other hand, have lower escape fractions, 
meaning that their intrinsic \lya\ luminosity is orders of magnitude higher 
than their observed luminosity. This implies that a \lya\ emitter with a 
given intrinsic \lya\ luminosity could have either a high or low value of 
\fesc, depending on its physical characteristics 
(galaxy size, metallicity, circular velocity, SFR, etc.).
 In the constant escape fraction scenario, on the other hand, 
 all galaxies at a given observed \lya\ luminosity have the same intrinsic 
 luminosity, which means that galaxies need only to have a high 
 intrinsic \lya\ luminosity to be observed as \lya\ emitters.

We conclude from Fig. \ref{fig.fesc_llya} that a high production of \lya\ 
photons (or, equivalently, high intrinsic Lyman continuum luminosity) does not 
guarantee that a galaxy is visible in \lya. Our modelling of
the escape fraction selects a particular population of galaxies 
to be observed as \lya\ emitters, which in general is found to have 
lower metallicities, lower SFRs and larger half-mass radii than the 
bulk of the galaxy population. 
This is discussed in more detail in Section \ref{sec.nature}.

\begin{figure}
 \centering
\includegraphics[width=8cm,angle=90]{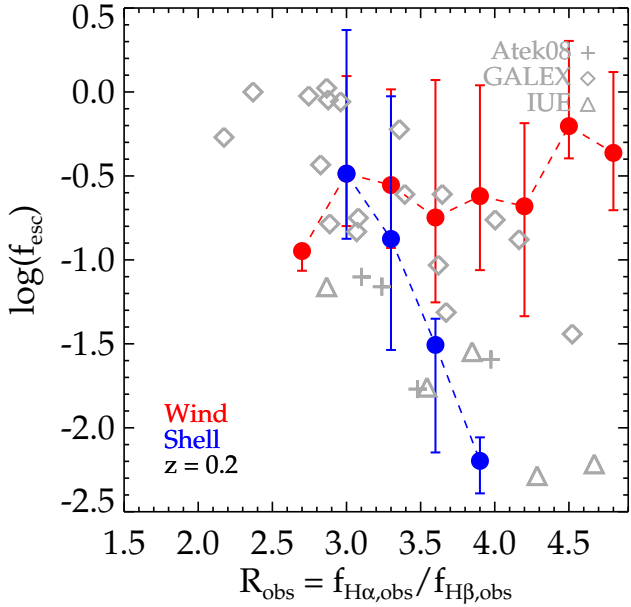}
\caption{The \lya\ escape fraction as a function of the observed (i.e. attenuated) 
ratio of H$\alpha$ to H$\beta$ flux at $z = 0.2$. 
Red circles show the median predictions using the Wind geometry, 
and blue circles show the median predictions of the Shell geometry. 
Error bars in both cases show the 10-90 percentiles of the distribution.
Grey symbols show observational measurements by 
\citet{atek08} (crosses), and GALEX and IUE samples (diamonds and 
triangles respectively) from \citet{atek09}.}
\label{fig.fesc_robs}
\end{figure}

It is worth asking at this point whether the predicted escape fractions 
are consistent with observational estimates. Observationally, \fesc\ 
is generally calculated either by inferring the SFRs from the \lya\ luminosity
and comparing to the SFR estimated from the UV continuum 
\citep[e.g. ][]{gawiser06,blanc11,guaita10}, 
or by using the ratio between \lya\ and another non-resonant 
hydrogen recombination line. The former requires assumptions about the stellar evolution 
model, the choice of the IMF and the modelling of dust extinction.
The second method only relies on the assumption that the 
comparison line is not affected by resonant scattering, and thus
the extinction can be estimated reliably from an extinction curve, 
and that the intrinsic line ratio corresponds to the ratio of the emission coefficients,
assuming case B recombination.

For this reason, we focus on the escape fraction measured using line ratios. 
These are often presented as a function of colour excess $E(B-V)$, which in 
turn is estimated from non-resonant recombination lines such as H$\alpha$ 
and H$\beta$ together with an assumed extinction curve, which describes 
foreground extinction \citep{atek08,atek09,ostlin09,hayes10b}. 
For the purposes of comparing the model predictions to the 
observationally-estimated values of \fesc, we convert the values of 
$E(B-V)$ quoted in the observations to 
$R_{\rm obs} \equiv f_{\rm H\alpha, obs}/f_{\rm H\beta,obs}$, 
the ratio between the observed fluxes of \ha\ and H$\beta$, since
this ratio is a direct prediction from \galform. To compute
$R_{\rm obs}$, we follow a standard relation \citep{atek08},
\begin{equation}
 \label{eq.ebv}
% E(B-V) = \frac{2.5 \times \log (2.86/R_{\rm obs})}{k(\lambda_{\alpha})- k(\lambda_{\beta})},
\log(R_{\rm obs}) = \log(R_{\rm int}) - E(B-V)\frac{k(\lambda_{\alpha}) - 
k(\lambda_{\beta})}{2.5},
\end{equation}
where $R_{\rm int}  = 2.86$ is the intrinsic line ratio between 
H$\alpha$ and H$\beta$ typically assumed under Case B recombination for
a medium at a temperature of $T = 10000$[K] \citep{osterbrock89}, 
and $k(\lambda_{\alpha}) = 2.63,$ and $k(\lambda_{\beta}) = 3.71$ 
are the values of the normalized extinction curve at each corresponding wavelength
from the extinction curve of \citet{cardelli89}. 

Fig. \ref{fig.fesc_robs} shows the predicted relation between the \lya\ 
escape fraction and $R_{\rm obs}$ compared to 
observational estimates from \citet{atek08} and an analysis of UV 
spectroscopic data from the GALEX and IUE surveys by \citet{atek09}.
\begin{figure*}
  \centering
 \includegraphics[width=12cm,angle=90]{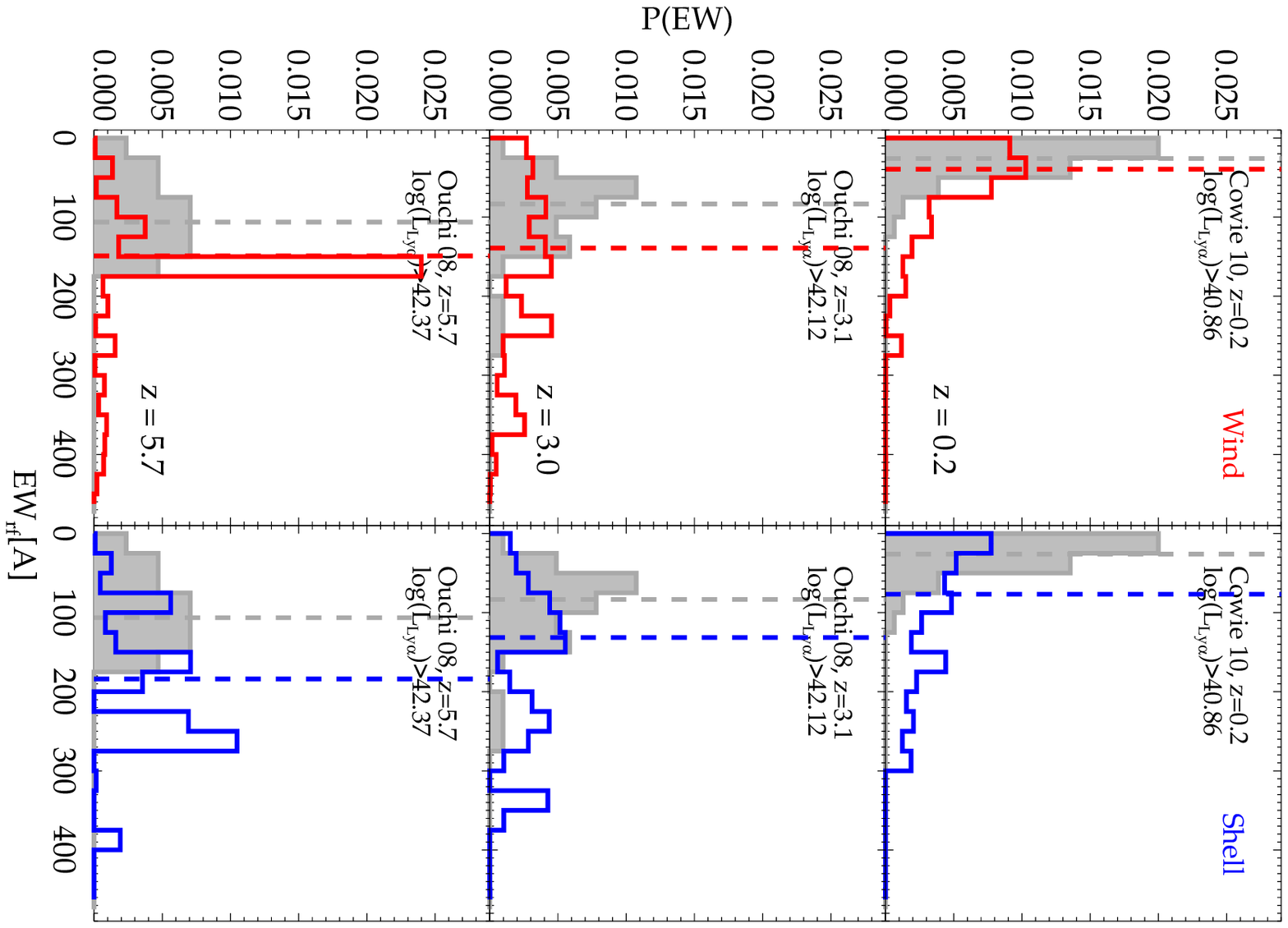}
\includegraphics[width=12cm,angle=90]{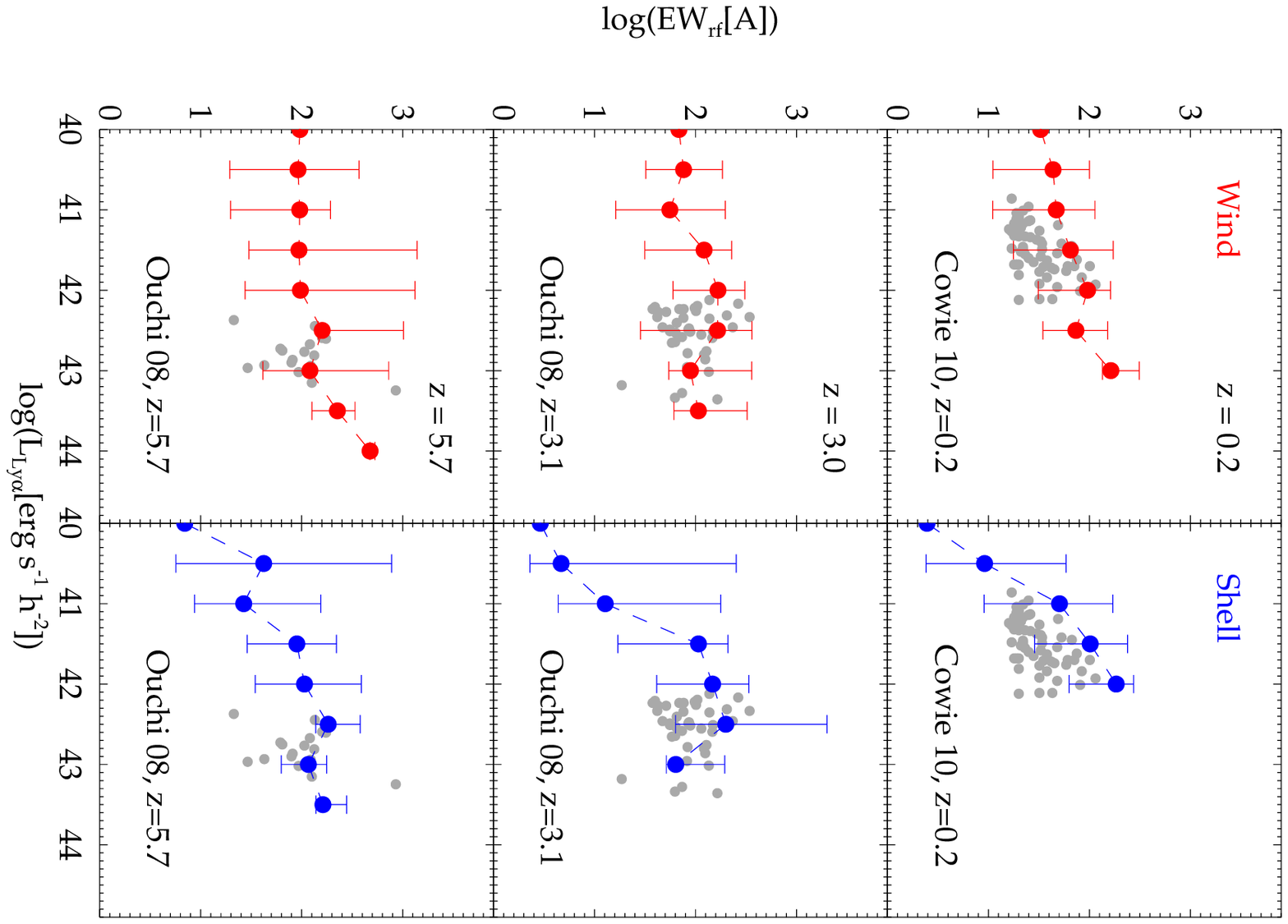}
 \caption{The (rest-frame) equivalent width distribtion at $z = 0.2, 3.0$ and $z=5.7$. 
The left panels show histograms of the distribution 
of EWs at different redshifts with the applied lower luminosity limit given in the boxes. 
The right panels show the median of the distribution of EWs as a function of \lya\ luminosity. 
In all cases, the Wind geometry
predictions are shown in the left column in red, and the Shell geometry in the right column in blue. 
The error bars on the right panel denote the 10-90 percentiles of the distribution of EWs.
Observational data (shown in gray) 
is taken from \citet{cowie10} at $z=0.2$ and \citet{ouchi08} for redshifts $z=3$ and $z=5.7$.
The vertical dashed lines in the left panel correspond to the median values of the distributions.}
\label{fig.ewdist}
\end{figure*}

The model predictions shown in Fig. \ref{fig.fesc_robs} include only 
galaxies with $\logllya > 41.5$, in order to approximately reproduce 
the selection of \lya\ emitters in the GALEX sample. Note, however, 
that the observational points shown in Fig. \ref{fig.fesc_robs} do 
not represent a complete statistical sample. Therefore it is 
not possible to perform a fair comparison, 
and the results shown here should be regarded as illustrative.

The Shell geometry shows remarkable agreement with the observational 
estimates of \fesc, reproducing the trend of lower escape fractions
 in galaxies with larger $R_{\rm obs}$. The Wind geometry, on the other hand, 
 is only partially consistent with the observational data, and it predicts a 
 rather flat relation between $\fesc$ and $R_{\rm obs}$.

%The range of $R_{\rm obs}$ covered by the observational 
%estimates ($2.0\lesssim R_{\rm obs} \lesssim 4.7$)
%is somewhat wider than that covered by {\color{magenta} the Shell geometry}
%($3.0\lesssim R_{\rm obs}\lesssim 4.0$). This difference 
%could be attributed to uncertainties in the estimates of 
%the \ha\ and $H\beta$ flux of the galaxies used in the 
%observational samples, or to the 
%calculation of the attenuation of emission lines by dust in \galform, 
%since this is computed as the extinction of the continuum at the wavelength
%of the  \ha\ and H$\beta$ lines. The details of the dust extinction calculation can be 
%found elsewhere \citep{cole00,baugh05,lacey08}.This model could be 
%extended to add an additional attenuation component of the emission lines, 
%although such modelling is beyond the scope of the present work.

\subsection{The \lya\ Equivalent width distribution}

The equivalent width $EW$ measures the strength of the line with respect to the continuum around it. 
We compute the EW simply by taking
the ratio of the predicted \lya\ luminosity of galaxies and the stellar continuum around the \lya\ line as 
computed by \galform, including attenuation by dust. 
Fig. \ref{fig.ewdist} shows a comparison of the 
EW distribution measured at different redshifts with the predictions from our outflow geometries. 

Overall, both outflow geometries predict EW distributions broader than the observational samples for all redshifts
studied. To characterise the predicted EWs, we compute the median of the distributions.
This will depend on the \lya\ luminosity limit applied to the sample to make a fair comparison 
with observed data. In Fig. \ref{fig.ewdist} we 
compare our model predictions with observational data from \citet{cowie10} at $z=0.2$, \citet{ouchi08}
at $z=3.1$ and also at $z=5.7$. 

The comparison between the observed and predicted EW by the Wind geometry median values 
(shown by vertical dashed lines in the left panel of Fig. \ref{fig.ewdist}) 
is encouraging at $z=0.2$, where the EW distributions have a median value of $\sim 30$ \AA{}.
However, both geometries give consistently higher median values of the EW when 
compared to the observational data at higher redshift. In addition, the disagreement in the median 
values of the EW distribution between observational data and the Shell geometry predictions becomes larger as we 
go to higher redshifts. 

At $z=0.2$, the EW distribution predicted using both outflow geometries is broader 
than the observational sample of \citet{cowie10} and reaches values of $EW \approx 300$\AA{}, 
whereas the observational sample only reaches $EW \approx 130$\AA{}.

At $z=3.0$, the observational sample
seems to peak around an $EW \approx 80$\AA{} and then declines until reaching 
$EW \approx 150$ \AA{}. The predicted EWs are consistent with the observational 
distributions, although the former reach values as high as $EW \approx 400$ \AA{}.
A similar disagreement between model predictions and observational data is found at $z=5.7$

\begin{figure}
\centering
\includegraphics[width=11cm,angle=90]{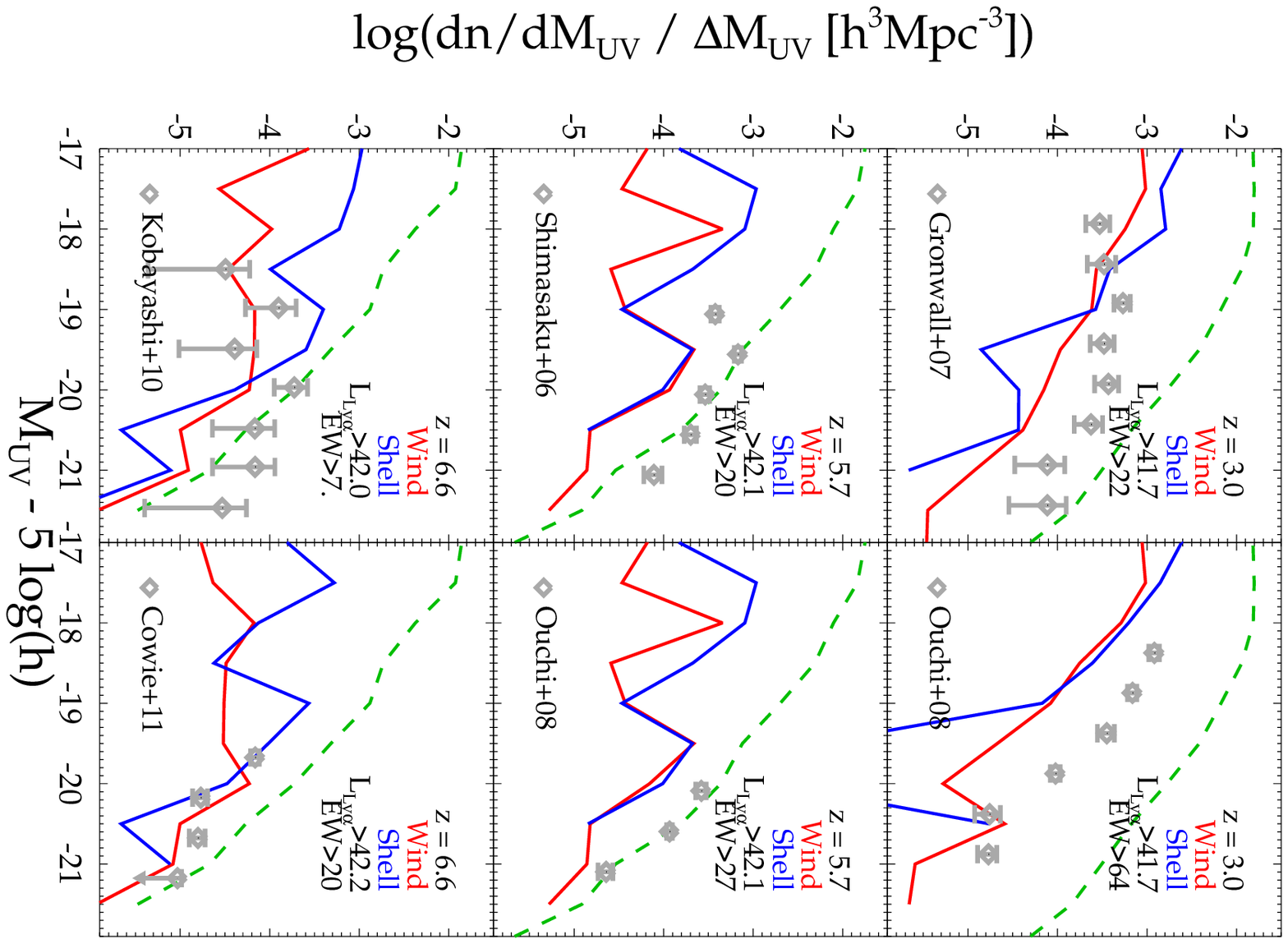}
\caption{The UV luminosity function of \lya\ emitters at redshifts $z=3.0$ 
(top), $z=5.7$ (middle) and $z=6.6$ (bottom). The Wind geometry predictions 
are shown in red, and those of the Shell geometry in blue. Observational 
data, shown with gray symbols, is taken from \citet{ouchi08} and 
\citet{gronwall07} at $z=3$, \citet{shimasaku06}
and \citet{ouchi08} at $z=5.7$ and \citet{kobayashi10} at $z=6.6$.
The limiting \lya\ luminosity and EW used to construct the models UV LF 
are shown in each panel, in units of $\log (\lunits)$ and [\AA{}], respectively.
The dashed green curves show the UV LF of all galaxies, as predicted by \galform, 
without imposing any \lya\ selection.
}
\label{fig.uvlf}
\end{figure} 
 
%Interestingly, we find that the predicted EWs at $z=5.7$ do not reproduce the observational estimates
%of \citet{ouchi08}. At this redshift, the outflow models present a rather noisy 
%distribution without any well defined peak, caused by the small 
%number of galaxies left to compute the EW distribution after the observational 
%constraints are applied. However, the bulk of the EW values are 
%consistent with the observational
%data of \citet{ouchi08}. 

The disagreement found between the EW values in the model predictions 
and the observations is difficult to understand from studying only the EW distributions.
Therefore, we perform a more detailed comparison by 
studying the relation between the median $EW$ and the \lya\ luminosity, 
as shown in the right panels of Fig. \ref{fig.ewdist}. 

Fig. \ref{fig.ewdist} shows that there is good agreement between the model 
predictions and the observational data at $z=0.2$, where both models
seem to reproduce the range and scatter of the observations. Overall, both 
models predict an increase of EW towards brighter \lya\ emiters:
The Wind geometry predicts that the median EW increases from $\approx 30$\AA{} to 
$\approx 100$\AA{} for \lya\ luminosities $\llya \approx 10^{40}-10^{42} \lunits$, whereas
the Shell geometry predicts a much steeper increase of the median EWs, from
$\approx 3$\AA{} to $\approx 100$\AA{} over the same \lya\ luminosity range.

At $z=3$, the EW values from the observed data are also consistent with both model 
predictions. The Shell geometry predicts a steep increase in the EW values with increasing \lya\ luminosity, 
whereas the Wind geometry shows a rather flat relation.
At $z=5.7$ the models are also consistent with most of the observed EW values 
within the predicted range of the EW distributions. 

\subsection{UV continuum properties of \lya\ emitters}
 
The variant of \galform\ used in this work has been previously shown to match the abundance
of LBGs (characterised by their UV luminosities) over a wide range of redshifts \citep{baugh05,lacey11,gonzalez11}. 
Therefore, a natural prediction to study with our model is the UV LF of a \lya-selected sample. 
 
%Observationally, the UV luminosity of galaxies with observed \lya\ emission has been used to put constraints on the reionization epoch
%\citep[see ][ and references therein]{kashikawa06,ouchi09,ouchi10,stark10,stark11}.

Fig. \ref{fig.uvlf} shows the UV ($1500$ \AA{}) LF of \lya\ emitters at redshifts $3.0, 5.7$ and $6.6$. 
To compare the model predictions 
with observational data, we mimic the UV selection applied to each sample. 
These correspond 
to constraints on the limiting \lya\ luminosity and the minimum EW. 
Moreover, we have chosen to compare our model predictions with two 
observational samples at each redshift. Although 
the observational samples show similar limiting \lya\ luminosity at each redshift, 
the value of the minimum EW, which is different among observational samples, 
has an important impact on the predicted 
UV LF of \lya\ emitters, as shown in Fig. \ref{fig.uvlf}.

At $z=3$, we compare our model predictions with the UV LF from 
\citet{ouchi08} and \citet{gronwall07}. Overall, our models are found to 
underpredict the observational estimates, although they are consistent 
with the faint end of the LF measured by \citet{gronwall07},  
and in reasonable agreement at the bright end with the observational sample of
\citet{ouchi08}.
Likewise, at $z=5.7$ both outflow geometries are found to undepredict
the UV LF of \lya\ emitters measured by \citet{shimasaku06} and 
\citet{ouchi08}.

The situation is somewhat different at $z=6.6$. Both outflow
geometries are consistent with the faint end of the UV LF of \lya\ 
emitters shown in \citet{kobayashi10}, who recalculated
the UV LF of \lya\ emitters from the sample of \citet{kashikawa06} 
after applying a 
brighter \lya\ luminosity cut, to ensure that a more complete
sample was analysed. Also, both outflow geometries are 
consistent with the UV LF of \lya\ emitters measured by \citet{cowie11}.

For comparison, we show in Fig. \ref{fig.uvlf} the UV LF of all galaxies
predicted by \galform, i.e. without applying any selection. Note this was already 
shown to agree remarkably well with observed LFs of LBGs in \citet{lacey11}.
This {\it total} UV LF is always above the UV LF of 
\lya\ emitters predicted by the outflow geometries.

At $z=3$, we notice that the UV LF of all galaxies is above the 
observed LFs of \lya\ emitters. This is consistent with the idea that \lya\ emitters
constitute a sub-sample of the galaxy population.  However, 
at $z=5.7$ and $z=6.6$ this UV LF matches remarkably well 
the observed UV LFs of \lya\ emitters. 
A similar finding has also been reported in 
observational papers \citep[e.g. ][]{kashikawa06,ouchi08} when 
comparing the UV LF of LBGs to that of \lya\ emitters.
Interestingly, only the observational LF from \citet{cowie11}
falls significantly below the predicted total UV LF of all galaxies, and is at the
same time consistent with the predicted UV LF of \lya\ emitters for both 
outflow geometries.

It has been argued that attenuation by the IGM might play a significant role in 
shaping the LF of \lya\ emitters at $z>5$. \citet{kashikawa06} 
interpret their measured LF of \lya\ emitters as evidence of an 
abrupt decrease in the amplitude of the bright end of the \lya\ 
LF at $z=6.6$ compared to $z=5.7$. Since the 
corresponding UV LFs of \lya\ samples do not seem to evolve 
in the same way, this has been suggested as a result of a change in the 
ionization state of the IGM (for which \lya\ photons are more 
sensitive than continuum photons).

Our model does not compute any attenuation by the IGM. 
The effect of reionization in \galform\ is modeled simply 
by preventing the cooling of gas in haloes of a given circular 
velocity ($V_{\rm circ}=30 {\rm km/s}$ 
in our \galform\ variant), when their redshift is smaller than 
a redshift $z_{\rm reion} = 10$ \citep[see][for a more detailed model]{benson02}.

Some theoretical work exploring the abundances of \lya\ emitters 
at these high redshifts has attempted to take into account the attenuation 
of the \lya\ line due to absorption by the IGM 
\citep{kobayashi07,dayal11,kobayashi10}. 
In our model, a simple mechanism to improve the agreement 
between the observed and predicted UV LFs of \lya\ emitters at high redshift would 
be to add a constant attenuation factor of the \lya\ luminosity by the IGM at high redshifts, 
set to match the UV LF of \lya\ emitters.
We defer doing this to a future paper.\\

\begin{figure}
 \centering
 \includegraphics[width=7cm,angle=90]{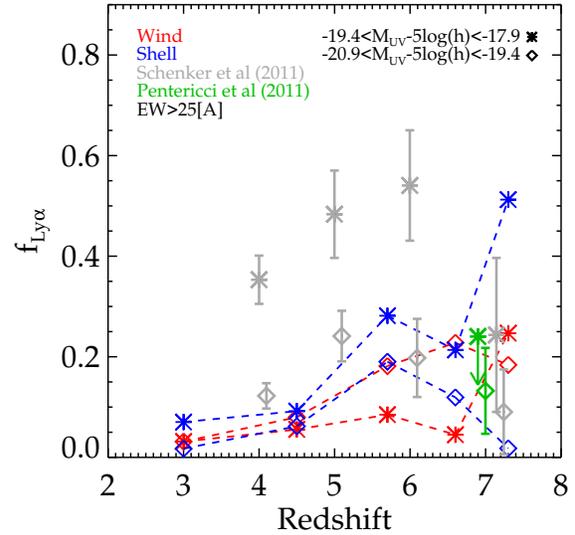}
\caption
{The fraction of \lya\ emitters with rest-frame equivalent width 
$EW_{\rm rf}> 25 $\AA{} for galaxies with $-19.4<M_{\rm UV}-5\log(h)<-17.9$ 
(asterisks) and $-20.9 < M_{\rm UV} - 5 \log(h)  <  -19.4$ (diamonds)  as a function of redshift. 
Predictions of the Wind and Shell geometries are 
shown in red and blue, respectively. 
The observational measurements of \citet{schenker11} and \citet{pentericci11} 
are shown in gray and green, respectively.}
\label{fig.xlya_z}
\end{figure}

We now focus on the 
fraction of galaxies exhibiting \lya\ in emission predicted by our models. 
As described in Appendix \ref{appendix.grid}, in our radiative transfer model 
we follow a maximum of 1000 photons per galaxy, and so we can compute a 
minimum value for the escape fraction of $10^{-3}$. 
However, in a significant fraction of galaxies, none of the photons escape 
from the outflows, resulting in them being assigned $\fesc = 0$, 
and thus having no \lya\ emission at all. 
These galaxies could be related to the observed population of galaxies 
showing \lya\ in absorption \citep[e.g. ][]{shapley03}, 
which we will examine in a future paper.  

Fig. \ref{fig.xlya_z} compares our model predictions with the observed fraction of 
\lya\ emitters found in Lyman-break galaxies (LBGs) at $z \sim 4-7$ by 
\citet{schenker11} and by \citet{pentericci11} at $z \sim 7$. 
\lya\ emitters are defined here as galaxies with a \lya\
$EW_{rf}\ge 25$\AA{}. The samples are split according to two different rest-frame $UV$ 
magnitudes ranges, as shown in Fig. \ref{fig.xlya_z}.
For simplicity, we define $UV$ magnitudes at rest-frame 
$1500$\AA{}.

\begin{figure*}
\centering
 \includegraphics[width=6cm,angle=90]{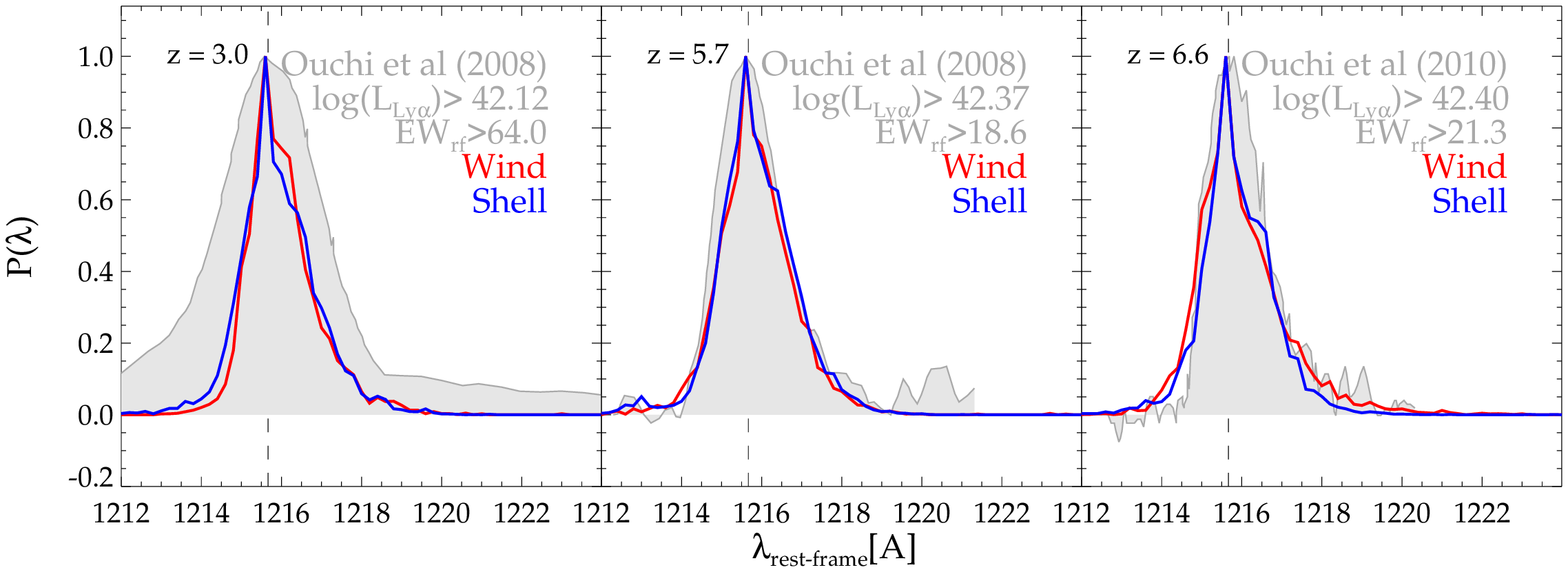}
 \includegraphics[width=6cm,angle=90]{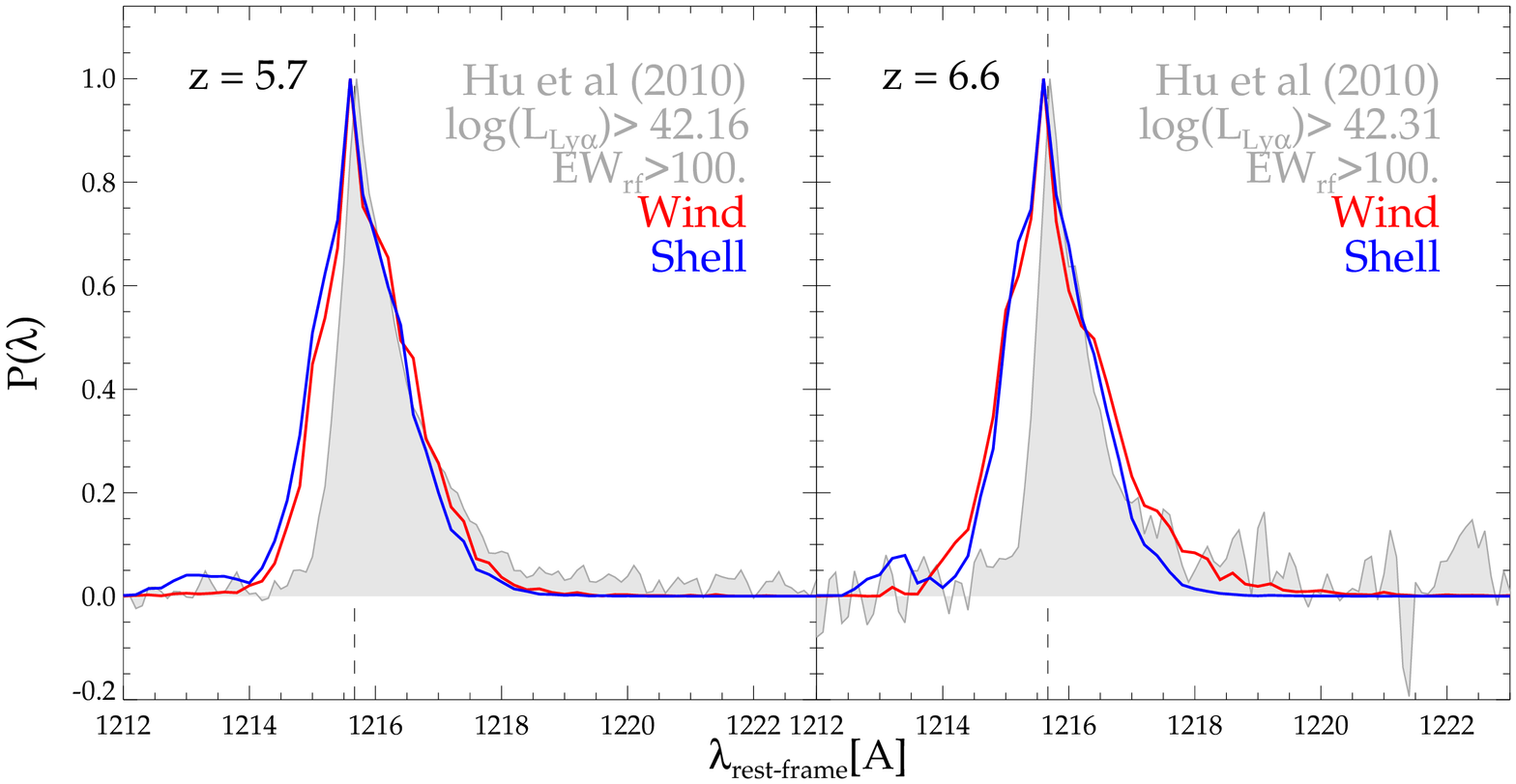}
\caption{Composite \lya\ line profiles of samples at different redshifts as 
indicated in each panel. 
The gray shaded regions show the \lya\ line profiles from composite
observational samples taken from \citet{ouchi08,ouchi10} and 
\citet{hu10} (as indicated in the legend of each box), whereas the 
solid curves show the model predictions using the Wind geometry (red) 
and the Shell geometry (blue). Galaxies in the models used
to construct the composite spectra have been selected in the same 
way as the observed spectra, with the criteria shown in the legend of each box, 
where $L_{Ly\alpha}$ is shown in units of \lunits\ and the EW in \AA{}.}
\label{fig.lyaprofiles}
\end{figure*}

For the two ranges of UV-magnitudes shown, the Shell geometry
appears to have a larger fraction of \lya\ emitters as a function of
redshift than the Wind geometry. 
When comparing to the observational data of \citet{schenker11}, 
we find that overall both outflow geometries underpredict the fraction of \lya\
emitters measured observationally. In particular, only at $z>5$ are the 
Wind and Shell geometries consistent with the fraction of \lya\ 
emitters measured in the UV magnitude range $-20.9<M_{UV}-5\log(h)<-19.4$.

Despite the differences, both outflow geometries predict an increase
in the fraction of \lya\ emitters with redshift, which is qualitatively consistent with
the observations up to $z\sim 6$. 
At higher redshifts, the observational data of \citet{schenker11} and \citet{pentericci11}
suggests a decline in the fraction of \lya\ emitters at $z\sim 7$. They interpret
this decline as the impact of the neutral IGM attenuating the \lya\ luminosity from 
galaxies.

Our models, on the other hand, show a trend consistent to the observations in the 
magnitude range $-19.4<M_{UV}-5\log(h)<-17.9$, except at the highest redshift, $z=7.3$  
where both models increase their fraction of \lya\ emitters instead of decreasing it, as observations
do. However, 
it is worth noticing that in the UV magnitude range $-20.9<M_{UV}-5\log(h)<-19.4$
both outflow geometries predict a decline of the \lya\ fraction similar to the one
found observationally. Therefore, Fig. \ref{fig.xlya_z} shows that our model 
predictions imply that the decline in the \lya\ fraction in LBG samples found 
at high redshifts is not conclusively driven by the presence of neutral HI in 
the IGM attenuating \lya\ photons.

%Our models predict that, given the constraints defined before, both outflow
%models predict no galaxies 
%are found to have strong \lya\ emission at $z=3$.
%
%By $z=4.5$, the fraction grows to $\sim 10\%$ and $\siin the Wind model, but 
%remains in zero in the Shell model. At $z=5.7$, $\sim 10\%$ and 
%$\sim 20\%$ are predicted by the Shell and Wind models, respectively.
%By $z=6.6$, the Wind model predicts no galaxies are found to be \lya\ 
%emitters, whereas the Shell model reaches $\sim 20\%$ at this redshift.

%When comparing to the observational sample of \citet{stark10}, we find 
%a reasonable agreement between the model predictions and the 
%observational data. Considering the uncertainties in the observational 
%measurement, both models are found to be consistent with the 
%fractions measured observationally.

\subsection{Observed \lya\ line profiles}

Observational measurements of individual and stacked line profiles 
of \lya\ suggest the presence of outflows in galaxies \citep[e.g. ][]{shapley03,kashikawa06,dawson07,ouchi08,hu10,kornei10,ouchi10,steidel10,kulas11}. 
\lya\ emitters can be characterised by studying the spectral features of 
the composite spectrum from a set of spectroscopic observations. 
The most prominent feature observed are asymmetric 
peaks, where the line is extended towards the red side. Other common 
spectral features are the appearance of a secondary peak and  P-Cygni 
absorption features \citep[see, e.g ][]{shapley03}.

In this section we compare our model predictions with the composite 
spectra of high redshift samples of \lya\ emitters
studied by \citet{ouchi08}, \citet{ouchi10} and \citet{hu10}. 
Based on the method used to construct composite spectra in the 
observational studies, we construct composite spectra in the model 
as follows. First, the \lya\ profiles are normalised to their peak values. 
Then the spectrum is shifted so that the peaks coincide with the 
\lya\ line centre, $\lambda_{\rm Ly\alpha} = 1216$\AA{}. 
Finally, the spectra are averaged at each wavelength bin.

The above method for constructing a composite spectrum has some 
important drawbacks. Since the redshift of \lya\ emitters is computed
from the wavelength of the peak of the \lya\ line, any offset of the 
peak due to radiative transfer effects is removed (see Fig. 
\ref{fig.comp_spectra_nh} and Fig. \ref{fig.prop_profiles}). 
On the other hand, the normalisation 
of the line profiles to the peak value can 
help enhance certain spectral features inherent to \lya\ emitters by 
removing any dependence of the composite spectrum on \lya\ 
luminosity. This also means that spectral features characteristic of 
a particular \lya\ luminosity could be difficult to spot.

The top panels in Fig. \ref{fig.lyaprofiles} show a comparison between composite 
spectra at redshifts $z=3, 5.7$ and $z=6.6$ from \citet{ouchi08,ouchi10} 
with the predictions from both outflow geometries. 
Overall, both outflow geometries show very similar composite line profiles, 
regardless of redshift or limiting luminosity. A hint of a secondary peak redward of the line centre 
is weakly displayed for some configurations, but it is not strong enough to make a clear 
distinction between the line profiles predicted by both geometries. 
%Moreover, since both outflow models predict similar column densities, 
%which in turn controls the width of the \lya\ profiles \citep[see also ][]{verhamme08}, it is not 
%surprising that the profiles predicted are similar for both models.

The \lya\ line in the composite spectrum estimated at $z=3.1$ by \citet{ouchi08} 
is broader than our model predictions. This could suggest that these galaxies
have a larger column density or expansion velocity than our model predicts, as 
Fig. \ref{fig.prop_profiles} also shows.
At $z=5.7$ and $z=6.6$, Fig. \ref{fig.lyaprofiles}
shows remarkable agreement between the models 
and the observations by \citet{ouchi08} and \citet{ouchi10}, respectively.

The situation is different when comparing with the composite spectra at $z=5.7$ 
and $z=6.6$ estimated by \citet{hu10}. Their composite spectrum is 
constructed from samples of \lya\ emitters of similar limiting \lya\ luminosity
but with $EW_{\rm rf}>100$\AA{}, significantly greater than the 
EW limit in the \citet{ouchi08} samples at $z>5$, which have 
($EW_{\rm rf}>20$\AA{}).
  
In this case, the observational composite spectra are narrower than our 
model predictions. Moreover, the observed
spectra appear to be more asymmetric than their counterparts in the 
\citet{ouchi08} and \citet{ouchi10} samples. 
The asymmetry of the \lya\ line 
varies with the outflow expansion velocity, as discussed in Fig. 
\ref{fig.prop_profiles}.

%In our model, we could increase the value of $f_V$ (the proportionality 
%between the circular velocity and the outflow expanding velocity) 
%by a suitable value to match the assymetry
%of the lines in the \citet{hu10} sample. However, this would force 
%to reduce the value of $f_R$, implying the column densities 
%would be larger than what we obtain with our standard model parameters, 
%lowering the escape fractions and broadening the \lya\ line profiles. 

It is interesting to notice that the composite spectra of \citet{hu10} show 
a sharp cut-off on their blue side, which is not well reproduced by our outflow geometries. 
This lack of photons in the blue side of the spectra could be interpreted
as the impact of the IGM at these high redshifts removing the blue-side 
of the \lya\ spectrum, as discussed 
by \citet{dijkstra07} and more recently by \citet{laursen11}. Since our 
outflow geometries do not show this feature, it seems likely to be caused by the 
presence of a neutral IGM.
However, this does not 
explain the overall good agreement with the \citet{ouchi08} composite
spectra at $z=5.7$ and the \citet{ouchi10}
composite spectra at $z=6.6$.

The predicted \lya\ profiles shown in Fig. \ref{fig.lyaprofiles} show reasonable 
agreement with the data, thus supporting the idea that \lya\ photons escape 
mainly through galactic outflows. It is worth reminding the reader that 
our outflow geometries are not tuned to reproduce the observed line shapes, 
and therefore these represent genuine predictions of the outflow geometries.

\section{The nature of \lya\ emitters}
\label{sec.nature}

\begin{figure*}
\includegraphics[width=16cm,angle=90]{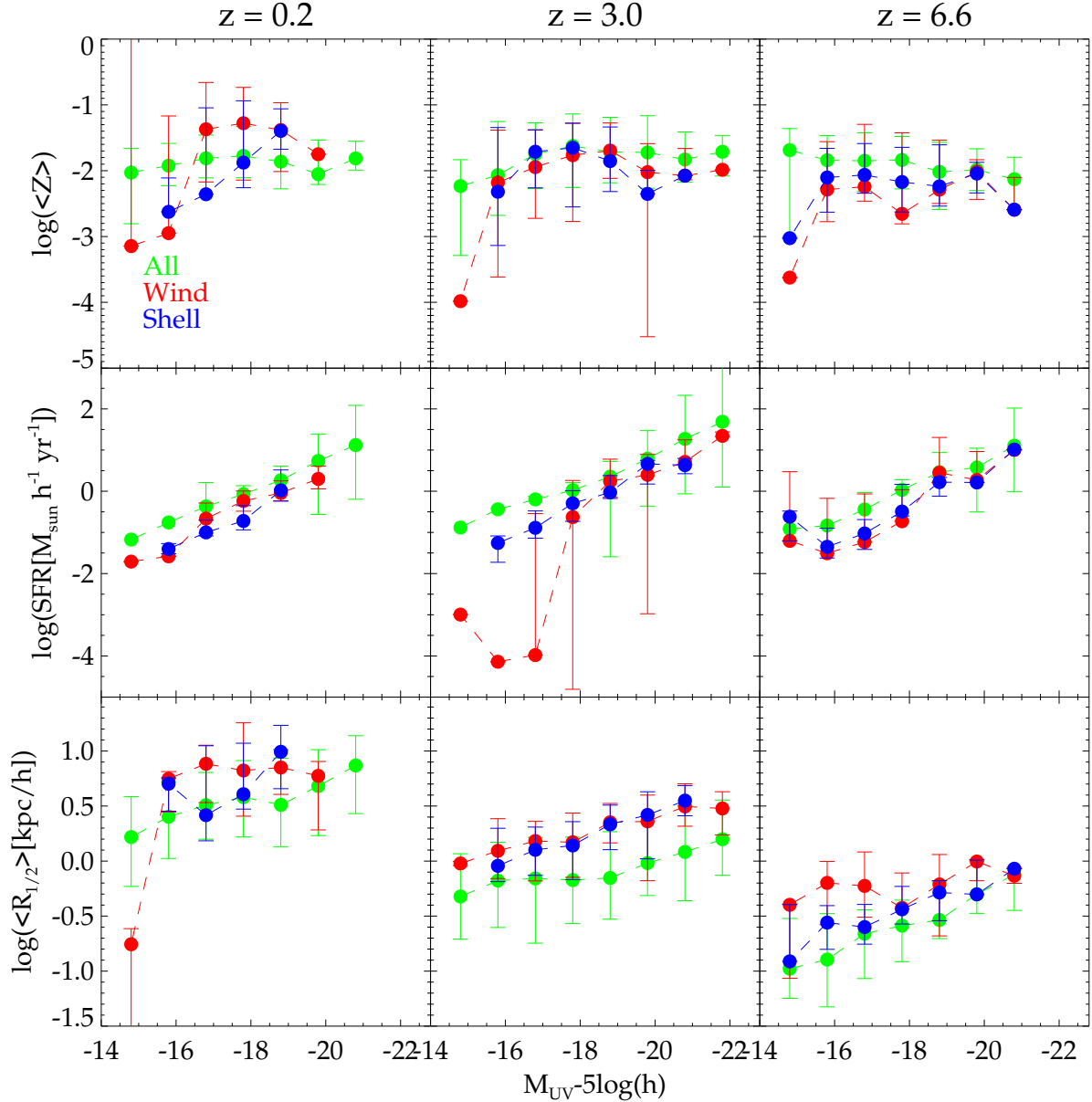}  
\caption{Galaxy properties as a function of extincted $UV$ magnitude for redshifts 
$z=0.2$ (left), $z=3.0$ (middle) and $z=6.6$ (right). Each row shows a 
different property. The top row shows the weighted metallicity of
the cold gas, the middle row the instantaneous SFR, and the bottom row the half-mass radius. 
Green circles denote the median of the distribution including all galaxies per 
magnitude bin. Red and blue circles denote the median of the distribution 
of galaxies selected as {\it typical} \lya\ emitters (i.e. with $\logllya \geq 41.5$ 
and $EW \geq 20$ \AA{}) in the Wind and Shell geometries, respectively. 
Error bars denote the 10-90 percentiles 
of the corresponding distribution.}
\label{fig.lyanature}
\end{figure*}

After performing the detailed comparison between observational data 
and model predictions in the previous section, we conclude
that our outflow geometries reproduce at some level the general features of 
the observations. We now turn to the question of what are the physical conditions that 
make a galaxy observable through its \lya\ emission.
As previously shown in Fig. \ref{fig.xlya_z}, only a fraction of $UV$-selected 
galaxies have detectable \lya\ emission, 
implying that this selection targets galaxies with particular characteristics. 
To reveal the properties of \lya-selected galaxies, we show in Fig. \ref{fig.lyanature} 
a comparison between all galaxies and what we define here as a {\it typical} 
\lya\ emitter, i.e. a galaxy with $\logllya \geq 41.5$ and $EW_{\rm rf} \geq 20$\AA{}. 
This corresponds to a \lya\ luminosity limit where \lya\ emitters are abundant 
over the redshift range $0.2<z<6.6$, and the EW limit corresponds to a 
typical EW limit in observational samples.
We use a top-hat 
filter centered on a rest-frame wavelength of 
$\lambda = 1500$\AA{} to ensure that the rest frame $UV$ magnitude is the 
same for all redshifts. 

The top row of Fig. \ref{fig.lyanature} shows that, in general, 
both outflow geometries predict \lya\ emitters at high redshift to have similar or somewhat lower metallicities 
than the bulk of the galaxy population at the same $UV$ magnitude. At 
$z=0.2$, however, the metallicity of \lya\ emitters is larger than that of 
the overall galaxy population for a range of UV magnitudes.

The result that \lya\ emitters have lower metallicities at higher redshifts than the overall 
galaxy population may not be surprising, 
since to first order we expect galaxies with low
metallicities to have a low amount of dust and thus to be less attenuated in 
\lya\ emission. However, according to Eq. \eqref{eq.taud}, metallicity is 
not the only factor controlling the amount of dust, which is why our 
models predict that at some magnitudes, typical \lya\ 
emitters can have the same (or even higher) metallicities than 
the bulk of the galaxy population.

Observational studies of the \lya\ emitter population at $z\sim 0.2$ 
have found that these galaxies have in general lower metallicities than 
the bulk of the galaxy population at the same stellar masses 
instead of $UV$ magnitude
\citep[e.g.][]{finkelstein11,cowie10}. 
At higher redshifts the same conclusion is drawn from the observational samples 
\citep[see, e.g.][]{gawiser06,pentericci07,finkelstein09b}.
The metallicities our outflow geometries predict at different stellar masses
cannot be directly compared with observational estimates, since these compute
the stellar mass from SED fitting assuming a Salpeter or similar IMF. Our 
model, on the other hand, assumes a top-heavy IMF in starbursts, which 
are the dominant component of the \lya\ emitter population in our model (see 
Fig. \ref{fig.cumlf}).
 
%Despite the different galaxy selection, these results are consistent with 
%the conclusion that \lya\ emitters are in general
%galaxies with low metallicities.

Historically, it was thought that metallicity was the main parameter 
driving the observability of the \lya\ line, implying that
\lya\ emitters should be essentially metal-free galaxies 
\citep{meier81,hartmann84,hartmann88}. Our model predictions, on 
the other hand, imply that the observability of the \lya\ line depends 
on the interplay between more physical properties. 
Nevertheless, it is not the metallicity itself that determines the amount 
of dust that photons need to cross through when escaping. 
In our models, the dust optical depth depends both on the metallicity 
of the galaxies and on the hydrogen column density of the outflows, 
which is in turn a function of several properties, such as 
size, mass ejection rate, expansion velocity and cold gas content.
Therefore, it is not surprising that \lya\ emitters can be either metal-poor or 
metal-rich compared to the bulk of the galaxy population.

Fig. \ref{fig.lyanature} also reveals that typical \lya\ emitters in the Wind geometry
have overall lower metallicities than their counterparts in the Shell geometry. 
This result is consistent with the higher sensitivity to dust in the Shell geometry 
compared with the Wind geometry, shown in Fig. \ref{fig.comp_spectra_nh}. 
Due to the different response to dust of each model, it is natural that the 
predicted metallicities are correspondingly different. 

Another clear difference between the overall galaxy population and 
\lya\ emitters is seen when looking 
at the instantaneous star formation rate at different UV magnitudes, 
shown in the middle panel of Fig. \ref{fig.lyanature}. 
Here it is clear that regardless of redshift or UV magnitude, both models 
predict that \lya\ emitters should have smaller SFRs than typical galaxies with
the same UV magnitude.

The instantaneous SFR has an important impact on the escape of \lya\ photons 
in both models. In the Wind geometry, the mass ejection rate of the outflow depends 
directly on the SFR, as is clear from Eq. \eqref{eq.m_ej}. The mass ejection rate is 
in turn directly proportional to the hydrogen column density of the outflows, according
to Eq. \eqref{eq.nh_wind}, which in turn affects the path length of photons, the 
number of scatterings, and the amount of dust, as discussed previously. Hence, a galaxy 
with a low SFR could have a high \lya\ escape fraction, making it observable. In the Shell 
geometry the SFR is also important, since it correlates with the cold gas mass, which in turn 
determines the hydrogen column density of the outflow.

Our outflow geometries are consistent with the observational evidence for \lya\ emitters 
having modest star formation rates \citep[e.g.][]{gawiser06,gronwall07,guaita10}. 
However, it is worth remarking that our model predictions do not imply that \lya\ emitters 
have low SFRs, but instead have lower SFRs than the bulk of the galaxy population at the 
same $UV$ magnitude. As shown in the previous section, \lya\ emitters at high 
redshifts are predicted to be mainly starbursts. 

Finally, we also study the difference in the size of galaxies, since our outflow geometries 
depend strongly on the galaxy half-mass radius. In the Wind geometry, the hydrogen 
column density scales as $\sim \langle R_{1/2} \rangle^{-1}$, and in the Shell geometry 
the hydrogen column density scales as $\sim \langle R_{1/2} \rangle^{-2}$. In order to 
obtain a high \lya\ escape fraction, $N_H$ has to be low in general, and this could 
be attained by having a large $\langle R_{1/2} \rangle$. Hence, not surprisingly, our 
outflow geometries predict that \lya\ emitters typically have larger sizes than the 
bulk of the galaxy population. This result is independent of redshift and UV magnitude, 
although in detail the difference in sizes may vary with these quantities.
A similar result is reported by \citet{bond09} at $z=3$. They find the sizes of 
\lya\ emitters (in \lya\ emission) to be always smaller than $\sim 3 {\rm kpc/h}$, 
consistent with the half-mass radius of our typical \lya\ emitters at this redshift.

\section{Summary and conclusions}
\label{sec.conclusions}

In this paper we couple the galaxy properties predicted by the \citet{baugh05} 
version of the semianalytical model \galform\ with a Monte Carlo 
radiative transfer model of the escape of \lya\ photons to study the properties 
of \lya\ emitters in a cosmological context.

Motivated by observational evidence that galactic outflows shape the 
asymmetric profiles of the \lya\ line, we developed two different outflow geometries,
each defined using the predicted properties of galaxies in \galform\ in a 
slightly different way. Our Shell geometry, which consists of an expanding 
thin spherical shell, has a column density $N_H$ proportional to the cold gas mass in the 
ISM of galaxies. Our Wind geometry, on the other hand, consists of a spherical 
expanding wind with number density that decreases with increasing radius. 
The column density in the Wind geometry is related to the mass ejection rate from 
supernovae, which is computed by \galform.

%The different geometry and number densities in the outflows for each model 
%translate into different \lya\ profiles and escape fractions, as shown in Fig. 
%\ref{fig.comp_spectra_nh}. 

We study in detail the \lya\ line profiles and characterise them in terms of 
their width, asymmetry and offset with respect to the line centre, as a function
of the outflow column density, expansion velocity and metallicity.
The \lya\ properties of the outflow geometries we study are found to be 
sensitive to the column density, expansion velocity and the geometry
of the outflows, as shown in Fig. \ref{fig.prop_profiles}. 
Metallicity is found to have a smaller impact 
on the line profiles, although it has a great impact on the \lya\ escape fraction, as
shown in Fig. \ref{fig.prop_fesc}.

Both the width and offset from the line centre are found to increase for 
outflows with increasing column densities and expansion velocities. In both cases, 
the \lya\ line profiles in the Shell geometry are more affected by changes of these
properties than the Wind geometry.

The \lya\ escape fraction is found to decrease with metallicity and column density, since
those two properties are directly proportional to the optical depth of absorption. Also,
higher values of the expansion velocity tend to increase the escape fraction.
This rather complicated interplay between the column density, expansion velocity, 
metallicity and geometry justifies our choice for computing the \lya\ escape
fraction from a fully-fledged \lya\ radiative transfer model instead of 
imposing a phenomenological model. 

The drawback is, of course, the 
difficulty in coupling both models in order to obtain a \lya\ \fesc\ for each
galaxy predicted by \galform. One critical step is to choose the free parameters
in our models.
In order to set these for each model, we attempt to fit the observed 
cumulative luminosity function of \lya\ emitters over the redshift range $0<z<7$. 
We find that a single choice of parameters applied at all redshifts cannot reproduce 
the observed CLFs. Motivated by the observed difference in sizes between the star 
forming regions in local and high redshift starbursts \citep[e.g.][]{rujopakarn11}, 
we allow the outflow radii in our models to evolve with redshift when the galaxies 
are starbursts. By doing this we can find a suitable combination of parameters to 
match the measured \lya\ CLF. 

It is worth pointing out that the need to invoke a redshift dependence in the outflow 
sizes in starbursts could suggest that other important physical processes 
which determine the escape of \lya\ photons  may not be included in this work. 
Although outflows have been proposed in the past as a mechanism to boost the 
otherwise very low escape fraction of \lya\ photons, other scenarios have been proposed. 
A clumpy ISM could boost the escape of \lya\ photons with respect to that of
Lyman continuum photons, since the former have a probability of bouncing off a dust 
cloud, and therefore have more chances to escape, whereas in the latter case photons 
travel through dust clouds, and hence have more chances of being absorbed 
\citep{neufeld91,hansen06}. 
%\citet{finkelstein11} suggested that \lya\ emitters at $z\sim 0.2$ have an ISM that 
%is somewhat less clumpy than at higher redshifts (\citealt{finkelstein09b}; but see 
%\citealt{atek09}). This would make low redshift \lya\ emitters fainter than their high 
%redshift counterparts, perhaps explaining partly the difficulty we find when 
%trying to model the observed abundances.
 
Another physical effect not included in our modelling is the attenuation of 
\lya\ radiation due to the scattering of \lya\ photons when crossing regions of 
neutral hydrogen in the IGM. This effect has been shown to be important 
at $z\gtrsim 6$ when the fraction of neutral hydrogen in the IGM is thought 
to have been significant \citep[see, e.g. ][]{dijkstra07,dayal10a,dijkstra11,laursen11}. 
Hence, we do not expect this to significantly change our model predictions at 
low redshifts.
 
A more fundamental uncertainty lies in the calculation of the {\it intrinsic} \lya\ 
luminosity of galaxies, which does not depend upon the radiative transfer 
modelling. We can assess whether \galform\ computes the correct intrinsic \lya\ 
luminosity by studying the intrinsic (unattenuated) \ha\ luminosity function,
since both emission-line luminosities are directly related to the production rate 
of Lyman continuum photons \citep[see ][]{orsi10}, and differ only by 
their case B recombination emission coefficient \citep{osterbrock89}. 
Moreover, the \ha\ emission from galaxies is 
less sensitive to dust than \lya\, since these photons do not undergo 
multiple scatterings in the ISM, making their path lengths 
shorter than the typical path lengths \lya\ photons experience, and thus 
making it easier to estimate their attenuation by dust. 

The attenuation by dust can be estimated by computing the ratio of 
the intensity of two or more emission lines and comparing with
the expectation for case B recombination
\citep[see, for example, ][]{kennicutt83,kennicutt98b}. 
By comparing observed dust-attenuated \ha\ LFs with \galform\ 
predictions \citep{orsi10} we have found that, at $z\sim 0.2$,  
\galform\ roughly overestimates the intrinsic \ha\ luminosities by a 
factor $\sim 3$. The scatter in the observed \ha\ luminosity functions is 
large, making it difficult to estimate this accurately. Nevertheless, the intrinsic \lya\ 
luminosities predicted at $z = 0.2$ are a factor $\sim 10$ or more brighter
than what is needed to reproduce the observational results, if we fix the 
free parameters of the outflow geometries, meaning that the uncertainty in 
the intrinsic \lya\ luminosity is not solely responsible for the discrepancy 
between the observed and predicted LFs.

Despite this, our simple outflow geometries are found to be in agreement 
with a set of different observations, implying that our modelling does 
reproduce the basic physical conditions determining the escape of \lya\ 
photons from galaxies. 

A direct prediction of our outflow geometries is the distribution of hydrogen 
column densities of \lya\ emitters. We find that both models feature column densities with 
values ranging from $N_H \sim 10^{19}-10^{23} [{\rm cm^{-2}}]$, 
which is consistent with observational estimates shown in \citet{verhamme08}. 
The column density distribution is closely related to the predicted \lya\ escape 
fractions. We find that bright \lya\ emitters generally have high escape fractions, 
and faint \lya\ emitters have low escape fractions. An important consequence 
of this is that certain galaxy properties which correlate with the intrinsic \lya\ 
luminosity (such as the instantaneous SFR) do not correlate in a simple way with 
the observed \lya\ luminosity. In other words, galaxies with the same intrinsic \lya\ 
luminosity (or, say, the same SFR) could be observed with different \lya\ 
luminosities, making the interpretation of the properties of \lya\ emitters complicated.

The predicted escape fractions in the models are remarkably consistent with observational 
measurements (Fig. \ref{fig.fesc_robs}), giving further support to the scenario of \lya\ 
photons escaping through galactic outflows. Although the Wind geometry is only 
partially consistent with the observed escape fractions, it is worth pointing out that
the observational data used to make the comparison in Fig. \ref{fig.fesc_robs} does
not constitute a representative sample of the galaxy population at this redshift.

Since our Monte Carlo radiative transfer model records the frequency with which 
photons escape from an outflow, we make use of this information to study the predicted 
\lya\ line profiles from our outflow geometries and compare them with observational 
measurements. In some cases the agreement is remarkably good, but in others we 
find significant differences (see Fig. \ref{fig.lyaprofiles}). A detailed study of the line 
profiles could reveal important information about the galaxy properties. We plan to 
undertake such a study of \lya\ line profiles in the 
context of a galaxy formation model in a future paper.

Finally, our models predict that only a small fraction of galaxies should be selected 
as \lya\ emitters. We illustrate in Fig. \ref{fig.lyanature} that \lya\ emitters are 
found in general to have low metallicities  (except at $z=0.2$), 
low instantaneous SFR and large sizes, 
compared to the overall galaxy population. These constraints arise naturally 
as a consequence of the radiative transfer modelling incorporated in \galform\ in
this paper. Galaxies need to have low star formation rates and large sizes in order 
to display an outflow with a small column density. Since the dust content depends 
on the optical depth of dust and not the metallicity alone, \lya\ emitters are not 
necessarily low-metallicity galaxies in our models.

The models presented here for the emission of \lya\ represent an important step 
towards a detailed understanding
of the physical properties of these galaxies. With the advent of large observational campaigns 
in the forthcoming years focusing on detecting \lya\ emitters at high redshifts, 
new data will help us refine and improve
our physical understanding of these galaxies, and thus, enable us to improve our knowledge
of galaxy formation and evolution, particularly in the high redshift Universe.

\section*{Acknowledgenemts}

AO acknowledges a STFC-Gemini scholarship. This work was supported in part by Proyecto 
Gemini 320900212 and an 
STFC rolling grant. 
We thank Peter Laursen for making data from the \mocalata\ radiative transfer code
available for us to test our code, and Nelson Padilla for useful comments and discussions.
Part of the calculations for this work were performed using the Geryon supercomputer at AIUC.

%\begin{thebibliography}{0}
\bibliographystyle{mn2e}
%\bibliography{bibliography}
%\bibliographystyle{plainnatJEG}
%\bibliographystyle{nature}
% \bibliographystyle{apj}
\bibliography{outflows_lya}
%\end{thebibliography}

\appendix

\section{Description of the Monte Carlo radiative transfer code}
\label{app.lyart}

In the context of \lya\ radiative transfer, photon frequencies, $\nu$, are 
usually expressed in terms of Doppler units $x$, given
by Eq. \eqref{eq.x}. The  thermal velocity dispersion of 
the gas, $v_{\rm th}$, is given by
\begin{equation}
 v_{\rm th} = \left(\frac{2k_BT}{m_p}\right)^{1/2},
\label{eq.vth}
\end{equation}
where $k_B$ is the Boltzmann constant, $T$ is the gas temperature, 
$m_p$ is the proton mass and $\nu_0$ is the central frequency of 
the \lya\ line, $\nu_0 = 2.47 \times 10^{15} {\rm Hz}$. 

When a \lya\ photon interacts with a hydrogen atom, the scattering 
cross section in the rest frame of the atom is given by
\begin{equation}
 \sigma_{\nu} = f_{12} \frac{\pi e^2}{m_ec}\frac{\Gamma/4\pi^2}
 {(\nu - \nu_0)^2 + (\Gamma/4\pi)^2},
\end{equation}
where $f_{12}=0.4162$ is the \lya\ oscillator frequency, and 
$\Gamma = A_{12} = 6.25\times 10^8 {\rm s}^{-1}$ is 
the Einstein coefficient for the \lya\ transition ($n=2$ to $n=1$).

The optical depth of a \lya\ photon with frequency $\nu$ is determined 
by convolving this cross section with the velocity distribution of the gas,
\begin{equation}
 \tau_{\nu}(s) = \int_0^s \int_{-\infty}^{+\infty}n(V_z)\sigma(\nu,V_z)\ 
 {\rm d}V_z {\rm d}l,
\end{equation}
where $V_z$ denotes the velocity component along the photon's direction. 
Atoms are assumed to have a Maxwell-Boltzmann velocity distribution in the rest
frame of the gas. In Doppler units, the optical depth can be written as
\begin{equation}
 \tau_x(s) = \sigma_H(x)n_Hs = 5.868\times 10^{-14}T_4^{-1/2}
 N_H\frac{H(x,a)}{\sqrt{\pi}},
\label{eq.tau_x}
\end{equation}
where $n_H$ is the hydrogen density, $N_H$ the corresponding 
hydrogen column density, $T_4$ the temperature in units
of $10^4 {\rm K}$ and $a$ is the Voigt parameter, defined as
\begin{equation}
 a = \frac{\Gamma/4\pi}{\Delta \nu_D} = 4.7\times 10^{-4}T_{4}^{-1/2}
\end{equation}

 The Hjerting function $H(x,a)$ \citep{hjerting38} describes the Voigt scattering 
 profile,
\begin{equation}
\label{eq.H}
 H(x,a) = \frac{a}{\pi}\int_{-\infty}^{+\infty}\frac{e^{-y^2}\ {\rm d}y}
 {(y - x)^2 + a^2},
\end{equation}
which is often approximated by a central resonant core and power-law 
``damping wings" for frequencies  $|x|$ below/above 
a certain boundary frequency $x_c$, which typically ranges between 
$2.5<x_c<4$. As a consequence, photons with frequencies close to the line
centre have a large scattering cross section compared to those with frequencies 
in the wings of the profile. Hence, photons will be more likely
to escape a medium when they have a frequency away from the line centre.

Scattering events are considered to be {\it coherent} (the 
frequency of the photon is the same before and after the scattering event) only 
in the rest frame of the atom, but not in the observer's frame.
Thus, the thermal motion of the atom, plus any additional bulk motion of the 
gas, will potentially Doppler shift the frequency of the photons, giving them
the chance to escape from the resonant core. 

In the following, $\xi_1, \xi_2, \xi_3, ...$ are different random numbers in the range
$[0,1]$.

The location of the interaction (with either a dust grain or a hydrogen atom) 
is calculated as follows. The optical depth $\tau_{\rm int}$ the photon will 
travel is determined by sampling the probability distribution

\begin{equation}
 P(\tau) = 1 - e^{-\tau},
\end{equation}
and so
\begin{equation}
\label{eq.tau_i}
 \tau_{\rm int} = -\ln(1 - \xi_1).
\end{equation}
This optical depth corresponds to a distance travelled $s$ given by
\begin{equation}
\label{eq.tau_s}
\tau(s) = \tau_x(s) + \tau_d(s),
\end{equation}
where $\tau_x(s)$ and $\tau_d(s)$ are the optical depths due to 
hydrogen atoms and dust grains respectively. The length
of the path travelled is determined by finding the distance $s$ 
where $\tau(s) = \tau_{\rm int}$ by setting
\begin{equation}
s = \frac{\tau_{\rm int}}{n_H\sigma_x + n_d\sigma_d},
\end{equation}
where $n_d$ and $\sigma_d$, the number density of dust grains 
and cross-section for interaction with dust, are described below. 

The new location of the photon corresponds to the point where it interacts 
with either a hydrogen atom or a dust grain.
To find out which type of interaction the photon experiences, we compute 
the probability $P_H(x)$ of being scattered 
by a hydrogen atom, given by
\begin{equation}
 P_H(x) = \frac{n_H\sigma_H(x)}{n_H\sigma_H(x) + n_d\sigma_d}.
\end{equation}
We generate a random number $\xi_2$ and compare it to $P_H$. If $\xi_2<P_H$, 
then the photon interacts with the hydrogen atom, otherwise, it interacts with 
dust.

When interacting with a dust grain, a \lya\ photon can be either absorbed or scattered. This depends on the
 albedo of dust particles. At the wavelength of \lya, the albedo is $A \sim 0.4$, depending
on the extinction curve used. If the
 \lya\ photon is absorbed, then it is lost forever. If not, then it will be scattered. The new direction will
 depend on a probability distribution for the elevation angle $\theta$, whereas for the azimuthal angle $\phi$
the scattering will be uniformly distributed. 
The scattering angle $\theta$ can be obtained from the \citet{henyey41} 
phase function
\begin{equation}
\label{eq.dust_angle}
 P_{HG}(\mu) = \frac{1}{2}\frac{1 - g^2}{(1 + g^2 - 2g\mu)^{3/2}},
\end{equation}
where $\mu = \cos\theta$ and $g = \langle \mu \rangle$ is the asymmetry parameter. If $g = 0$, Eq.\eqref{eq.dust_angle}
reduces to isotropic scattering.  $g=1(-1)$ implies complete forward (backward) scattering. In general $g$ depends on 
the wavelength. For \lya\ photons, $g=0.73$.

If the photon is interacting with dust, then we generate a random number $\xi_3$ to determine whether it is going
to be absorbed or scattered, comparing this number to $A$. 
If the photon is absorbed, then it is lost. If it is scattered, then a new direction must be drawn. 

The interaction of photons with hydrogen atoms is more complicated.
Inside an HI region, atoms move in random directions with velocities
given by the Maxwell-Boltzmann distribution. Each of these atoms will 
{\it see} the same photon moving with a different
frequency, due to the Doppler shift caused by their velocities. Since the 
cross section for scattering depends on the 
frequency of the photon, the probability for an atom to interact with 
a photon will depend on a combination
of the frequency of the photon and the velocity of the atom.

To compute the direction of the photon after the scattering event, 
a Lorentz transformation to the frame of the atom must be made.
The direction of the photon in this frame, $\vect{n'_o}$, is given 
by a dipole distribution, with the symmetry axis defined by the incident
direction $\vect{n'_i}$
\begin{equation}
 P(\theta) = \frac{3}{8}(1 + \cos^2\theta),
\label{eq.dipolar}
\end{equation}
where $\theta$ is the polar angle to the direction $\vect{n'_i}$. The 
azimuthal angle of the outgoing 
photon is uniformly distributed in the range $0\le \phi <2\pi$. With a Lorentz 
transformation back to the frame of the medium we obtain 
$\vect{n_o}$, the new direction of the photon.
Finally, the new frequency of the photon is given by
\begin{eqnarray}
 x_f & = & x - \vect{n_i} \cdot \vect{u} + \vect{n_o}\cdot \vect{u}, \\
    & = & x - u_{\parallel} + \vect{n_o}\cdot \vect{u},
\label{eq.xf}
\end{eqnarray}
where $\vect{u} = \vect{v}/v_{\rm th}$ is the velocity of the hydrogen atom in units of the thermal 
velocity, and $u_{\parallel}$ is the atom's velocity component along the photon's direction prior to the 
scattering event.

%\subsubsection{Accelerating the code}
The algorithm described above will follow the scattering of a photon 
until it escapes (or is absorbed), and then the process
starts again with a new photon traveling on a different path, and so on 
until we are satisfied with the number of photons
generated. In practice, for the runs shown in this paper the number of 
photons generated varies between a few thousand up to several hundred 
thousand, depending on the accuracy of the result we wish to achieve. 

For the typical HI regions studied here, the number of scatterings that 
photons undergo before escaping could be as high as several
tens or hundreds of millions. If we want to model several thousand photons, 
then the total number of calculations grows enormously
and the task becomes computationally infeasible.
However, most of the scattering events will occur when the photon is at the 
line centre, or very close to it, where the cross section for scattering
peaks. Whenever the photon falls near the centre it will experience so many 
scatterings that the actual distance travelled
between each scattering event will be negligible, since in this case it will
most likely be scattered by an atom with a velocity close to zero. Hence, the 
frequency after such scattering will remain in the resonant core.
This motivates the possibility of accelerating the code performance 
by skipping such {\it inconsequential} scattering events.

Following \citet{dijkstra06}, a critical frequency, $x_{\rm crit}$, defines a transition 
from the resonant core to the wing. Whenever a photon
is in the core (with $|x|<x_{\rm crit}$) we can push it to the wings by allowing the 
photon to be scattered only by a rapidly moving atom. We do this
by modifying the distribution of perpendicular velocities by a {\it truncated} 
Gaussian, i.e. a distribution which is a Gaussian for $|u|>x_{\rm crit}$ but zero 
otherwise. The modified perpendicular velocities are then drawn from
\begin{eqnarray}
 u_{\perp 1} & = & \sqrt{x_{\rm crit}^2-\ln(\xi_4)}\cos(2\pi\xi_2) \\
u_{\perp 2} & = & \sqrt{x_{\rm crit}^2-\ln(\xi_4)}\sin(2\pi\xi_2).
\end{eqnarray}
When doing this, we allow the photon to redshift or blueshift away from 
the line centre, thus reducing the cross section for scattering and
increasing the path length. For the configurations studied here, we found 
that a value of $x_{\rm crit} = 3$ provides a good balance between accuracy and
efficiency of the code, reducing the execution time by a factor 100 or more with 
respect to the non-accelerated case.

\subsection{Validation of the radiative transfer code}
\label{appendix.validation}

The flexibility of our Monte Carlo radiative transfer code allows us to reproduce 
configurations for which analytical solutions are available. Hence, these 
are ideal to test the performance and accuracy of the code. In the following 
we describe the tests we have performed on our code, where each comparison 
with an analytical solution tests a different aspect of the code.

\begin{figure}
\centering
\includegraphics[width=8cm,angle=90]{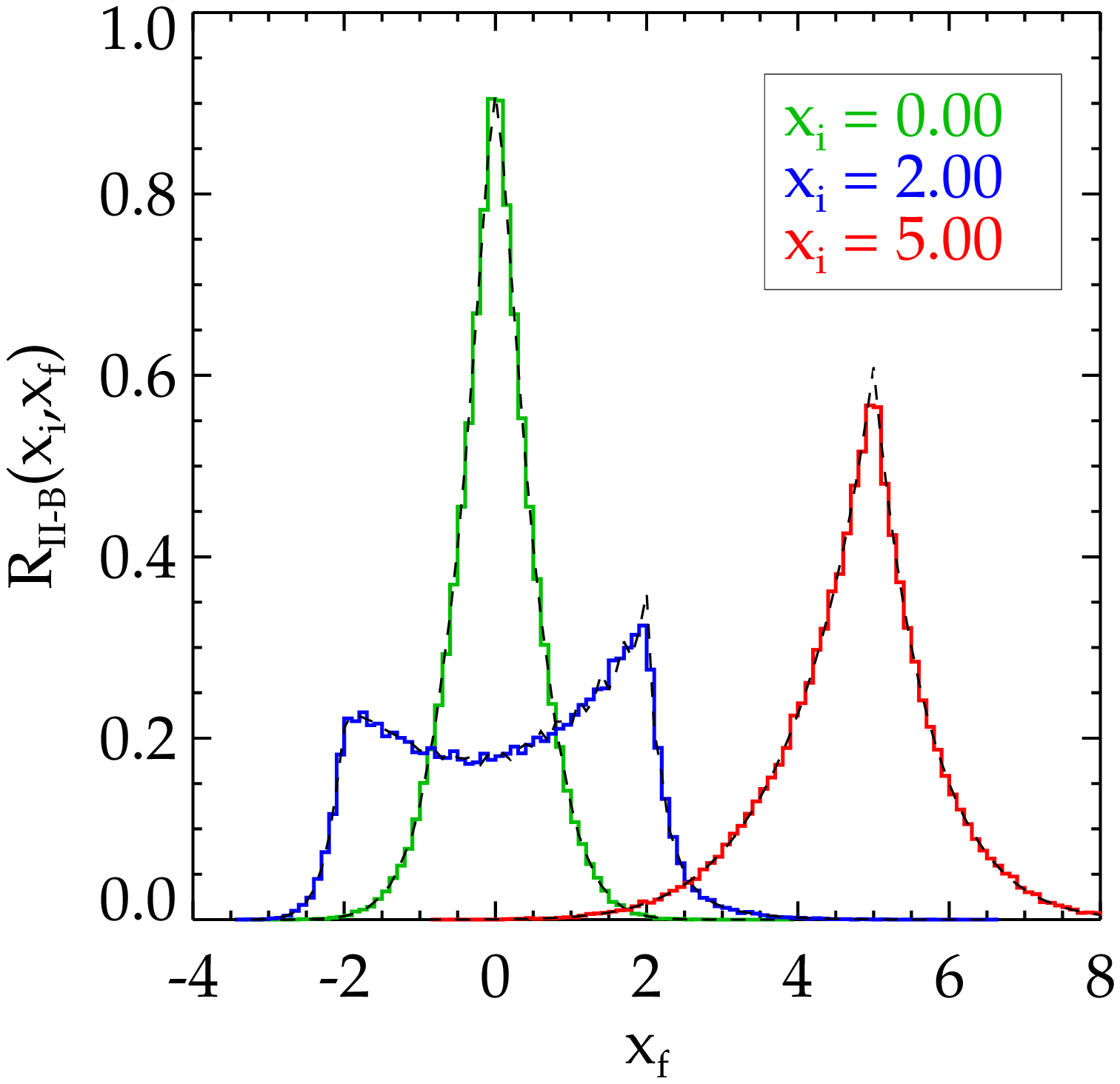}
\caption[The redistribution function of \lya\ photons scattered by hydrogen atoms]
{The redistribution function of \lya\ photons scattered by hydrogen atoms for different initial frequencies.
The histograms show the resulting frequency distribution from the Monte Carlo code, whereas the dashed lines show
a numerical integration of the analytical solution of \citet{hummer62}.}
\label{fig.RII}
\end{figure}
 
Fig. \ref{fig.RII} shows the resulting redistribution function for 3 different initial 
frequencies using $\sim 10^5$ photons. There is
a remarkably good agreement
between the Monte Carlo code and the analytical expression of \citet{hummer62} 
for coherent scattering with a dipolar angular distribution, including radiation damping. 

\begin{figure}
\centering
\includegraphics[width=6.5cm,angle=90]{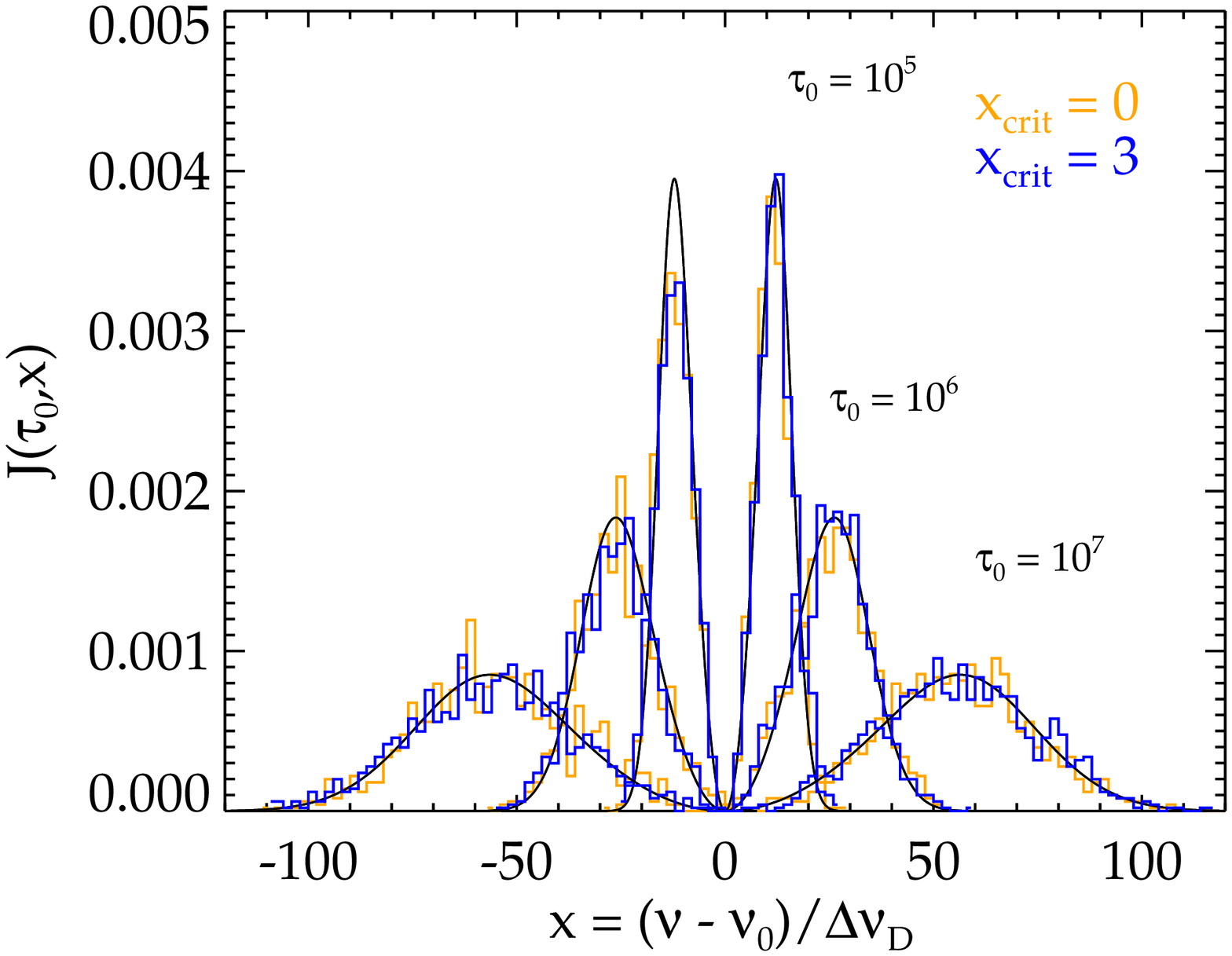}
\caption[\lya\ spectrum emerging from a homogeneous static slab]
{\lya\ spectrum emerging from a homogeneous static slab at $T = 10 $[K], for optical depths at the line centre
of $\tau_0 = 10^5, 10^6$ and $10^7$, as shown in the plot. 
The profiles are symmetric around $x=0$. The more optically thick the medium, 
the farther from the line centre the resulting peaks of each profile are found. The solid lines
show the analytical solution by \citet{harrington73}, and the orange and blue histograms show the results from the Monte Carlo
code for a choice of $x_{crit} = 0$ and $3$, respectively.}
\label{fig.spec_homslab}
\end{figure}

The emergent spectrum from an optically thick, homogeneous static slab 
with photons generated at the line centre was first
calculated by \citet{harrington73}, and the result was generalised by 
\citet{neufeld90}, allowing the generated photons to have any
frequency. 
% Harrington's solution for the emergent \lya\ spectrum for a slab with thickness characterised by its optical depth at the line centre
% $\tau_0$, is
% \begin{equation}
% \label{eq.harrington}
% J(\pm \tau_0,x) = \frac{\sqrt{6}}{24}\frac{x^2}{\sqrt{\pi}a\tau_0}\frac{1}{\cosh\left[\sqrt{\pi^3/54}x^3/(a\tau_0)\right]}.
% \end{equation}
% Both expressions are valid when $a\tau_0 \geqslant 10^3/\sqrt{\pi}$, or for $\tau_0 \geqslant 1.2\times 10^6$
% when $T = 10^4 {\rm K}$. \\

Fig. \ref{fig.spec_homslab} shows the emergent spectrum from a 
simulated homogeneous slab. The temperature of the medium was chosen to 
be $T=10$K, since in this regime the analytical expression is accurate for optical 
depths down to $\tau_0 \sim 10^4$, which is faster 
to compute with the code.

The typical \lya\ line profile is double peaked, and is symmetrical with respect to 
the line centre. The centres of the peaks are displaced
away from $x=0$ by a value determined by $\tau_0$. The higher the optical depth,
 the farther away from the line centre and the wider
the profile will be. Fig. \ref{fig.spec_homslab} compares the analytic solution 
of \citet{harrington73} with the ouput from the basic code
(orange histogram), and the accelerated version (blue histogram).
Overall, it is clear that the non-accelerated version of the code reproduces 
the analytical formula over the range of optical depths shown here.
When $x_{\rm crit} = 3$ (the blue histogram in Fig. \ref{fig.spec_homslab}), 
the output is virtually indistinguishable from the non-accelerated 
version, but the running time has been decreased by a factor $\sim 200$. 
Therefore, Fig. \ref{fig.spec_homslab} confirms that the choice of $x_{\rm crit} = 3$ 
does not compromise the accuracy of the results.

\begin{figure}
\centering
 \includegraphics[width=8cm,angle=90]{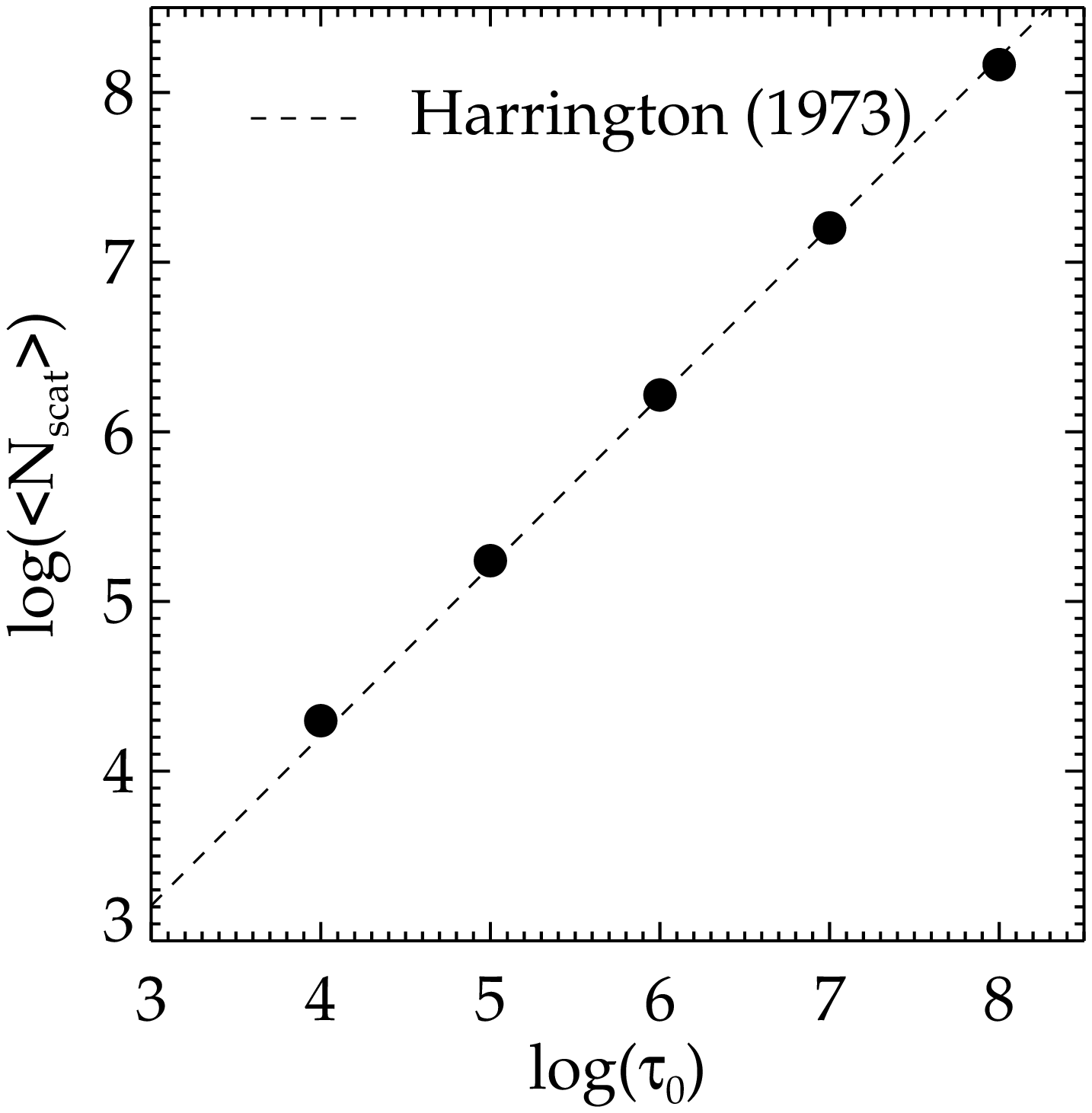}
 \caption[Mean number of scatterings as a function of the optical depth in the line centre of the medium.]
{Mean number of scatterings as a function of the optical depth in the line centre of the medium. The
circles show the results from the Monte Carlo code for configurations with different $\tau_0$. The dashed line
shows the analytical solution of \citet{harrington73}. }
\label{fig.mean_nscat}
\end{figure}

\citet{harrington73} also computed the mean number of scatterings 
expected before a \lya\ photon escapes from an optically
thick medium for the homogeneous slab. He found 
\begin{equation}
\label{eq.nscat}
 \langle N_{scat} \rangle = 1.612 \tau_0.
\end{equation}
Fig. \ref{fig.mean_nscat} shows a comparison between 
the mean number of scatterings computed using our code with the 
analytical prediction of \citet{harrington73}. The agreement is remarkably good.

Following closely the methodology of \citet{harrington73} and 
\citet{neufeld90}, \citet{dijkstra06} computed the emergent spectrum from a
static sphere. 
% Their result is
% \begin{equation}
% \label{eq.dijkstra}
%  J(x,\tau_0) = \frac{\sqrt{\pi}}{\sqrt{24}a\tau_0} \left[\frac{x^2}{1 + \cosh\left[ \sqrt{2\pi^3/27}(|x^3|/a\tau_0)\right]}\right],
% \end{equation}
 \begin{figure}
\centering 
\includegraphics[width=6.5cm,angle=90]{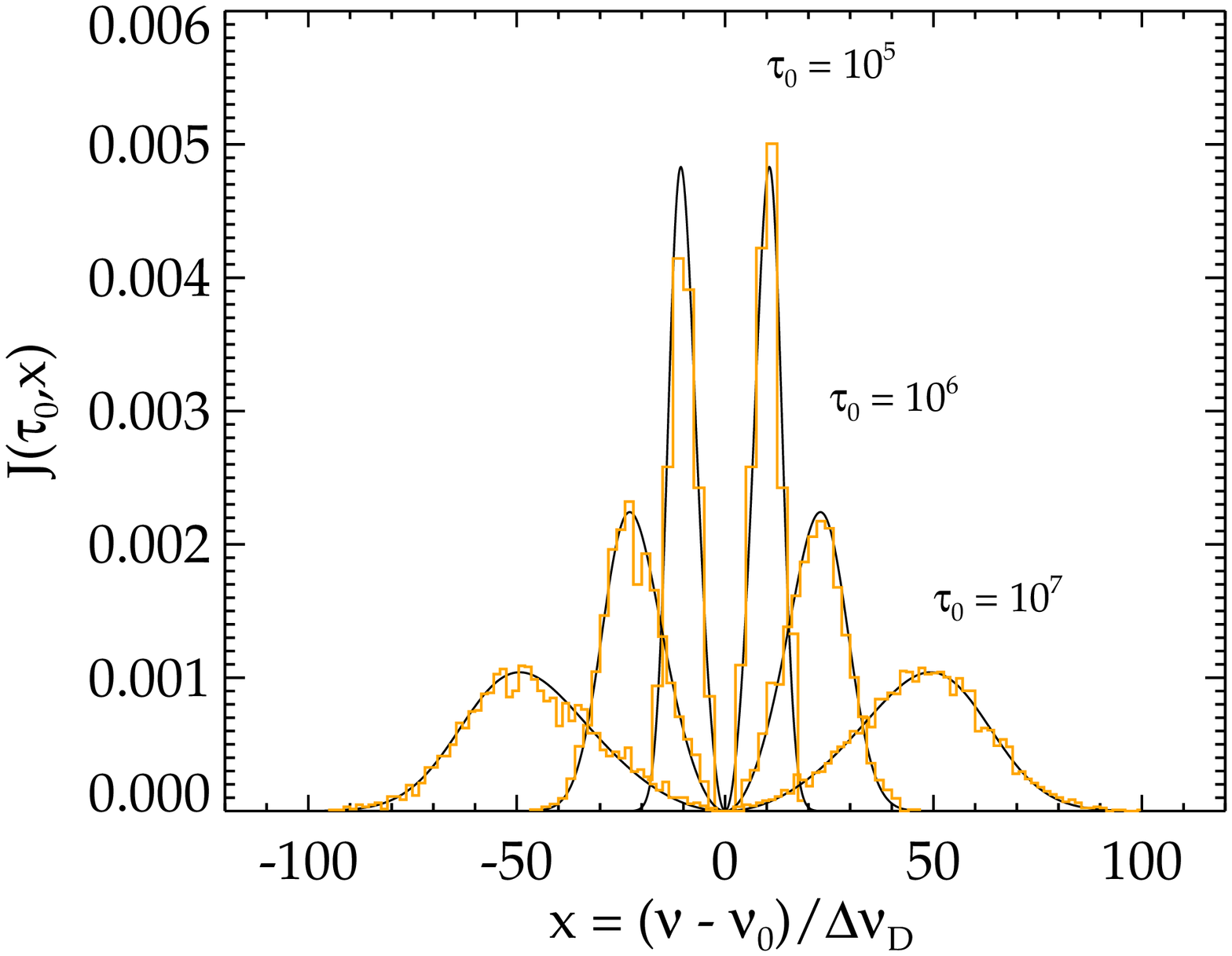}
\caption[\lya\ spectrum emerging from an homogeneous static sphere]
{\lya\ spectrum emerging from a homogeneous static sphere at $T = 10 $K, for optical depths at the line centre of
$\tau_0 = 10^5, 10^6$ and $10^7$. The profiles are symmetric around $x=0$. The thicker the medium, 
the farther from the line centre the resulting peaks of each profile are found. The solid lines
show the analytical solution of \citet{dijkstra06} and the histograms show the results from our Monte Carlo
code.}
\label{fig.sphere}
\end{figure}
Fig. \ref{fig.sphere} shows a comparison between the analytic prediction 
and the output from the code at different optical depths.
Again, there is a very good agreement between the two. 
The optical depths shown in Fig.\ref{fig.sphere} were chosen to be different from those
in Fig. \ref{fig.spec_homslab} to show that the code is following closely 
the expected emergent spectrum for a range of optical depths 
spanning several orders of magnitude.

\citet{neufeld90} computed an analytical expression for the escape 
fraction of photons emitted from an homogeneous, dusty slab. 
%The 
%solution is valid for very high optical depths ($a\tau_0 > 10^3$), 
%and in the limit $(a\tau_0)^{1/3} \gg \tau_a$, where $\tau_a$ is the optical
%depth of absorption by dust.
% \begin{equation}
%  \label{eq.fesc_homslab}
% f_{esc} = \frac{1}{\cosh\left[\zeta' \sqrt{(a\tau_0)^{1/3} \tau_a}\right] },
% \end{equation}
% where $\zeta' \equiv \sqrt{3}/\zeta\pi^{5/12}$, and $\zeta \approx 0.525$ is a fitting parameter.\\

\begin{figure}
 \centering
\includegraphics[width=8cm,angle=90]{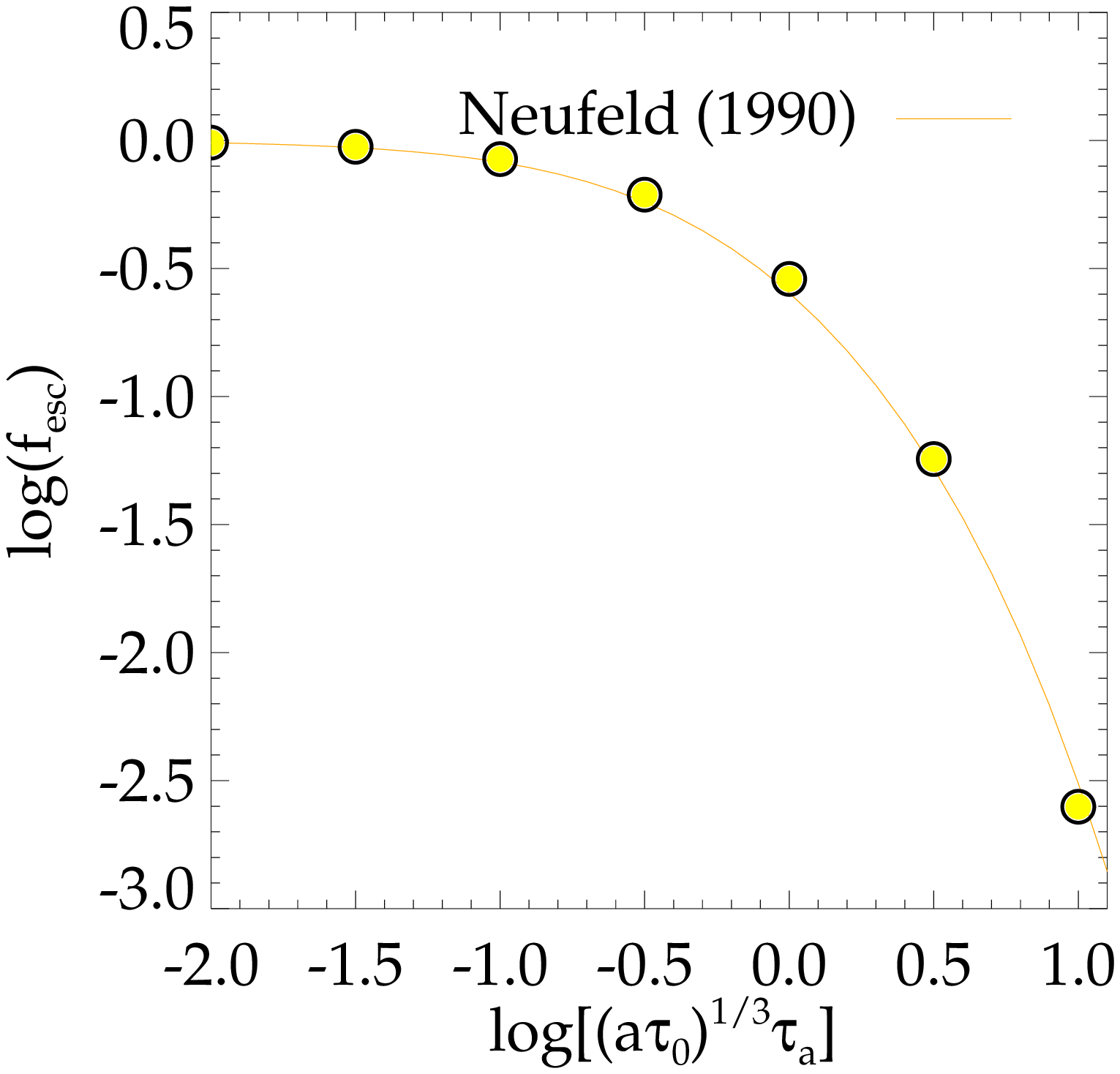}
\caption[The escape fraction of \lya\ photons from an homogeneous dusty slab]
{The escape fraction of \lya\ photons from an homogeneous dusty slab. The optical depth of hydrogen scatterings at the line centre
$\tau_0$ is held constant at $\tau_0 = 10^6$, and different values of the optical depth of absorption $\tau_a$ are chosen.
Circles show the output from the code, and the solid orange curve shows the analytical prediction of \citet{neufeld90}.}
\label{fig.fesc_homslab}
\end{figure}
Fig. \ref{fig.fesc_homslab} shows a comparison between 
the escape fraction obtained from a series of simulations,
keeping $\tau_0$ fixed and varying $\tau_a$, with the analytical 
solution of \citet{neufeld90}. We find a remarkable agreement between the analytical
solution and our code. The escape fraction, as expected, decreases 
rapidly for increasing $\tau_a$, which, for a fixed $\tau_0$, translates into 
having a higher concentration of dust in the slab. 

\begin{figure}
\centering
\includegraphics[width=6.5cm,angle=90]{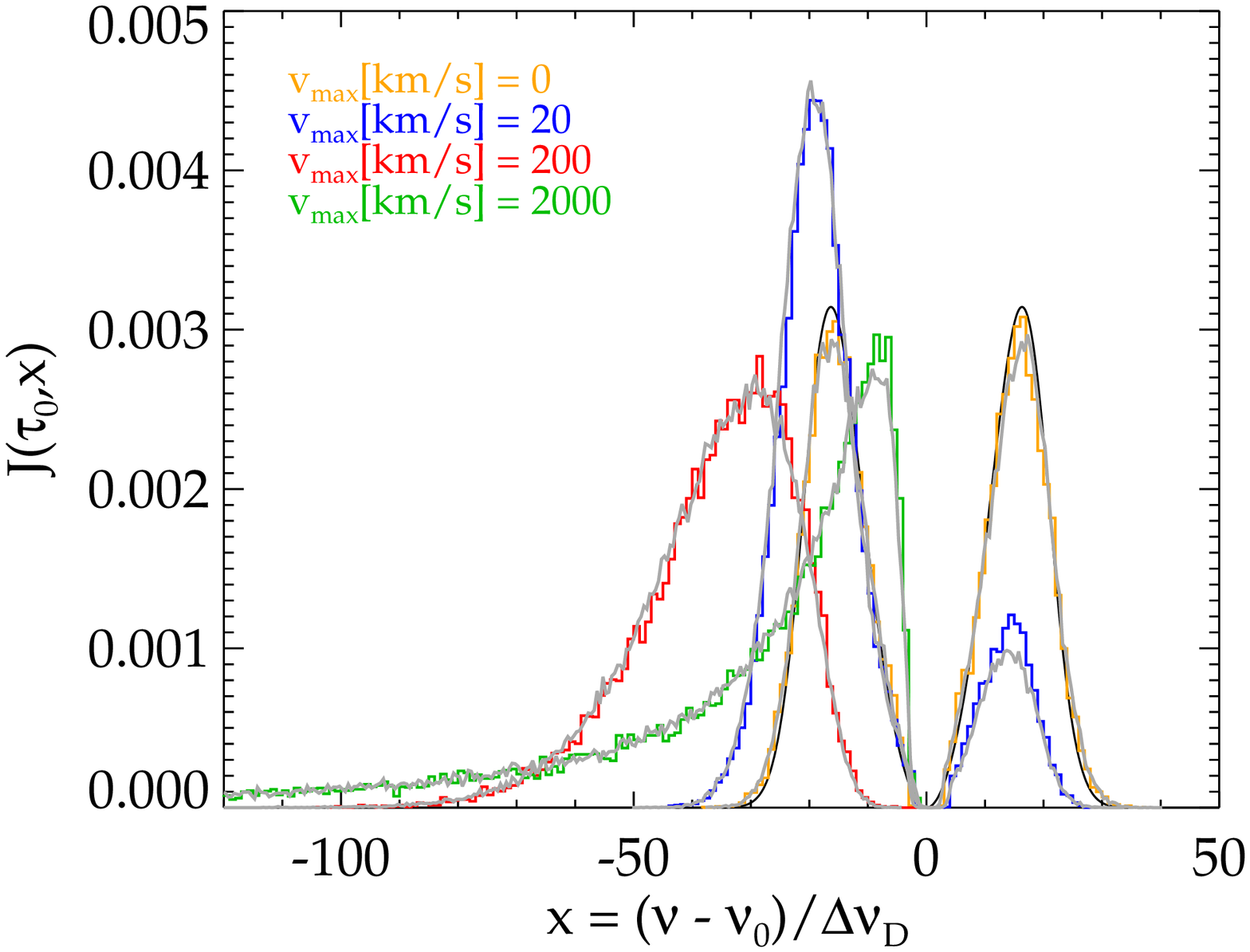}
\caption[The emergent \lya\ spectrum from an expanding sphere]
{The emergent \lya\ spectrum from a linearly expanding sphere with velocity zero at the centre and velocity at the edge 
$v_{max} = 0,20,200$ and $2000 {\rm km/s}$ shown in orange, blue, red and green respectively. The optical depth at the 
line centre is kept fixed at $\tau_0 = 10^{7.06}$. The analytical solution of \citet{dijkstra06} for the static case is 
shown in black. The coloured histograms show the output from the code. The gray solid curves show the results obtained with
the \citet{laursen09a} code \mocalata\ (their Fig. 8).} 
\label{fig.expanding_sphere}
\end{figure}

To validate the effect of bulk motions in the gas, we model the case of 
an expanding homogeneous sphere, with a velocity
at a distance $\vect{r}$ from the centre given by
\begin{eqnarray}
 v_{bulk} & = & H \vect{r}, \\ 
 H & = & \frac{v_{max}}{R},
\end{eqnarray}
where $v_{max}$ is the velocity of the sphere at its edge, and $R$ is the radius of the sphere.

There is no analytical solution for this configuration 
(except when $T=0$, see \citet{loeb99}), so 
we decided to compare our results to those found by a similar Monte Carlo code.
We perform this comparison with media at $T = 10^4[K]$. 
Fig. \ref{fig.expanding_sphere} shows a comparison between 
our code and the results obtained with the \mocalata\ Monte Carlo code
\citep{laursen09a}. Both codes agree very well. Moreover, 
the figure helps us understand the effect of bulk motions of the gas 
on the emergent spectrum. First, when $v_{max} = 0$ we
recover the static solution, \citep{dijkstra06}. When $v_{max} = 20 {\rm km/s}$, 
the velocity of the medium causes photons to have a higher probability of 
being scattered by atoms with velocities dominated by the velocity of the medium. 
These atoms {\it see} the photons as being redshifted, and hence the 
peak of the spectrum is shifted slightly towards the red part of the spectrum, although
still a fraction of photons appear to escape blueshifted.
When $v_{max} = 200 {\rm km/s}$, 
the blue peak is completely erased, and the peak is shifted even 
further to the red side. For very high velocities, such as 
$v_{max} = 2000 {\rm km/s}$, the velocity gradient makes the medium 
optically thin, and the average number of scatterings 
decreases drastically, and consequentially the photons
have less chance of being redshifted far into the wings, thus shifting 
the peak back to the centre, but still with no photons in the blue 
side of the spectrum.

\section{The effect of the UV background}
An additional feature of the Wind geometry is the option of computing 
the resulting ionization of the medium by photons in 
the intergalactic UV background.
The Wind geometry, as described in detail in section \ref{sec.wind}, is assumed to be completely neutral, 
but photoionization from the UV background could have the effect of modifying
the density profile of the neutral gas.

% \begin{figure}
% \centering
% \includegraphics[width=8cm,angle=90]{figs/plot_nh.eps}
% \caption[The effect of the UV background on the number density profiles at two different redshifts]
% {The effect of the UV background on the number density profiles for two randomly selected galaxies at $z=0.2$ (red curves)
% and $z=3.0$ (blue curves). The solid curves show the initial, unattenuated profiles. The dashed lines show the resulting
% profiles when including the UV background but without including self-shielding. The dotted-dashed lines show the effect
% of the UV background when self-shielding is included.}
% \label{fig.nh_uv}
% \end{figure}

It is generally believed that the extragalactic UV background is dominated by 
radiation from quasars and massive young stars in star forming galaxies \citep{haardt96,haardt01,meiksin09}. 
The mean intensity of the UV background at 
the observed frequency $\nu_0$ and redshift $z_0$ is defined as 
\begin{equation}
 J_0(\nu_0,z_0) = \frac{1}{4\pi} \int_{z_0}^{\infty} {\rm d}z 
 \frac{{\rm d}l}{{\rm d}z}\frac{(1 + z_0)^3}{(1+z)^3}\epsilon(\nu,z)
e^{-\tau_{\rm eff}(\nu_0,z_0,z)},
\label{eq.j0}
\end{equation}
where $z$ is the redshift of emission, $\nu = \nu_0(1+z)/(1+z_0)$, ${\rm d}l/{\rm d}z$ 
is the line element in a Friedmann cosmology, $\epsilon$ is the proper space-averaged
volume emissivity and $\tau_{\rm eff}$ is an effective optical depth due to absorption by the IGM. 
There is no explicit solution of equation \eqref{eq.j0} 
since it must be computed iteratively by solving the cosmological 
radiative transfer equation \citep{peebles93}.
For our analysis we use the values of $J_0(\nu,z)$ tabulated by \citet{haardt01}. 
Notice, however, that more recent calculations of the UV background 
flux \citep{bolton07,meiksin09} show that the photoionizing background predicted 
by the \citet{haardt01} model may be an underestimate at $z>5$. 

The fraction of ionized hydrogen $x \equiv n_{HII}/n_H$ varies 
according to a balance between radiative and collisional ionizations and recombinations
inside the cloud:
\begin{equation}
 \alpha_A n_e x = (\Gamma_H + \beta_H n_e) (1 -x),
\end{equation}
where $\alpha_A = 4.18\times 10^{-13}{\rm [cm^3 s^{-1}]}$ is 
the case A recombination coefficient at $T = 10^4K$ \citep{osterbrock89},
the photoionization rate $\Gamma_H(z)$ from the UV background is given by
\begin{equation}
\label{eq.gamma}
 \Gamma_H(z) = \int_{\nu_0}^{\infty} \frac{4\pi J_0(\nu,z)}{h\nu} \sigma_{\nu}(H){\rm d} \nu,
\end{equation}
and the collisional ionization rate at $T = 10^4K$, is $\beta_H = 6.22 
\times 10^{-16}{\rm [cm^{3} s^{-1}]}$\citep{cen92}.

As the UV flux penetrates the outflow, it will be attenuated by the outer 
layers of neutral hydrogen. 
The UV flux reaching an inner layer of the HI region is attenuated by this 
{\it self-shielding} process according to
\begin{equation}
\label{eq.jnu}
 J(\nu) = J_0(\nu) e^{-\tau(\nu)},
\end{equation}
where $J_0(\nu)$ is the original, un-shielded UV flux, and the optical 
depth $\tau(\nu)$ when UV photons travel a distance 
$d$ inside the HI region (coming from outside) is given by
\begin{equation}
\label{eq.taunu}
 \tau(\nu) = \sigma_{\nu}(H) \int_{R_{\rm out}}^d n_H(r) {\rm d}r.
\end{equation}

The photoionization rate is computed from the outer radius inwards. 
For each shell inside the outflow, $\Gamma_H$ is computed taking
into account the attenuation given by equations \eqref{eq.jnu} and 
\eqref{eq.taunu}, making the photoionization
rate smaller as photons penetrate inside the HI region.

We have found that the result of this calculation modifies strongly 
the outer layers of the wind, but since the number density of atoms increase 
rapidly when going inwards, the self-shielding effect effectively suppresses the 
UV radiation for the inner layers of the wind. 
For winds with low neutral hydrogen column densities the UV background 
is found to penetrate deeper into the wind, but in general 
the innermost region is left unchanged. 

In principle, we could compute the effect of the UV background on the Shell 
geometry as well. However, we do not perform this calculation since, 
in this case, the number density 
inside the outflow layers depend strongly on the physical 
thickness of the Shell, which in turn depends on the parameter $f_{th}$. 
As discussed in the previous section, this parameter is considered to have 
an arbitrary value provided that $f_{th} \gtrsim 0.9$. If we include the UV 
background in the Shell geometry, the \lya\ properties 
would depend on the value of $f_{th}$ assumed, which is an unnecessary 
complication to the model.

\section{A grid of configurations to compute the escape fraction}
\label{appendix.grid}

\galform\ typically generates samples numbering many thousands 
of galaxies brighter than a given flux limit at a number of redshifts.
The task of running the radiative transfer 
code for each galaxy is infeasible considering the time 
it takes the Monte Carlo code to reach completion, 
which varies from a few seconds up to several hours for some extreme configurations.
Therefore, this motivates the need to 
develop an alternative method to assign a \lya\ escape fraction for each 
galaxy predicted by \galform. Instead of running the radiative transfer
code to each galaxy, we construct a grid of configurations spanning the
 whole range of galaxy properties, as predicted by \galform.

The first step to construct the grid is to choose which parameters will be used. 
In principle, each outflow geometry (Wind or Shell) requires
4 input parameters from \galform: three of these, $\langle V_{\rm circ} 
\rangle, \langle R_{\rm 1/2} \rangle$ and $\langle Z_{\rm gas} \rangle$ are 
used by both geometries. In addition, $\dot{M}_{ej}$ is required in the Wind 
geometry, and $\langle M_{\rm gas} \rangle$ in the Shell geometry. 

However, a grid of models using four parameters becomes rapidly inefficient 
when trying to refine the grid. A grid with an appropriate
binning of each parameter can have as many elements as the number of 
galaxies for which the grid was constructed, and hence 
also becomes prohibitively expensive.

Therefore, we look for degeneracies in the escape fraction when using 
combinations of the input parameters 
from \galform, in order to reduce the number of parameters 
for the construction of the grid. The idea is to find a combination of parameters
which, when kept fixed while varying its individual components, 
gives the same escape fraction.

The natural choice for this is to use the column density 
$N_H$ as one parameter. \citet{neufeld90}
found that the escape fraction from a homogeneous, dusty slab is a 
function of the optical depth at the line centre $\tau_0$ and the 
optical depth of absorption $\tau_a$. Both quantities are, in turn, a function of 
the column density $N_H$. In the Shell geometry, 
$N_H \propto M_{\rm gas}/\langle R_{\rm 1/2}\rangle^2$, whereas in the 
Wind geometry $N_H \propto \dot{M}_{ej}/(\langle R_{1/2} \rangle V_{\rm exp})$.
Although promising, we find that we do not recover a constant escape fraction in 
the Wind geometry when the column density is kept fixed while 
varying its individual terms. The reason is that the expansion velocity plays 
a more complicated role when computing the escape fraction, 
with the escape fraction increasing rapidly with increasing velocity 
regardless of the other parameters of the medium.

\begin{figure}
 \centering
\includegraphics[width=8cm,angle=90]{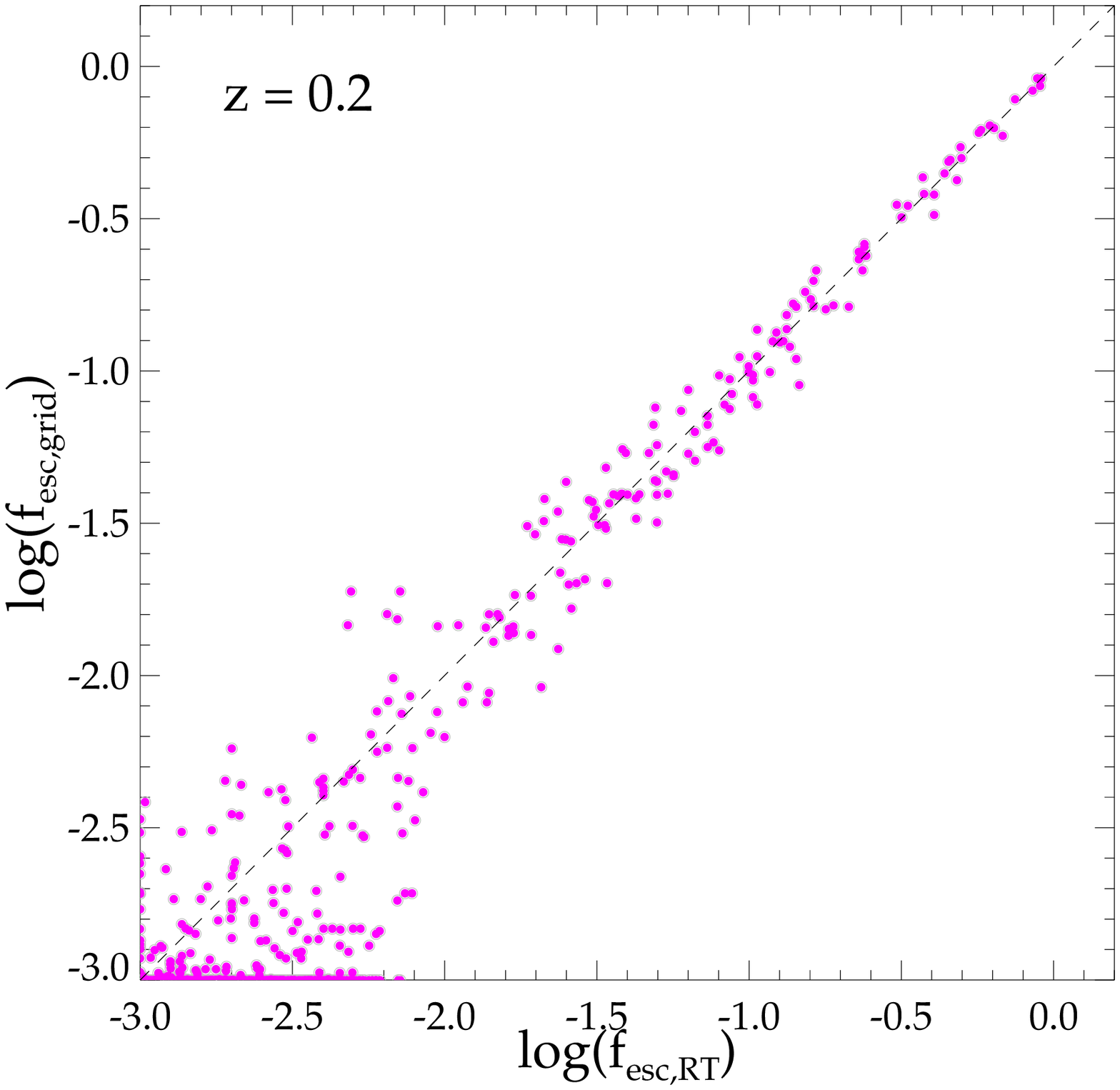}
\caption[Comparison of the escape fraction obtained using a direct calculation or interpolating from a grid]
{Comparison of the escape fraction obtained using a direct calculation 
($f_{\rm esc,RT}$) or interpolating from a grid ($f_{\rm esc,grid}$).
Points represent a subsample of $\sim 1000$ galaxies selected from \galform\ at $z= 0.2$.}
\label{fig.test_fesc}
\end{figure}

Therefore, we construct three-dimensional grids for each outflow geometry. 
We define $C_{\rm wind} \equiv \dot{M}_{\rm ej}/R_{\rm inn}$, and 
$C_{\rm shell} \equiv M_{\rm shell}/R_{\rm shell}^2$. Then, 
in the Wind geometry the parameters are $C_{\rm wind},V_{\rm exp}$ and 
$Z_{\rm gas}$, whereas in the Shell geometry we use $C_{\rm shell}, V_{\rm exp}$ and 
$Z_{\rm gas}$. We choose to cover each parameter with a 
bin size appropriate to recover the expected escape 
fraction with a reasonable accuracy when interpolating in the 
grid, but also ensuring that the number of grid elements 
to be computed is significantly smaller than the 
total number of galaxies in the sample. 

We find that, when covering each parameter in logarithmic bins of $0.1$ 
we get escape fractions that are accurate enough, and the number of 
elements of the grid we need to compute is usually a factor $\sim 20$ smaller than the 
total number of galaxies in the sample.

We fix the number of photons to run for each grid point to compute the 
escape fraction, since this will determine
the speed with which each configuration will be completed. By studying the 
resulting luminosity function of galaxies, we find
that running the code with a  maximum number of photons $N_p = 1000$ 
gives results which have converged over the range of luminosities 
observed. This means that the minimum escape fraction
we are able to compute is $\fesc = 10^{-3}$. Although there are configurations 
where \fesc\ can be lower than this, 
they are found not to contribute significantly to the luminosity functions.

Fig. \ref{fig.test_fesc} shows an example of the performance 
of the grid we use to compute the escape fraction in the Shell 
geometry using a sub-sample of galaxies from \galform\ at $z= 0.2$, 
chosen in a way to cover the entire range of intrinsic \lya\ luminosities. 
The accuracy of the grid gets progressively worse with lower
escape fractions, since these have intrinsically larger errors due to 
the constraint on the maximum number of photons used to compute \fesc.
However, as discussed previously, we found that to reproduce accurately the luminosity functions 
there is no need to reduce the size of the parameter bins or increase the number of photons
used.
\end{document}